\documentclass[aps,superscriptaddress,floats,twocolumn,epsf,prr,nofootinbib]{revtex4-2} 
\usepackage[dvipsnames]{xcolor}
\usepackage{float}
\usepackage{graphicx}
\usepackage{epstopdf}
\usepackage[colorlinks=true,urlcolor=blue,citecolor=blue,linkcolor=blue,breaklinks=true]{hyperref}
\usepackage{color}
\usepackage{colortbl}
\usepackage{setspace}
\usepackage[utf8]{inputenc}
\usepackage{amsmath}
\usepackage{placeins}
\usepackage{mathtools}
\usepackage{hyperref}
\usepackage{changes}
\usepackage{tikz-feynman}
\usepackage{tikz-feynhand}
\usepackage{amssymb}
\usepackage{bm}
\usepackage[export]{adjustbox}
\usepackage{bbold}
\usepackage{changes}
\usepackage[normalem]{ulem}
\definecolor{TUgreen}{RGB}{162, 198, 194}

\usepackage{hyperref}
\hypersetup{colorlinks=true,breaklinks,linkcolor=blue,urlcolor=blue,citecolor=blue}


\begin{document}

\title{Non-perturbative effects of short-range spatial correlations at the two-particle level}

\author{Michael Meixner}
\thanks{Both authors contributed equally.}
\affiliation{Max-Planck-Institut f{\"u}r Festk{\"o}rperforschung, Heisenbergstra{\ss}e 1, 70569 Stuttgart, Germany}

\author{Matthias Reitner}
\thanks{Both authors contributed equally.}
\affiliation{Institute of Solid State Physics, TU Wien, 1040 Vienna, Austria}

\author{Thomas Sch{\"a}fer}
\email{t.schaefer@fkf.mpg.de}
\affiliation{Max-Planck-Institut f{\"u}r Festk{\"o}rperforschung, Heisenbergstra{\ss}e 1, 70569 Stuttgart, Germany}
\affiliation{Dipartimento di Fisica, Università di Trieste, Strada Costiera 11, I-34151 Trieste, Italy}

\author{Alessandro Toschi}
\affiliation{Institute of Solid State Physics, TU Wien, 1040 Vienna, Austria}

\definecolor{Goldenrod}{cmyk}{0, 0.24, 0.85, 0.15}
\definecolor{colorB}{cmyk}{0, 0.20, 0.80, 0.07}
\definecolor{colorA}{cmyk}{0, 0.84, 0.85, 0.25}
\definecolor{colorF}{cmyk}{0, 0.1,0.1,0}
\definecolor{Shade1}{cmyk}{0.1, 0, 0.03, 0}
\definecolor{Shade2}{cmyk}{0.3, 0.05, 0.1, 0}
\definecolor{Shade3}{cmyk}{0.35, 0.15, 0, 0}
\definecolor{Shade4}{cmyk}{0.3, 0.2, 0.1, 0}

\renewcommand{\i}{{\mathrm{i}}}
\newcommand{\e}{{\mathrm{e}}}
\newcommand{\up}{\uparrow}
\newcommand{\down}{\downarrow}
\renewcommand{\d}{\mathrm{d}}
\newcommand{\T}{\mathrm{T}}
\renewcommand{\part}{\mathcal{Z}}
\newcommand{\vk}{\ensuremath{\mathbf{k}}}
\newcommand{\vq}{\ensuremath{\mathbf{q}}}
\newcommand{\vK}{\ensuremath{\mathbf{K}}}
\newcommand{\vQ}{\ensuremath{\mathbf{Q}}}
\newcommand{\vR}{\ensuremath{\mathbf{R}}}
\newcommand{\Gn}{G^{0^{-1}}}
\newcommand{\tikzx}[2]{%
    \draw (#1-0.1, #2-0.1) -- (#1+0.1, #2+0.1); 
    \draw (#1-0.1, #2+0.1) -- (#1+0.1, #2-0.1); 
}

\date{ \today }

\begin{abstract} 
By means of cellular dynamical mean-field theory (CDMFT) we study  how  short-range correlations drive the breakdown of the self-consistent perturbation theory in two-dimensional systems and the most relevant physical consequences associated to it. 
To this aim, we first derive in a structured and consistent way the Bethe-Salpeter equation (BSE) formalism at the CDMFT level in all physical channels, explicitly addressing the important aspect of the related Ward identities.
In this context, we perform systematic calculations of the BSE for the two-dimensional Hubbard model at half-filling at intermediate coupling. Our study illustrates how the divergence of a fundamental building block of the BSE in the charge channel, the two-particle irreducible vertex, systematically occurs at lower interactions than in the (purely local) DMFT case, due to short-range antiferromagnetic fluctuations. 
Further, the change of sign of the eigenvalues of the generalized charge susceptibility associated to the vertex divergences is identified as the essential prerequisite to drive, at larger interaction values, the physics of the Mott transition in two dimensions, as well as of the adjacent phase-separation instabilities.
\end{abstract}
\maketitle

\section{Introduction}

\label{sec:introduction}

The lack of an intrinsically small interaction scale in the many-electron problem poses a major challenge to the condensed matter theory, as it makes it necessary, for several physically relevant cases (e.g., Mott-metal insulator transitions, unconventional superconductivity, quantum criticality, etc), to perform calculations in parameter regimes where a many-body perturbative expansion does no longer hold.
This theoretical challenge is further exacerbated in the significant case of layered/two-dimensional systems such as cuprates and nickelates, due to the systematic enhancement of non-local spatial correlations originated by their reduced dimensionality.

In the last decade, numerous studies have been carried out, which  investigated both formal aspects \cite{Schaefer2013,Janis2014,Kozik2015,Stan2015,Rossi2015,Rossi2016,Schaefer2016c,Gunnarsson2017,Tarantino2018,Chalupa2017,Thunstrom2018,Vucicevic2018,Kim2020b,Reitner2024,Essl2024,Essl2025} and physical  consequences \cite{Gunnarsson2016,Springer2020, Melnick2020,Reitner2020,Chalupa2021,Mazitov2022,Mazitov2022b,Pelz2023,Adler2024,Kowalski2024,Reitner2025,Moghadas2025} of the breakdown of the self-consistent perturbation expansions.
In particular, through a systematic analysis on the one- and the two-particle level of basic many-electron systems (ranging from the zero-point model, the Falicov Kimball, the Hubbard atom, the Anderson impurity model, and the DMFT solution of the Hubbard model), some relevant aspects of this problems have been already unveiled. On the formal level, it has been rigorously demonstrated how the breakdown of the self-consistent perturbation theory can be signaled by a branching of the multivalued Luttinger-Ward functional \cite{Kozik2015,Schaefer2016c,Rossi2015,Stan2015,Tarantino2018,Thunstrom2018}, which occurs simultaneously \cite{Gunnarsson2017,Essl2025} with the divergence of  the two-particle 
irreducible vertex in the charge-channel \cite{Schaefer2013,Janis2014,Schaefer2016c,Chalupa2017, Vucicevic2018, Thunstrom2018, Springer2020, Pelz2023}. Aside from the specific, but quite significant algorithmic consequences \cite{Kozik2015,Vucicevic2018,Badr2024, Essl2025} of these issues, it has recently been discussed how, at the level of a purely on-site description, such formal/algorithmic  aspects of the perturbative breakdown are  linked to important  features of the correlated electron physics. In particular, they could be directly related \cite{Gunnarsson2016, Gunnarsson2017,Chalupa2021, Mazitov2022, Mazitov2022b,Adler2024} to the localization of the electronic charge and the formation of a local magnetic moment, both driven by increasing on-site electronic correlations. Even more importantly, in the purely local DMFT realm, it has been proven (analytically and numerically) how the breakdown of the perturbative theory (and its physical manifestations) constitute an essential prerequisite for the Mott-Hubbard transition \cite{Pelz2023} and the associated phase-separation \cite{Reitner2020, Reitner2024,Kowalski2024} to occur in the limit of infinite dimensions. 

The major limitation of most previous studies on this subject is their restriction\footnote{Among the very few exceptions we note the study of the parquet equation in dynamical cluster approximation \cite{Gunnarsson2016} and the CDMFT investigation of the vertex divergences in the context of a comparison with the misleading convergence of the nested cluster scheme \cite{Vucicevic2018}. In both cases, however, the main focus of the study was not the one of our paper, and the related cluster calculations were performed either sporadically for specific parameter point or for a single fixed temperature.}  to purely local/on-site  correlations.
This raises the legitimate question to what extent the conclusions drawn in the specific realm of purely local correlations can still hold in the realistic case of finite dimensional systems, where  non-local correlations in space might be no longer negligible. 
In particular, addressing this question is certainly \emph{compelling} in the important case, mentioned above, of two-dimensional (2D) systems, where spatial correlations beyond (dynamical) mean-field are expected to play a major role.

In this work, we investigate this issue,  performing a thorough study of the two-dimensional Hubbard model by means of a technique able to \emph{non-perturbatively} include \emph{short-range} non-local correlations in space
in a systematic fashion: the cellular DMFT (CDMFT) \cite{Kotliar2001,Lichtenstein2000}.
 While most of existing CDMFT studies are performed at the one-particle level, by computing one-particle Green's functions and self-energies, crucial information about the perturbative breakdown, as mentioned above, is typically encoded in the two-particle generalized susceptibilities and vertices, related one another through the Bethe-Salpeter equation (BSE). 
For this reason, before showing and analyzing our numerical results, we will present a consistent derivation of the explicit BSE expressions valid for the CDMFT in all physical scattering channels and discuss in detail some of their most fundamental properties. 
This specific theoretical and numerical framework will allow us to unveil how the non-local structure of the two-particle vertex function affects the breakdown of perturbation theory, and, at the same time, how it may drive, in general, the onset of thermodynamic instabilities at intermediate-to-strong coupling. Specifically, when including short range correlations, the Mott metal-insulator  transition (MIT) occurs at significantly lower interactions than in the DMFT. We trace back this change in the thermodynamic instability of the charge response in CDMFT to short range nearest-neighbor spin correlations.

The manuscript is organized as follows: In Sec.~\ref{sec:modelandmethods}, after providing some basic details about the model and the CDMFT scheme considered, we explicitly introduce the  corresponding Bethe-Salpeter equations and real-space Ward identities in {\sl all} the relevant (i.e., charge, spin and particle-particle) channels. In Sec.~\ref{sec:resultsVertex}, we illustrate and discuss our numerical results on the non-perturbative breakdown, which are encoded in a relevant four-point quantity of CDMFT, namely the two-particle irreducible charge vertex of the impurity cluster.
Sec.~\ref{sec:resultsMIT} then analyzes, on the two-particle level, the occurrence of thermodynamic instabilities at the MIT, unveiling their link to the underlying non-perturbative nature of the BSE vertices. Conclusions and outlook are eventually presented in Sec.~\ref{sec:conclusion}.

\section{Model and Methods}
\label{sec:modelandmethods}
In this work, we study the paramagnetic (SU(2)-symmetric) phase of the particle-hole symmetric, non-frustrated two-dimensional Hubbard model \cite{Hubbard1963,Hubbard1964,Gutzwiller1963,Kanamori1963,Qin2022,Arovas2022} on the square lattice 
\begin{equation} \label{eq:Hamiltonian}
H=-t\sum_{\langle \mathrm{i,j} \rangle,\sigma}c^\dagger_{\mathrm{i},\sigma}c_{\mathrm{j},\sigma}+U\sum_{\mathrm{i}}n_{\mathrm{i},\uparrow}n_{\mathrm{i},\downarrow}-\frac{U}{2} \sum_{\mathrm{i},\sigma}n_{\mathrm{i},\sigma},
\end{equation}
where $c_{\mathrm{i},\sigma}^\dagger$ ($c_{\mathrm{i},\sigma}$) represent the fermionic creation (annihilation) operators of an electron with spin $\sigma\in\{\uparrow,\downarrow\}$ on site $\mathrm{i}$, $t$ the nearest-neighbor (n) hopping amplitude and $U$ the Coulomb repulsion. We set $t=1$ throughout the paper and give all energies in this unit.

\subsection{Cellular dynamical mean-field theory}
\begin{figure}

    \centering
\begin{tikzpicture}[scale=0.7]
        \begin{feynhand}
            \draw [dashed,gray] (-0.75,-1.7) -- (-0.75,-0.75);
            \draw [dashed,gray] (-1.7,-0.75) -- (-0.75,-0.75);
            \draw [dashed,gray] (-1.7,2.25) -- (-0.75,2.25);
            \draw [dashed,gray] (-0.75,3.2) -- (-0.75,2.25);          
            \draw [dashed,gray] (2.25,3.2) -- (2.25,2.25);
            \draw [dashed,gray] (3.2,2.25) -- (2.25,2.25);
            \draw [dashed,gray] (2.25,-1.7) -- (2.25,-0.75);
            \draw [dashed,gray] (3.2,-0.75) -- (2.25,-0.75);         
            \draw [gray] (-0.75,-0.75) -- (-0.75,2.25);
            \draw [gray] (-0.75,2.25) -- (2.25,2.25);
            \draw [gray] (2.25,2.25) -- (2.25,-0.75);
            \draw [gray] (2.25,-0.75) -- (-0.75,-0.75);
            \draw[fill=colorA] (0,0) circle (0.2);
            \draw[fill=colorA] (3,0) circle (0.2);
            \draw[fill=colorA] (0,3) circle (0.2);
            \draw[fill=colorA] (3,3) circle (0.2);
            \draw[fill=colorB] (-1.5,-1.5) circle (0.2);
            \draw[fill=colorB] (-1.5,1.5) circle (0.2);
            \draw[fill=colorB] (-1.5,0) circle (0.2);
            \draw[fill=colorB] (1.5,0) circle (0.2);
            \draw[fill=colorB] (0,-1.5) circle (0.2);
            \draw[fill=colorB] (0,1.5) circle (0.2);
            \draw[fill=colorB] (3,1.5) circle (0.2);
            \draw[fill=colorB] (3,-1.5) circle (0.2);
            \draw[fill=colorB] (1.5,-1.5) circle (0.2);
            \draw[fill=colorB] (1.5,1.5) circle (0.2);
            \draw[fill=colorB] (-1.5,3) circle (0.2);
            \draw[fill=colorB] (1.5,3) circle (0.2);
            \node at (0,-0.4) {$0$};
            \node at (1.5,-0.4) {$1$};
            \node at (1.5,1.1) {$3$};
            \node at (0,1.1) {$2$};
            \propag[plain, with arrow=0.99] (1.7,0.2) to [quarter left, edge label =$\quad \quad \quad t \, \e^{\mathrm{i} 2 k_x }$] (2.8,0.2);
            \propag[plain, with arrow=0.99] (1.7,-0.2) to [quarter left, edge label = $t \, \e^{-\mathrm{i}2 k_y}$] (1.7,-1.3);
            \draw[->] (-1.9,-1.9) -- (-1.9,-0.7);
            \draw[->] (-1.9,-1.9) -- (-0.7,-1.9);
        
            \node at (-2.2,-1.) {$y$};
            \node at (-1.,-2.2) {$x$};
            \fill[fill=Shade2] (4.5, -1.7) rectangle (5.825,-0.475);
            \fill[fill=Shade2] (4.5, 3.2) rectangle (5.825,1.975);
            \fill[fill=Shade2] (9.4, 3.2) rectangle (8.175, 1.975);
            \fill[fill=Shade2] (9.4, -1.7) rectangle (8.175,-0.475);
            \fill[fill=Shade3] (5.825, -1.7) rectangle (8.175,-0.475);
            \fill[fill=Shade3] (5.825, 1.975) rectangle (8.175,3.2);
            \fill[fill=Shade4] (5.825, -0.475) rectangle (4.5,1.975);
            \fill[fill=Shade4] (8.175, -0.475) rectangle (9.4,1.975);
            \fill[fill=Shade1] (5.825, -0.475) rectangle (8.175,1.975);
            \draw [dashed,gray] (4.5,-1.7) -- (4.5,3.2);
            \draw [dashed,gray] (4.5,-1.7) -- (9.4,-1.7);
            \draw [dashed,gray] (4.5,3.2)  -- (9.4,3.2);
            \draw [dashed,gray] (9.4,-1.7) -- (9.4,3.2);
            \draw [gray] (5.825,-0.475) -- (5.825,1.975);
            \draw [gray] (5.825,1.975) -- (8.175,1.975);
            \draw [gray] (8.175,1.975) -- (8.175,-0.475);
            \draw [gray] (8.175,-0.475) -- (5.825,-0.475);
            \node[font=\sffamily] at (7,0.75){\textbf{x}};
            \node[font=\sffamily] at (7,3.2){\textbf{x}};
            \node[font=\sffamily] at (9.4,3.2){\textbf{x}};
            \node[font=\sffamily] at (9.4,0.75){\textbf{x}};
            \propag[plain, with arrow=0.92] (7,0.75) to [ edge label = $Q$] (9.4,0.75);
            \propag[plain, with arrow=0.92] (9.4,0.75) to [ edge label = $\tilde{q}$] (8.5,-0.2);
            
            \draw[->] (4.5,-1.7) -- (4.5,-0.5);
            \draw[->] (4.5,-1.7) -- (5.7,-1.7);
            \node at (4.2,-1.) {$k_y$};
            \node at (5.4,-2.2) {$k_x$};

            \node at (4.1,-1.7) {$-\pi$};
            \node at (4.2,3.1) {$\pi$};
            \node at (4.5,-2) {$-\pi$};
            \node at (9.3,-2) {$\pi$};
            \node at (5.425,-0.345) {$-\frac{\pi}{2}$};
            \node at (5.525,1.875) {$\frac{\pi}{2}$};
            \node at (5.845,-0.775) {$-\frac{\pi}{2}$};
            \node at (8.075,-0.775) {$\frac{\pi}{2}$};
            \node at (6.96,-0.2){RBZ};
        \end{feynhand}
\end{tikzpicture}
\caption{The unit cell of the superlattice with a four-atomic basis on the left-hand side. The four sites are labelled from $0$ to $3$. Red indicates specific geometrically equivalent sites in different unit cells. On the right-hand side, the reciprocal superlattice basis is indicated by x-markers, where the shading gives the reduced Brillouin zones ($\mathrm{RBZ}$) in comparison to the Brillouin zone of the initial one-atomic basis Hubbard model \cite{Ayral2017}. Any vector in the initial Brillouin zone can be decomposed by the closest reciprocal lattice vector $\vQ$ and a vector $\tilde{\vq} \in \mathrm{RBZ}$. }
\label{fig:Sektch}
\end{figure}
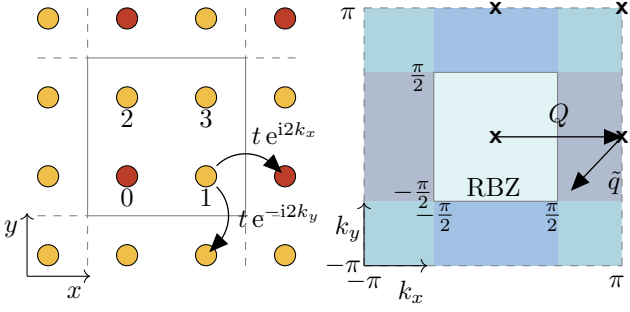

We compute the single particle observables and Green's function of the Hubbard model via the cellular dynamical mean-field theory (CDMFT) \cite{Kotliar2001,Lichtenstein2001}, a cluster extension \cite{Maier2005,Rohringer2018} of the dynamical mean-field theory \cite{Metzner1989,Georges1992a,Georges1996}.  As indicated in Fig.~\ref{fig:Sektch}, the original lattice problem is understood as a superlattice of $N_c=2\times2$-site unit cells, where the circles indicate lattice sites and the red circles sketch the new periodicity of the superlattice. The $2\times2$ sites in a unit cell form a four-atomic basis. The right-hand side of Fig.~\ref{fig:Sektch} sketches the corresponding reduced Brillouin zone (RBZ). CDMFT then maps the superlattice onto an cluster impurity problem, where a $N_c$-site cluster is solved exactly, while the remaining system is represented by a frequency-dependent fermionic mean-field $\mathcal{G}$. To that end, a Dyson equation on the superlattice is formulated \cite{Ayral2017}, where site indices $\mathrm{i,j}$ are treated as orbital-like quantum numbers, leading to a superlattice Green function \cite{Klett2020}
\begin{equation}
\label{Eq:G}
    G^{\mathrm{ij}}_{\vk\sigma}(\nu)=\left[\left[(\i\nu+\mu) \mathbb{1}-\varepsilon_{\vk}-\Sigma_{\sigma}(\nu)\right]^{-1}\right]^{\mathrm{ij}},
\end{equation}
where the matrix $\varepsilon_{\vk}$ ($\mathbf{k} \in$ RBZ) introduces the hopping processes in the unit cell and over the edge of the unit cell as in Fig.~\ref{fig:Sektch} with the hopping matrix 
\begin{equation}
\begin{split}
\label{eq:hopping}
\varepsilon_{\vk}&=
    -t\begin{pmatrix}
        0& 1+\mathrm{e}^{\i 2 k_x}&1+\mathrm{e}^{+\i 2 k_y}&0\\
        1+\mathrm{e}^{-\i 2 k_x}&0&0&1+\mathrm{e}^{\i 2 k_y}\\
        1+\mathrm{e}^{-\i 2 k_y}&0&0&1+\mathrm{e}^{\i 2 k_x}\\
        0&1+\mathrm{e}^{-\i 2 k_y}&1+\mathrm{e}^{-\i 2 k_x}&0\\
    \end{pmatrix},
\end{split}
\end{equation}
and $\Sigma^{\mathrm{ij}}_{\sigma}(\nu)$ represents the cluster impurity self-energy matrix. To map the lattice onto the impurity problem, after integrating out the momentum-dependent degrees of freedom of the lattice, one requires that:
\begin{equation}
    G^{\mathrm{ij}}_{\mathrm{loc},\sigma}(\nu)=\int_{\vk\in\mathrm{RBZ}}G^{\mathrm{ij}}_{\vk \sigma}(\nu)\overset{!}{=}G^{\mathrm{ij}}_{\sigma}(\nu),
    \label{eq:approx_1P_Gf}
\end{equation} with  $G_\sigma^{\mathrm{ij}}(\tau)=-\langle T_\tau c_{\mathrm{i}\sigma}(\tau)c^\dagger_{\mathrm{j}\sigma}\rangle$ being the impurity Green's function and  $T_\tau$  the imaginary-time ordering operator, respectively, see App.~\ref{App:Gf}. Then, by exploiting the inverse Dyson equation for the impurity cluster, one can derive the explicit expression for the Weiss field of CDMFT:
\begin{equation}
\label{Eq:Weissfield}
   \mathcal{G}^{-1}_{\sigma}(\nu)=\left[\int_{\vk\in\mathrm{RBZ}}\mathrm{d}\vk\,G_{\vk\sigma}\right]^{-1}(\nu)+\Sigma_{\sigma}(\nu),
\end{equation} where cluster indices have been omitted here for readability. Throughout the manuscript we enforce the paramagnetic solution by setting $\mathcal{G}_{\uparrow} = \mathcal{G}_{\downarrow}$. From the updated cluster model, a new self-energy can be computed by utilizing an impurity solver. This cycle is iterated until convergence is reached.
Further details of the algorithm are given in Ref.~\cite{Maier2005,Meixner2024}. In our case, the $N_c=2\times 2$ auxiliary cluster model is solved via the interaction expansion continuous-time quantum Monte-Carlo (CT-INT) solver \cite{Rubtsov2005,Gull2008a} provided as an application of the TRIQS library \cite{TRIQS}.

\subsection{Bethe-Salpeter equation for CDMFT}
CDMFT approximates the superlattice self-energy $\Sigma$ as the one locally defined on the cluster (``cluster-local'') by requiring Eq.~(\ref{eq:approx_1P_Gf}). Thus, $\Sigma$ is constant for $\vk\in\mathrm{RBZ}$:
\begin{equation}
    \Sigma^{\mathrm{ij}}_{\vk\sigma}(\nu)=\Sigma^{\mathrm{ij}}_{\sigma}(\nu).
    \label{eq:approx_1P}
\end{equation}
In the action formalism, this approximation is equivalent to the assumption that both the superlattice and the cluster impurity problem share the same Baym-Kadanoff functional $\Phi$ \cite{Baym1962}, which ensures  thermodynamic consistency. Furthermore, due to its conserving nature ~\cite{Hettler2000,Potthoff2003,Senechal2004Book}, our one- and two-particle quantities fulfill Ward identities, see Sec.~\ref{Sec:Benchmark}.

For this $\Phi$-derivable approximation, the self-energy is the first functional derivative with respect to the Green function $\Sigma^{\mathrm{ji}}_{\sigma}(\nu)=(1/T)\delta\Phi/\delta G^{\mathrm{ij}}_{\sigma}(\nu)$ for a given temperature $T$. The corresponding two-particle quantity is the irreducible vertex $\Gamma^{\mathrm{ijhl}}_{\sigma\sigma'}(\nu,\nu')=(1/T) \delta\Sigma^{\mathrm{ji}}_{\sigma}(\nu)/\delta G^{\mathrm{hl}}_{\sigma'}(\nu')$ on the cluster impurity, which enters the Bethe-Salpeter equation (BSE) of the two-particle superlattice quantities [see Eq.~(\ref{Eq:Chi_c_q}) below]. The corresponding approximation of the two-particle irreducible vertex $\Gamma$ is referred to as the cluster-local approximation \cite{Potthoff2018} on the two-particle level \cite{Georges1996,Jarrell1992,Musshoff2021}.

In practice, the irreducible vertex  $\Gamma^{\mathrm{ijhl}}_{\sigma\sigma'}(\nu,\nu')$ is calculated from the two-particle generalized susceptibility $\chi^{\mathrm{ijhl}}_{\sigma\sigma'}$ on the cluster impurity via the respective (inverse) cluster BSE. While in this work, we focus on the particle-hole sector, in the following, we state the BSEs in both particle-hole (ph) and particle-particle (pp) channels, since an explicit expression for the latter is, to the best our knowledge, so far missing in the literature for real-space clusters.

\paragraph{The particle-hole sector: Charge and magnetic channels.}

We define the two-particle generalized susceptibility for the ph channel following Refs.~\cite{Bickers04,Tagliavini2018,Rohringer2012}:

\begin{equation}
\begin{split}
\chi^{\alpha \beta}_{\mathrm{ph},\sigma\sigma'}&(\omega)
\coloneq\chi^{\mathrm{ijhl}}_{\mathrm{ph},\sigma\sigma'}(\omega,\nu,\nu')\\=&\int_0^\beta \mathrm{d}\tau_1 \mathrm{d}\tau_2 \mathrm{d}\tau_3  \, \mathrm{e}^{-\i\nu\tau_1} \mathrm{e}^{\i(\omega+\nu)\tau_2}\mathrm{e}^{-\i(\omega+\nu')\tau_3} \\&\left[\langle T_{\tau}c_{\mathrm{i},\sigma}^\dagger (\tau_1)c_{\mathrm{j},\sigma}(\tau_2)c^\dagger_{\mathrm{h},\sigma'}(\tau_3)c_{l,\sigma'}(0)\rangle \right]\\&-\frac{1}{T}\delta(\omega)G^{\mathrm{ji}}_{\sigma}(\nu)G^{\mathrm{lh}}_{\sigma'}(\nu'),
\end{split}
\end{equation}
where we introduce a left- and right-sided multi-index notation by the Greek letters $\alpha=(i,j,\nu)$, $\beta=(h,l,\nu')$ \cite{Musshoff2021}. The generalized particle-hole susceptibility can be expressed as Feynman diagrams in the following way: \begin{equation}
    \begin{split}
&\chi^{\textcolor{Cerulean}{\alpha},\textcolor{BrickRed}{\beta}}_{\mathrm{ph},\sigma\sigma'}(\omega)\\&=\chi^{\mathrm{ijhl}}_{\mathrm{ph},\sigma\sigma'}(\omega,\nu,\nu') \\&=
    \begin{tikzpicture}[baseline=0.cm]
        \begin{feynhand}
            \node at (-1.4,1) {$\mathrm{j},\sigma,\nu+\omega$};
            \node at (-1.4,-0.8) {$\mathrm{i},\sigma,\nu$};
            \node at (1.4,-0.8) {$\mathrm{l},\sigma',\nu'$};
            \node at (1.4,1) {$\mathrm{h},\sigma',\nu'+\omega$};
            \vertex (a) at (-1.2,0.7);
            \vertex (b) at (-1.2,-0.5);
            \vertex (alpha) at (-1.2,0.1) {$\mathbf{\textcolor{Cerulean}{\alpha}}$};
            \vertex (c) at (1.2,-0.5); 
            \vertex (d) at (1.2,0.7); 
            \vertex (beta) at (1.2,0.1) {$\mathbf{\textcolor{BrickRed}{\beta}}$};
            \vertex (u) at (-0.6,0.7);
            \vertex (v) at (-0.6,-0.5); 
            \vertex (w) at (0.6,-0.5); 
            \vertex (x) at (0.6,0.7); 
            \fill[TUgreen] (v) rectangle (x);
            \node at (0,0.1) {$G^{(2)}$};
            \propag [fer] (a) to (u);
            \propag [fer] (x) to (d);
            \propag [antfer] (b) to (v);
            \propag [antfer] (w) to (c);
            \propag (u) to (x);
            \propag (v) to (w);
        \end{feynhand}
    \end{tikzpicture}
    -
    \begin{tikzpicture}[baseline=0.cm]
        \begin{feynhand}
            \node at (-1,0.7) {$\mathrm{j}$};
            \node at (-1,-0.5) {$\mathrm{i}$};
            \node at (1,-0.5) {$\mathrm{l}$};
            \node at (1,0.7) {$\mathrm{h}$};
            \node at (0,-0.8) {$1/T \delta(\omega)$};
            \vertex (a) at (-0.8,0.7);
            \vertex (u) at (0,0.1);
            \vertex (b) at (-0.8,-0.5); 
            \vertex (c) at (0.8,-0.5); 
            \vertex (d) at (0.8,0.7); 
            \propag [fer] (c) to (d);
            \propag [fer] (a) to (b);
        \end{feynhand}
    \end{tikzpicture}, \\
\end{split}
 \end{equation} where the left-hand side indices are denoted $\alpha$ and the right-hand side indices $\beta$, respectively. For the rest of this work, we set $\i\omega=0$, considering only the static case, and omit this variable.

In the $\mathrm{SU}(2)$-symmetric case considered here, the BSE can be separated into the charge ($\mathrm{ch}$) and spin ($\mathrm{sp}$) channels \cite{Bickers2004,Rohringer2012}. The BSE for the generalized susceptibility on the impurity cluster $\chi_{\mathrm{\mathrm{ch}/\mathrm{sp}}}=\chi_{\mathrm{ph},\uparrow\uparrow}\pm\chi_{\mathrm{ph},\uparrow\downarrow}$ reads \cite{Musshoff2021}:
\begin{equation}
\label{Eq:BSE}
\begin{split}
    \chi&_{\mathrm{ch}/\mathrm{sp}}^{\alpha,\beta}=\chi_{0}^{\alpha,\beta}-T^2\chi_{0}^{\alpha,\gamma}\Gamma^{\overline{\gamma},\overline{\delta}}_{\mathrm{ch}/\mathrm{sp}}\chi^{\delta,\beta}_{\mathrm{ch}/\mathrm{sp}},\\
\end{split}
\end{equation}
where repeated indices are to be summed over, $\Gamma^{\overline{\gamma},\overline{\delta}}_{\mathrm{ch}/\mathrm{sp}}$ represents the respective two-particle irreducible vertex for left- and right sided multi-indices $\overline{\gamma},\overline{\delta}$ with pairwise swapped cluster indices in comparison to $\gamma,\delta$, and 
\begin{equation}
\label{eq:loc_bubble}
\chi_{0}^{\alpha,\beta}(\omega)\coloneq\chi_{0}^{\mathrm{ijhl}}(\omega,\nu)=-\frac{1}{T} G^{\mathrm{li}}(\nu)G^{\mathrm{jh}}(\nu+\omega)\delta_{\nu,\nu'}
\end{equation}
is the bubble contribution to the generalized susceptibility in the ph-channel \cite{vanLoon2024-2}. We can write the BSE Eq.~(\ref{Eq:BSE}) in matrix representation, which can be then inverted to obtain \cite{Vucicevic2018} an explicit expression for
\begin{equation}
\label{Eq:Gamma}
\begin{split}
\Gamma&_{\mathrm{ch}/\mathrm{sp}}^{\overline{\alpha},\overline{\beta}}=\frac{1}{T^2}\left( \left[\chi^{-1}_{\mathrm{ch}/\mathrm{sp}}\right]^{\alpha,\beta}-\left[\chi^{-1}_{0}\right]^{\alpha,\beta}\right).
\end{split}
\end{equation}
Within a cluster-local approximation \cite{Georges1996,Jarrell1992,Musshoff2021}, the momentum dependent charge/spin response of the lattice $\chi_{\mathrm{ch}/\mathrm{sp}}^{\vq,\alpha,\beta}$ can then be obtained from the following lattice BSE
\begin{equation}
\label{Eq:Chi_c_q}
\begin{split}
    \chi&_{\mathrm{ch}/\mathrm{sp}}^{\vq,\alpha,\beta}=\left( T^2\Gamma_{\mathrm{ch}/\mathrm{sp}}^{\overline{\alpha},\overline{\beta}}+\left[\chi^{\vq,\alpha,\beta}_{0}\right]^{-1}\right)^{-1},
\end{split}
\end{equation}
with the definition of the lattice bubble
\begin{equation}
\label{eq:latt_bubble}
    \chi^{\vq,\alpha,\beta}_{0}(\omega)=-\frac{\delta_{\nu\nu'}}{T N_\vk}\sum_{\vk}G^{\mathrm{li}}_{\vk}(\nu)G^{\mathrm{jh}}_{\vk+\vq}(\nu+\omega),
\end{equation}
where the fermionic momentum sum takes the cluster-locality of $\Gamma$ into account. Hitherto, we have formulated all BSEs on the real-space superlattice.
Here, we should recall that the mapping of the generalized susceptibilities defined on the cluster and the RBZ $\chi^{\vq,\mathrm{ijhl}}_r(\nu,\nu',\omega)$ onto the reciprocal lattice $\chi_{r,\vK,\vK'}^{\vq,\vQ}(\nu,\nu',\omega)$ \cite{Potthoff2018} via a Fourier transformation \cite{Senechal2011}, implicitly assumes an equivalence of each atom, which is strictly speaking an approximation, as the cluster Green's function is not translational invariant.

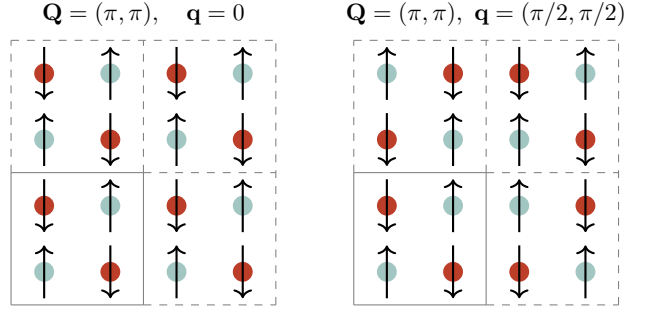
\begin{figure}
\centering
\begin{tikzpicture}[scale=0.88]
        \begin{feynhand}
            \node[scale=0.9] at (2,4.4) {$\vQ=(\pi,\pi),\quad\vq=0$};
            \draw [dashed,gray] (0,2) -- (0,4);
            \draw [dashed,gray] (2,0) -- (4,0);
            \draw [dashed,gray] (4,0) -- (4,4);
            \draw [dashed,gray] (0,4) -- (4,4);
            \draw [dashed,gray] (2,2) -- (2,4);
            \draw [dashed,gray] (2,2) -- (4,2);
            \draw [gray] (0,0) -- (0,2);
            \draw [gray] (0,0) -- (2,0);
            \draw [gray] (2,0) -- (2,2);
            \draw [gray] (0,2) -- (2,2);
            \def\al{0.8}
            \def\circsize{0.15}
            \draw[draw=none,fill=TUgreen] (0.5,0.5) circle (\circsize);
            \draw[draw=none,fill=colorA] (0.5,1.5) circle (\circsize);
            \draw[draw=none,fill=TUgreen] (0.5,2.5) circle (\circsize);
            \draw[draw=none,fill=colorA] (0.5,3.5) circle (\circsize);
            \draw[draw=none,fill=colorA] (1.5,0.5) circle (\circsize);
            \draw[draw=none,fill=TUgreen] (1.5,1.5) circle (\circsize);
            \draw[draw=none,fill=colorA] (1.5,2.5) circle (\circsize);
            \draw[draw=none,fill=TUgreen] (1.5,3.5) circle (\circsize);
            \draw[draw=none,fill=TUgreen] (2.5,0.5) circle (\circsize);
            \draw[draw=none,fill=colorA] (2.5,1.5) circle (\circsize);
            \draw[draw=none,fill=TUgreen] (2.5,2.5) circle (\circsize);
            \draw[draw=none,fill=colorA] (2.5,3.5) circle (\circsize);
            \draw[draw=none,fill=colorA] (3.5,0.5) circle (\circsize);
            \draw[draw=none,fill=TUgreen] (3.5,1.5) circle (\circsize);
            \draw[draw=none,fill=colorA] (3.5,2.5) circle (\circsize);
            \draw[draw=none,fill=TUgreen] (3.5,3.5) circle (\circsize);
            \draw[->, thick] (0.5,0.5 - (\al/2, 0) -- (0.5,0.5 +(\al/2, 0);
            \draw[->, thick] (1.5,0.5 + (\al/2, 0) -- (1.5,0.5 -(\al/2, 0);
            \draw[->, thick] (1.5,1.5 - (\al/2, 0) -- (1.5,1.5 +(\al/2, 0);
            \draw[->, thick] (0.5,1.5 + (\al/2, 0) -- (0.5,1.5 -(\al/2, 0);
            \draw[->, thick] (2.5,0.5 - (\al/2, 0) -- (2.5,0.5 +(\al/2, 0);
            \draw[->, thick] (3.5,0.5 + (\al/2, 0) -- (3.5,0.5 -(\al/2, 0);
            \draw[->, thick] (3.5,1.5 - (\al/2, 0) -- (3.5,1.5 +(\al/2, 0);
            \draw[->, thick] (2.5,1.5 + (\al/2, 0) -- (2.5,1.5 -(\al/2, 0);
            \draw[->, thick] (0.5,2.5 - (\al/2, 0) -- (0.5,2.5 +(\al/2, 0);
            \draw[->, thick] (1.5,2.5 + (\al/2, 0) -- (1.5,2.5 -(\al/2, 0);
            \draw[->, thick] (1.5,3.5 - (\al/2, 0) -- (1.5,3.5 +(\al/2, 0);
            \draw[->, thick] (0.5,3.5 + (\al/2, 0) -- (0.5,3.5 -(\al/2, 0);
            \draw[->, thick] (2.5,2.5 - (\al/2, 0) -- (2.5,2.5 +(\al/2, 0);
            \draw[->, thick] (3.5,2.5 + (\al/2, 0) -- (3.5,2.5 -(\al/2, 0);
            \draw[->, thick] (3.5,3.5 - (\al/2, 0) -- (3.5,3.5 +(\al/2, 0);
            \draw[->, thick] (2.5,3.5 + (\al/2, 0) -- (2.5,3.5 -(\al/2, 0);
        \end{feynhand}
\end{tikzpicture}$\qquad$
\begin{tikzpicture}[scale=0.88]
        \begin{feynhand}
            \node[scale=0.9] at (2,4.4) {$\vQ=(\pi,\pi),\,\,\vq=\left(\pi/2,\pi/2\right)$};
            \draw [dashed,gray] (0,2) -- (0,4);
            \draw [dashed,gray] (2,0) -- (4,0);
            \draw [dashed,gray] (4,0) -- (4,4);
            \draw [dashed,gray] (0,4) -- (4,4);
            \draw [dashed,gray] (2,2) -- (2,4);
            \draw [dashed,gray] (2,2) -- (4,2);
            \draw [gray] (0,0) -- (0,2);
            \draw [gray] (0,0) -- (2,0);
            \draw [gray] (2,0) -- (2,2);
            \draw [gray] (0,2) -- (2,2);
            \def\al{0.8}
            \def\circsize{0.15}
            \draw[draw=none,fill=TUgreen] (0.5,0.5) circle (\circsize);
            \draw[draw=none,fill=colorA] (0.5,1.5) circle (\circsize);
            \draw[draw=none,fill=colorA] (0.5,2.5) circle (\circsize);
            \draw[draw=none,fill=TUgreen] (0.5,3.5) circle (\circsize);
            \draw[draw=none,fill=colorA] (1.5,0.5) circle (\circsize);
            \draw[draw=none,fill=TUgreen] (1.5,1.5) circle (\circsize);
            \draw[draw=none,fill=TUgreen] (1.5,2.5) circle (\circsize);
            \draw[draw=none,fill=colorA] (1.5,3.5) circle (\circsize);
            \draw[draw=none,fill=colorA] (2.5,0.5) circle (\circsize);
            \draw[draw=none,fill=TUgreen] (2.5,1.5) circle (\circsize);
            \draw[draw=none,fill=TUgreen] (2.5,2.5) circle (\circsize);
            \draw[draw=none,fill=colorA] (2.5,3.5) circle (\circsize);
            \draw[draw=none,fill=TUgreen] (3.5,0.5) circle (\circsize);
            \draw[draw=none,fill=colorA] (3.5,1.5) circle (\circsize);
            \draw[draw=none,fill=colorA] (3.5,2.5) circle (\circsize);
            \draw[draw=none,fill=TUgreen] (3.5,3.5) circle (\circsize);
            \draw[->, thick] (0.5,0.5 - (\al/2, 0) -- (0.5,0.5 +(\al/2, 0);
            \draw[->, thick] (1.5,0.5 + (\al/2, 0) -- (1.5,0.5 -(\al/2, 0);
            \draw[->, thick] (1.5,1.5 - (\al/2, 0) -- (1.5,1.5 +(\al/2, 0);
            \draw[->, thick] (0.5,1.5 + (\al/2, 0) -- (0.5,1.5 -(\al/2, 0);
            \draw[->, thick] (2.5,0.5 + (\al/2, 0) -- (2.5,0.5 -(\al/2, 0);
            \draw[->, thick] (3.5,0.5 - (\al/2, 0) -- (3.5,0.5 +(\al/2, 0);
            \draw[->, thick] (3.5,1.5 + (\al/2, 0) -- (3.5,1.5 -(\al/2, 0);
            \draw[->, thick] (2.5,1.5 - (\al/2, 0) -- (2.5,1.5 +(\al/2, 0);
            \draw[->, thick] (0.5,2.5 + (\al/2, 0) -- (0.5,2.5 -(\al/2, 0);
            \draw[->, thick] (1.5,2.5 - (\al/2, 0) -- (1.5,2.5 +(\al/2, 0);
            \draw[->, thick] (1.5,3.5 + (\al/2, 0) -- (1.5,3.5 -(\al/2, 0);
            \draw[->, thick] (0.5,3.5 - (\al/2, 0) -- (0.5,3.5 +(\al/2, 0);
            \draw[->, thick] (2.5,2.5 - (\al/2, 0) -- (2.5,2.5 +(\al/2, 0);
            \draw[->, thick] (3.5,2.5 + (\al/2, 0) -- (3.5,2.5 -(\al/2, 0);
            \draw[->, thick] (3.5,3.5 - (\al/2, 0) -- (3.5,3.5 +(\al/2, 0);
            \draw[->, thick] (2.5,3.5 + (\al/2, 0) -- (2.5,3.5 -(\al/2, 0);
        \end{feynhand}
\end{tikzpicture}
\caption{Examples of lattice modulations encoded in the CDMFT scheme: The left-hand side plot displays commensurate [$\vq=(0,0)$] anti-ferromagnetism [$\vQ=(\pi,\pi)$], while the right-hand side displays different superlattice variations [$\vq=(\pi/2,\pi/2)$], from the same cluster modulations [$\vQ=(\pi,\pi)$].}
\label{fig:AFM-Sketch}
\end{figure}

For single-particle quantities on a $2 \times 2$ cluster like the self-energy, this represents, however, a common periodization scheme, see e.g.~Refs.~\cite{Parcollet2004,Maier2005}. 
In a similar spirit, we can employ the following Fourier transform \cite{Senechal2011} for two-particle quantities by mapping the superlattice expressions discussed above to the reciprocal vectors \cite{Potthoff2018}:
\begin{equation}
\begin{split} \label{Eq:ToReciprocal_ph}
    \chi_{r,\vK,\vK'}^{\vq,\vQ}(\nu,\nu',\omega)&=\frac{1}{N_c^2}\sum_{\mathrm{ijhl}}  \e^{\i(\vR_\mathrm{i}-\vR_\mathrm{j})\vK} \e^{\i(\vR_\mathrm{h}-\vR_\mathrm{l})\vK'}\\
    & \times \e^{\i(\vR_\mathrm{h}-\vR_\mathrm{j})\vQ}\chi_r^{\vq,\mathrm{ijhl}}(\nu,\nu',\omega),
\end{split}
\end{equation}
 where $\vK$, $\vK'$ and $\vQ$ run over \{$(0,0)$, $(\pi,0)$, $(0,\pi)$, $(\pi,\pi)$\} and  $N_c=2\times2$. Note that, in contrast to DCA~\cite{Maier2005}, the momentum conservation of the reciprocal vectors $\vK,\vK'$ is \emph{not} enforced for the momentum bubble $\chi_{0,\vK,\vK'}^{\vq,\vQ} \not\propto \delta_{\vK,\vK'}$, however, the Fourier transform diagonalizes the dispersion Eq.~(\ref{eq:hopping}) and hence, for the disconnected cluster-local (impurity) bubble we have $\chi_{0,\vK,\vK'}^{\vQ}\propto \delta_{\vK,\vK'}$.
 The physical response function in the specific channel on the reciprocal lattice can then be obtained as
 \begin{equation} \label{eq:chiqQ}    \chi_{\mathrm{ch}/\mathrm{sp}}^{\vq,\vQ}(\omega )=\frac{T^2}{N_c}\sum_{\vK,\vK'}\sum_{\nu,\nu'}\chi_{\mathrm{ch}/\mathrm{sp},\vK,\vK'}^{\vq,\vQ}(\nu,\nu',\omega),
 \end{equation}
 where $\vQ$ indicates the modulation of the specific response inside the cluster, while $\vq$ indicates the phase vector of the cluster in the reciprocal superlattice,corresponding to a specific periodicity of the cluster in the superlattice. An illustrative sketch of exemplary antiferromagnetic modulations are displayed in Fig.~\ref{fig:AFM-Sketch}.
 The formal validity and the numerical implementation of Eq.~(\ref{eq:chiqQ}) have been tested by benchmarking it against applied-field calculations, which are thermodynamically consistent with a BSE treatment \cite{Hafermann2014a}, for details see Sec.~\ref{Sec:Benchmark} \cite{Schaefer2021}.
 
\paragraph{The particle-particle sector.}
Similarly as above, we define the generalized particle-particle susceptibility [cf. Eq.~(7) of \cite{Rohringer2012}] in the longitudinal channel as:
\begin{equation}
\begin{split}
\label{Eq:Def_pp}
\chi^{\alpha|\beta}_{\mathrm{pp},\sigma\sigma'}&(\omega)
\coloneq\chi^{\mathrm{ih|jl}}_{\mathrm{pp},\sigma\sigma'|\sigma\sigma'}(\omega,\nu,\nu')\\=&\int_0^\beta \mathrm{d}\tau_1 \mathrm{d}\tau_2 \mathrm{d}\tau_3  \, \mathrm{e}^{-\i\nu\tau_1} \mathrm{e}^{\i(\omega-\nu')\tau_2}\mathrm{e}^{-\i(\omega-\nu)\tau_3} \\&\left[\langle T_{\tau}c_{\mathrm{i},\sigma}^\dagger (\tau_1)c_{\mathrm{j},\sigma}(\tau_2)c^\dagger_{\mathrm{h},\sigma'}(\tau_3)c_{l,\sigma'}(0)\rangle \right]\\&-\frac{1}{T}\delta(\omega-\nu-\nu')G^{\mathrm{ji}}_{\sigma}(\nu)G^{\mathrm{lh}}_{\sigma'}(\nu'),
\end{split}
\end{equation} where we use for the index notation the convention ``$\text{outgoing}|\text{incoming}$" lines, and therein place the indices in the order of their appearance in Eq.~(\ref{Eq:Def_pp}). 
The generalized susceptibility can be expressed as Feynman diagrams\footnote{Note that in the diagram we switch the legs $\mathrm{h}$ and $\mathrm{j}$ relative to the ph-notation (in contrast to Ref.~\cite{Rohringer2013}), resulting in a global minus sign.} in the following manner:
\begin{equation}
    \begin{split}
&\chi^{\textcolor{Cerulean}{\alpha}|\textcolor{BrickRed}{\beta}}_{\mathrm{pp},\sigma\sigma'}(\omega)\\&=\chi^{\mathrm{ih|jl}}_{\mathrm{pp},\sigma\sigma'|\sigma\sigma'}(\omega,\nu,\nu') \\&=
    \begin{tikzpicture}[baseline=0.cm]
        \begin{feynhand}
            \node at (-1.4,0.7) {$\mathrm{h}$};
            \node at (-1.4,-0.5) {$\mathrm{i}$};
            \node at (1.4,-0.5) {$\mathrm{l}$};
            \node at (1.4,0.7) {$\mathrm{j}$};
            \node at (0,-0.8) {$1/T \delta(\omega-\nu'-\nu)$};
            \vertex (a) at (-1.2,0.7);
            \vertex (u) at (0,0.1);
            \vertex (b) at (-1.2,-0.5); 
            \vertex (c) at (1.2,-0.5); 
            \vertex (d) at (1.2,0.7); 
            \propag [plain] (d) to (u);
            \propag [plain] (c) to (u);
            \propag [fer] (u) to (b);
            \propag [fer] (u) to (a);
        \end{feynhand}
    \end{tikzpicture}
    -
    \begin{tikzpicture}[baseline=0.cm]
        \begin{feynhand}
            \node at (-1.4,1) {$\mathrm{h},\sigma',\omega-\nu$};
            \node at (-1.4,-0.8) {$\mathrm{i},\sigma,\nu$};
            \node at (1.4,-0.8) {$\mathrm{l},\sigma',\nu'$};
            \node at (1.4,1) {$\mathrm{j},\sigma,\omega-\nu'$};
            \vertex (a) at (-1.2,0.7);
            \vertex (b) at (-1.2,-0.5); 
            \vertex (c) at (1.2,-0.5); 
            \vertex (alpha) at (-1.2,0.1) {$\mathbf{\textcolor{Cerulean}{\alpha}}$};
            \vertex (d) at (1.2,0.7); 
            \vertex (u) at (-0.6,0.7);
            \vertex (v) at (-0.6,-0.5); 
            \vertex (alpha) at (1.2,0.1) {$\mathbf{\textcolor{BrickRed}{\beta}}$};
            \vertex (w) at (0.6,-0.5); 
            \vertex (x) at (0.6,0.7); 
            \fill[TUgreen] (v) rectangle (x);
            \node at (0,0.1) {$G^{(2)}$};
            \propag [antfer] (a) to (u);
            \propag [antfer] (x) to (d);
            \propag [antfer] (b) to (v);
            \propag [antfer] (w) to (c);
            \propag (u) to (x);
            \propag (v) to (w);
        \end{feynhand}
    \end{tikzpicture}. \\
\end{split}
\end{equation}
 Similarly we define the generalized susceptibility in the pp transverse channel:
\begin{equation}
\begin{split}
\chi^{\alpha|\beta}_{\mathrm{pp},\overline{\sigma\sigma'}}&(\omega)
\coloneq\chi^{\mathrm{ih|jl}}_{\mathrm{pp},\sigma\sigma'|\sigma'\sigma}(\omega,\nu,\nu')\\=&\int_0^\beta \mathrm{d}\tau_1 \mathrm{d}\tau_2 \mathrm{d}\tau_3 \, \mathrm{e}^{-\i\nu\tau_1} \mathrm{e}^{\i(\omega-\nu')\tau_2}\mathrm{e}^{-\i(\omega-\nu)\tau_3} \\&\left[\langle T_{\tau}c_{\mathrm{i},\sigma}^\dagger (\tau_1)c_{\mathrm{j},\sigma'}(\tau_2)c^\dagger_{\mathrm{h},\sigma'}(\tau_3)c_{l,\sigma}(0)\rangle \right]\\&-\frac{1}{T}\delta_{\sigma,\sigma'}\delta(\omega-\nu-\nu')G^{\mathrm{ji}}_{\sigma}(\nu)G^{\mathrm{lh}}_{\sigma'}(\nu'),
\end{split}
\end{equation}
for which the real-space crossing relations 
\begin{equation}
\label{Eq:crossing}
\begin{split}
    &\chi_{\mathrm{pp},\overline{\sigma\sigma'}}^{\mathrm{ih|jl}}(\omega,\nu,\nu')-\chi_{0,\mathrm{pp},\overline{\sigma\sigma'}}^{\mathrm{ih|jl}}(\omega,\nu,\nu')\\&\quad=-\chi_{\mathrm{pp},\sigma\sigma'}^{\mathrm{ih|lj}}(\omega,\nu,\omega-\nu')+\delta_{\sigma\sigma'}\chi_{0,\mathrm{pp},\sigma\sigma'}^{\mathrm{ih|lj}}(\omega,\nu,\omega-\nu')\\
    &\Gamma_{\mathrm{pp},\overline{\sigma\sigma'}}^{\mathrm{ih|jl}}(\omega,\nu,\nu')=-\Gamma_{\mathrm{pp},\sigma\sigma'}^{\mathrm{hi|jl}}(\omega,\omega-\nu,\nu') 
\end{split}
\end{equation} hold, by swapping the annihilation operators in the definition \cite{Bickers2004}.
The disconnected parts contained in the generalized susceptibility read
\begin{equation}
\begin{split}
    \chi^{\mathrm{ih|jl}}_{0,\mathrm{pp},\sigma\sigma'}(\omega,\nu,\nu')&=-\frac{1}{T}\delta(\nu-\nu') G^{\mathrm{li}}_\sigma (\nu)G^{\mathrm{jh}}_{\sigma'}(\omega-\nu').
\end{split}
\end{equation}
In this frequency notation, the pp-BSE can not be brought into diagonal form. However, as established \cite{Rohringer2013a} for the purely local case, one can define:
 $\tilde{\chi}_{\mathrm{pp},\uparrow\downarrow}^{\mathrm{ih|jl}}(\omega,\nu,\nu')=\chi_{\mathrm{pp},\uparrow\downarrow}^{\mathrm{ih|lj}}(\omega,\nu,\omega-\nu')$ (and analogous relations for $\Gamma$). 
\begin{figure*}[t!]
    \centering
\begin{equation}
    \begin{split}
    \begin{tikzpicture}[baseline=0.cm]
        \begin{feynhand}
            \node at (-1.4,1) {$\mathrm{j},\sigma,\nu+\omega$};
            \node at (-1.4,-0.8) {$\mathrm{i},\sigma,\nu$};
            \node at (1.4,1) {$\mathrm{h},\sigma',\nu'+\omega$};
            \node at (1.4,-0.8) {$\mathrm{l},\sigma',\nu'$};
            \vertex (a) at (-1.2,0.7);
            \vertex (b) at (-1.2,-0.5); 
            \vertex (c) at (1.2,-0.5); 
            \vertex (d) at (1.2,0.7); 
            \vertex (u) at (-0.6,0.7);
            \vertex (v) at (-0.6,-0.5); 
            \vertex (w) at (0.6,-0.5); 
            \vertex (x) at (0.6,0.7); 
            \fill[TUgreen] (v) rectangle (x);
            \node at (0,0.1) {$\chi_{\text{ph}}$};
            \propag [fer] (a) to (u);
            \propag [fer] (x) to (d);
            \propag [antfer] (b) to (v);
            \propag [antfer] (w) to (c);
            \propag (u) to (x);
            \propag (v) to (w);
        \end{feynhand}
    \end{tikzpicture}&=\begin{tikzpicture}[baseline=0.cm]
        \begin{feynhand}
            \vertex (a) at (-1.5,0.7);
            \vertex (b) at (-1.5,-0.5); 
            \vertex (c) at (1.5,-0.5); 
            \vertex (d) at (1.5,0.7); 
            \vertex (u) at (-0.9,0.7);
            \vertex (v) at (-0.9,-0.5); 
            \vertex (w) at (0.9,-0.5); 
            \vertex (x) at (0.9,0.7); 
            \fill[white] (v) rectangle (x);
            \node at (0,0.1) {$\chi_{0,\mathrm{ph}}$};
            \propag [fer] (u) to (x);
            \propag [antfer] (v) to (w);
            \end{feynhand}
    \end{tikzpicture}
    -\begin{tikzpicture}[baseline=0.cm]
        \begin{feynhand}
            \begin{scope}[shift={(-3.3, 0)}]
            \vertex (a1) at (-1.5,0.7);
            \vertex (b1) at (-1.5,-0.5); 
            \vertex (c1) at (1.5,-0.5); 
            \vertex (d1) at (1.5,0.7); 
            \vertex (u1) at (-0.6,0.7);
            \vertex (v1) at (-0.6,-0.5); 
            \vertex (w1) at (0.6,-0.5); 
            \vertex (x1) at (0.6,0.7); 
            \fill[TUgreen] (v1) rectangle (x1);
            \node at (0,0.1) {$\Gamma_{\mathrm{ph}}$};
            \propag (u1) to (x1);
            \propag (v1) to (w1);
            \end{scope}            
            \vertex (k2) at (-4.8,0.1);
            \begin{scope}[shift={(-4.7, 0)}]
            \vertex (a2) at (-1.5,0.7);
            \vertex (b2) at (-1.5,-0.5); 
            \vertex (c2) at (1.1,-0.5); 
            \vertex (d2) at (1.1,0.7); 
            \vertex (u2) at (-0.6,0.7);
            \vertex (v2) at (-0.6,-0.5); 
            \vertex (w2) at (0.6,-0.5); 
            \vertex (x2) at (0.6,0.7); 
            \fill[white] (v2) rectangle (x2);
            \node at (0,0.1) {$\chi_{0,\mathrm{ph}}$};
            \propag (w1) to (v2);
            \propag (x1) to (u2);
            \propag [fer] (u2) to (x2);
            \propag [antfer] (v2) to (w2);
            \propag (x2) to (d2);
            \propag (w2) to (c2);
            \end{scope}
        \end{feynhand}
    \end{tikzpicture}
    \cdot
    \begin{tikzpicture}[baseline=0.cm]
        \begin{feynhand}
            \vertex (a) at (-1.5,0.7);
            \vertex (b) at (-1.5,-0.5); 
            \vertex (c) at (1.5,-0.5); 
            \vertex (d) at (1.5,0.7); 
            \vertex (u) at (-0.6,0.7);
            \vertex (v) at (-0.6,-0.5); 
            \vertex (w) at (0.6,-0.5); 
            \vertex (x) at (0.6,0.7); 
            \fill[TUgreen] (v) rectangle (x);
            \node at (0,0.1) {$\chi_{\mathrm{ph}}$};
            \propag [fer] (a) to (u);
            \propag [fer] (x) to (d);
            \propag [antfer] (b) to (v);
            \propag [antfer] (w) to (c);
            \propag (u) to (x);
            \propag (v) to (w);
        \end{feynhand}
    \end{tikzpicture}
    \\
    \begin{tikzpicture}[baseline=0.cm]
        \begin{feynhand}
            \node at (-1.4,1) {$\mathrm{h},\sigma',\omega-\nu$};
            \node at (-1.4,-0.8) {$\mathrm{i},\sigma,\nu$};
            \node at (1.4,1) {$\mathrm{j},\sigma',\omega-\nu'$};
            \node at (1.4,-0.8) {$\mathrm{l},\sigma,\nu'$};
            \vertex (a) at (-1.2,0.7);
            \vertex (b) at (-1.2,-0.5); 
            \vertex (c) at (1.2,-0.5); 
            \vertex (d) at (1.2,0.7); 
            \vertex (u) at (-0.6,0.7);
            \vertex (v) at (-0.6,-0.5); 
            \vertex (w) at (0.6,-0.5); 
            \vertex (x) at (0.6,0.7); 
            \fill[TUgreen] (v) rectangle (x);
            \node at (0,0.1) {$\tilde{\chi}_{\text{pp}}$};
            \propag [antfer] (a) to (u);
            \propag [antfer] (x) to (d);
            \propag [antfer] (b) to (v);
            \propag [antfer] (w) to (c);
            \propag (u) to (x);
            \propag (v) to (w);
        \end{feynhand}
    \end{tikzpicture}&=\left(
    \begin{tikzpicture}[baseline=0.cm]
        \begin{feynhand}
            \vertex (a) at (-1.5,0.7);
            \vertex (b) at (-1.5,-0.5); 
            \vertex (c) at (1.5,-0.5); 
            \vertex (d) at (1.5,0.7); 
            \vertex (u) at (-0.9,0.7);
            \vertex (v) at (-0.9,-0.5); 
            \vertex (w) at (0.9,-0.5); 
            \vertex (x) at (0.9,0.7); 
            \fill[white] (v) rectangle (x);
            \node at (0,0.1) {$\chi_{0,\mathrm{pp}}$};
            \propag [antfer] (u) to (x);
            \propag [antfer] (v) to (w);
            \end{feynhand}
    \end{tikzpicture} -
    \begin{tikzpicture}[baseline=0.cm]
        \begin{feynhand}
            \vertex (a) at (-1.5,0.7);
            \vertex (b) at (-1.5,-0.5); 
            \vertex (c) at (1.5,-0.5); 
            \vertex (d) at (1.5,0.7); 
            \vertex (u) at (-0.6,0.7);
            \vertex (v) at (-0.6,-0.5); 
            \vertex (w) at (0.6,-0.5); 
            \vertex (x) at (0.6,0.7); 
            \fill[TUgreen] (v) rectangle (x);
            \node at (0,0.1) {$\tilde{\chi}_{\mathrm{pp}}$};
            \propag [antfer] (a) to (u);
            \propag [antfer] (x) to (d);
            \propag [antfer] (b) to (v);
            \propag [antfer] (w) to (c);
            \propag (u) to (x);
            \propag (v) to (w);
        \end{feynhand}
    \end{tikzpicture}\right)
    \cdot
    \begin{tikzpicture}[baseline=0.cm]
        \begin{feynhand}
            \begin{scope}[shift={(3.3, 0)}]
            \vertex (a1) at (-1.5,0.7);
            \vertex (b1) at (-1.5,-0.5); 
            \vertex (c1) at (1.5,-0.5); 
            \vertex (d1) at (1.5,0.7); 
            \vertex (u1) at (-0.6,0.7);
            \vertex (v1) at (-0.6,-0.5); 
            \vertex (w1) at (0.6,-0.5); 
            \vertex (x1) at (0.6,0.7); 
            \fill[TUgreen] (v1) rectangle (x1);
            \node at (0,0.1) {$\tilde{\Gamma}_{\mathrm{pp}}$};
            \propag (u1) to (x1);
            \propag (v1) to (w1);
            \end{scope}            
            \vertex (k2) at (4.8,0.1);
            \begin{scope}[shift={(4.7, 0)}]
            \vertex (a2) at (-1.5,0.7);
            \vertex (b2) at (-1.5,-0.5); 
            \vertex (c2) at (1.1,-0.5); 
            \vertex (d2) at (1.1,0.7); 
            \vertex (u2) at (-0.6,0.7);
            \vertex (v2) at (-0.6,-0.5); 
            \vertex (w2) at (0.6,-0.5); 
            \vertex (x2) at (0.6,0.7); 
            \fill[white] (v2) rectangle (x2);
            \node at (0,0.1) {$\chi_{0,\mathrm{pp}}$};
            \propag (w1) to (v2);
            \propag (x1) to (u2);
            \propag [antfer] (u2) to (x2);
            \propag [antfer] (v2) to (w2);
            \propag (x2) to (d2);
            \propag (w2) to (c2);
            \end{scope}
        \end{feynhand}
    \end{tikzpicture}
\end{split}
\end{equation}
    \caption{Representation of the ph-BSE Eq.~(\ref{Eq:BSE}) and pp-BSE Eq.~(\ref{Eq:BSE_pp}) for the generalized susceptibilities $\chi_{\mathrm{sp}/\mathrm{ch}}^{\mathrm{ijhl}}$ and  $\tilde{\chi}_{\mathrm{pp},\uparrow\downarrow}^{\mathrm{ih|jl}}$, respectively.}
    \label{fig:BSE_pp}
\end{figure*}
This results in the desired diagonal form of the BSE for the cluster case:
\begin{equation}
\begin{split}
\label{Eq:BSE_pp}
    \tilde{\chi}_{\mathrm{pp},\uparrow\downarrow}^{\mathrm{ih|jl}}(\omega,\nu,\nu')&=\\-T^2 \sum_{\nu_1,\nu_2,\mathrm{abcd}}&\left[\chi^{\mathrm{ih|ba}}_{0,\mathrm{pp},\uparrow\downarrow}(\omega,\nu,\nu_1)-\tilde{\chi}_{\mathrm{pp},\uparrow\downarrow}^{\mathrm{ih|ba}}(\omega,\nu,\nu_1)\right]\\&\cdot\tilde{\Gamma}_{\mathrm{pp},\uparrow\downarrow}^{\mathrm{ab|cd}}(\omega,\nu_1,\nu_2)\chi_{0,\mathrm{pp},\uparrow\downarrow}^{\mathrm{dc|jl}}(\omega,\nu_2,\nu').
\end{split}
\end{equation}
A detailed derivation of the latter expressions can be found in App.~\ref{App:Def}, by extending the treatment of \cite{Rohringer2012,Rohringer2013a} to the non-local case.
By switching to multi-index notation:
\begin{equation}
\begin{split}
    \alpha&=\mathrm{i,h},\nu\\
    \beta&=\mathrm{a,b},\nu_1\\
    \gamma&=\mathrm{d,c},\nu_2\\
    \delta&=\mathrm{l,j},\nu'
\end{split}
\end{equation} we find the matrix representation
\begin{equation}
\begin{split}
    \tilde{\chi}_{\mathrm{pp},\uparrow\downarrow}^{\alpha|\delta}=-T^2\left[\chi_{0,\mathrm{pp},\uparrow\downarrow}^{\alpha|\beta}-\tilde{\chi}_{\mathrm{pp},\uparrow\downarrow}^{\alpha|\beta}\right]\cdot\tilde{\Gamma}_{\mathrm{pp},\uparrow\downarrow}^{\overline{\beta}|\overline{\gamma}}\,\chi_{0,\mathrm{pp},\uparrow\downarrow}^{\gamma|\delta},
\end{split}
\end{equation} where repeated indices are to be summed over. A Feynman-diagrammatic illustration of this equation can be found in Fig.~\ref{fig:BSE_pp}.
The corresponding inverted matrix equation, similarly to Eq.~(B.26) in \cite{Rohringer2013a}, reads:
\begin{equation}
\tilde{\Gamma}_{\mathrm{pp},\uparrow\downarrow}^{\overline{\beta}|\overline{\gamma}}=\frac{1}{T^2}\left[\left(\tilde{\chi}_{\mathrm{pp},\uparrow\downarrow}-\chi_{0,\mathrm{pp},\uparrow\downarrow}\right)^{-1}+\left(\chi_{0,\mathrm{pp},\uparrow\downarrow}\right)^{-1}\right]^{\beta|\gamma},
\end{equation}
The momentum dependent pp-response $\tilde{\chi}_{\mathrm{pp},\uparrow\downarrow}^{\vq,\alpha|\delta}$ on the superlattice is given by
\begin{equation}
\label{Eq:Chi_pp_q}
\begin{split}
\tilde{\chi}&_{\mathrm{pp},\uparrow\downarrow}^{\vq,\alpha|\delta}=\chi^{\vq,\alpha,\beta}_{0,\mathrm{pp},\uparrow\downarrow}+\left( T^2\tilde{\Gamma}_{\mathrm{pp},\uparrow\downarrow}^{\overline{\alpha},\overline{\delta}}-\left[\chi^{\vq,\alpha,\beta}_{0,\mathrm{pp},\uparrow\downarrow}\right]^{-1}\right)^{-1}
\end{split}
\end{equation}
where the disconnected pp-correlator reads:
\begin{equation}
    \chi^{\vq,\alpha|\delta}_{0,\mathrm{pp},\uparrow\downarrow}(\omega)=-\frac{\delta_{\nu\nu'}}{TN_{\vk}}\sum_{\vk}G^{li}_{\vk,\up}(\nu)G^{jh}_{\vq-\vk,\down}(\omega-\nu).
\end{equation}
The two-particle response Eq.~(\ref{Eq:Chi_pp_q}) can then be also mapped to the reciprocal space \cite{Potthoff2018} via the Fourier transform \cite{Senechal2011}:
\begin{equation}
\begin{split}
 \tilde{\chi}_{\mathrm{pp},\vK,\vK',\uparrow\downarrow}^{\vq,\vQ}(\nu,\nu',\omega)&=\frac{1}{N^2_c}\sum_{ijhl}  \e^{\i(\vR_i-\vR_h)\vK} \e^{\i(\vR_j-\vR_l)\vK'}\\
    & \times \e^{\i(\vR_j-\vR_h)\vQ}\tilde{\chi}^{\vq,\mathrm{ih|jl}}_{\mathrm{pp},\uparrow\downarrow}(\nu,\nu',\omega)
\end{split}
\end{equation}
and the physical pairing susceptibility can be thus obtained by integrating out the fermionic degrees of freedom
\begin{equation}
     \chi_{\mathrm{pair}}^{\vq,\vQ}(\omega )\coloneqq\frac{T^2}{N_c}\sum_{\vK,\vK'}\sum_{\nu,\nu'}\tilde{\chi}_{\mathrm{pp},\vK,\vK',\uparrow\downarrow}^{\vq,\vQ}(\nu,\nu',\omega).
 \end{equation}
 
\subsection{real-space Ward identities and benchmarks}
\label{Sec:Benchmark}
 \begin{figure}[!h]
    \centering
    \includegraphics[width=\linewidth]{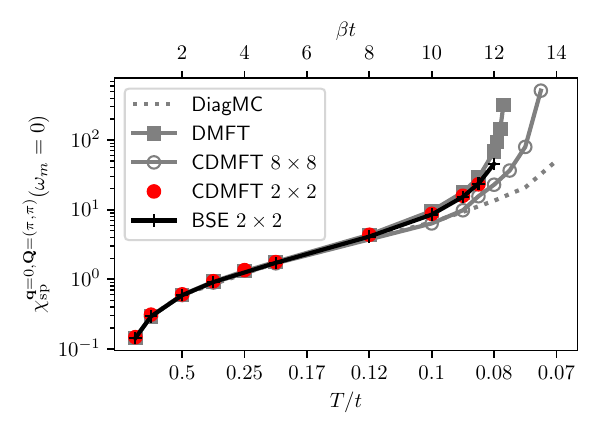}
    \includegraphics[width=\linewidth]{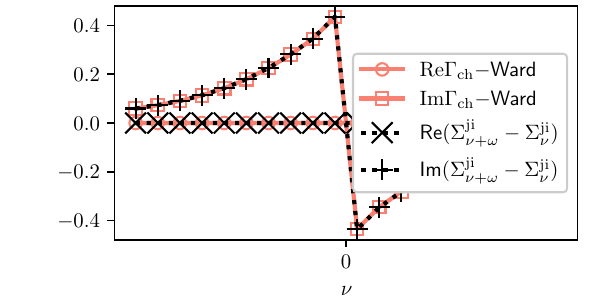}
    \caption{Benchmarking. Upper panel: Comparison of the anti-ferromagnetic response function $\chi_\mathrm{sp}^{\vq=0,\vQ=(\pi,\pi)}(\omega_m=0)$ of the $2\times2$-CDMFT from the BSE (black $+$-marker) and applied field calculations (red circle marker) to the DMFT \cite{Schaefer2021} (gray square marker) and $8\times8$-CDMFT \cite{Schaefer2021} (gray circle marker), at $U=2t$. The diagrammatic Monte-Carlo (DiagMC) calculations for the Hubbard model \cite{Schaefer2021} are given for comparison as dotted line. $\beta=1/T$ gives the inverse temperature. Lower panel: Exemplary Ward identity for the two particle irreducible vertex Eq.~(\ref{Eq:Ward-Gamma}) in the charge channel for $U=5.9t$ and $T=t/15$ for the orbital combination $\mathrm{i=0,j=3,h=0}$ and $\omega=0$ where each marker represents a fermionic Matsubara frequency. Red solid lines refer to the left, dotted black lines to the right-hand side of the equation, respectively.}
    \label{fig:BM}
\end{figure}
 In order to explicitly demonstrate the internal consistency and the conserving nature of our CDMFT expression for the BSEs, we derived Ward identities \cite{Ward1950,Behn1978,Krien2017,Ritz2025} for real-space and embedded real-space clusters via the action formalism. These are a direct manifestation of time-inversion invariance in the charge and spin channel of the $\mathrm{SU}(2)$-symmetric case. In particular,  for the generalized susceptibility $\chi$ in the ph channel we find:
\begin{equation}
\begin{split}
\label{Eq:Ward-Chi}
    G^{\mathrm{ji}}_{\uparrow}(\nu)\delta_{hj}-&G^{\mathrm{ji}}_{\uparrow}(\nu+\omega)\delta_{\mathrm{hi}}\\=T\sum_{\mathrm{l}}\sum_{\nu'}&\left(\chi^{\mathrm{ijhl}}_{\mathrm{ch}/\mathrm{sp}}(\omega,\nu,\nu') \mathcal{G}^{-1}_{\mathrm{hl},\uparrow}(\nu')\right.\\&\quad\left.-\chi^{\mathrm{ijlh}}_{\mathrm{ch}/\mathrm{sp}}(\omega,\nu,\nu') \mathcal{G}^{-1}_{\mathrm{lh},\uparrow}(\nu'+\omega)\right),
    \end{split}
 \end{equation}
 where $\mathcal{G}$ is the non-interacting Green function, see Eq.~(\ref{Eq:Weissfield}) for CDMFT. For the corresponding irreducible vertex we find:
\begin{equation}
\begin{split}
\label{Eq:Ward-Gamma}
    \Sigma^{\mathrm{ji}}(\nu+\omega)\delta_{\mathrm{hi}}-\Sigma^{\mathrm{ji}}&(\nu)\delta_{\mathrm{hj}}\\=T\sum_{\mathrm{l}}\sum_{\nu'}&\left(\Gamma^{\mathrm{ijlh}}_{\mathrm{ch}/\mathrm{sp}}(\omega,\nu,\nu') G^{\mathrm{lh}}_{\uparrow}(\nu'+\omega)\right.\\&\quad\left.-\Gamma^{\mathrm{ijhl}}_{\mathrm{ch}/\mathrm{sp}}(\omega,\nu,\nu') G^{\mathrm{hl}}_{\uparrow}(\nu')\right)   \\=T\sum_{\mathrm{l}}\sum_{\nu'}&\left(\Gamma^{\mathrm{ijlh}}_{\mathrm{ch}/\mathrm{sp}}(\omega,\nu,\nu') G^{\mathrm{lh}}_{\mathrm{loc},\uparrow}(\nu'+\omega)\right.\\&\quad\left.-\Gamma^{\mathrm{ijhl}}_{\mathrm{ch}/\mathrm{sp}}(\omega,\nu,\nu') G^{\mathrm{hl}}_{\mathrm{loc},\uparrow}(\nu')\right).
    \end{split}
 \end{equation}
 The second equality holds when inserting the CDMFT self-consistency condition Eq.~(\ref{eq:approx_1P_Gf}).
 
 For the normal state susceptibility in the pp channel, we have:
\begin{equation} \begin{split}\label{Eq:Ward-Chi-pp}
    G^{\mathrm{ji}}_{\uparrow}(\nu)\delta_{\mathrm{jh}}\delta_{\uparrow\sigma}-G^{\mathrm{ji}}_{\uparrow}&(\nu')\delta_{\mathrm{ih}}\delta_{\uparrow\sigma}\\=T\sum_{\mathrm{l}}\sum_{\omega}&\left(\tilde{\chi}^{\mathrm{ih|lj}}_{\mathrm{pp}, \sigma\uparrow}(\omega,\nu,\nu') \mathcal{G}^{-1}_{\mathrm{hl},\uparrow}(\omega-\nu')\right.\\&\left.-\tilde{\chi}^{\mathrm{il|hj}}_{\mathrm{pp}, \sigma\uparrow}(\omega,\nu,\nu') \mathcal{G}^{-1}_{\mathrm{lh},\uparrow}(\omega-\nu)\right).
\end{split}
\end{equation}
For details and derivations see App.~\ref{App:Ward}.
The Ward identities for $\Gamma$ are derived without the use of the BSE, and allow to benchmark, first, the BSE itself, and, second, to clarify, whether a sufficient number of Matsubara frequencies were used for a reliable matrix inversion in the BSE at the given temperature. 
An example for the Ward identity for $\Gamma_{\mathrm{ch}}$ Eq.~(\ref{Eq:Ward-Gamma}) is given in Fig.~\ref{fig:BM} which is perfectly fulfilled for the parameter regime under consideration in the main section of this manuscript, e.g., $U=5.9t$, $T=t/15$. 

The impurity $\Gamma$-vertex was compared to its weak coupling single boson exchange (SBE) decomposition \cite{Krien2020b} employed in real-space (App.~\ref{App:SBE_def}) successfully in all three channels (App.~\ref{App:SBE}).

In order to establish the reliability of our scheme, we compared the temperature dependence of the anti-ferromagnetic (AFM) lattice response to applied staggered field calculations, shown in Fig.~\ref{fig:BM}, top panel (for details, see App.~\ref{App:AFM}).

\section{The two-particle irreducible charge vertex in CDMFT}
\label{sec:resultsVertex}
\begin{figure*}
    \centering
    \includegraphics[width=0.497\linewidth]{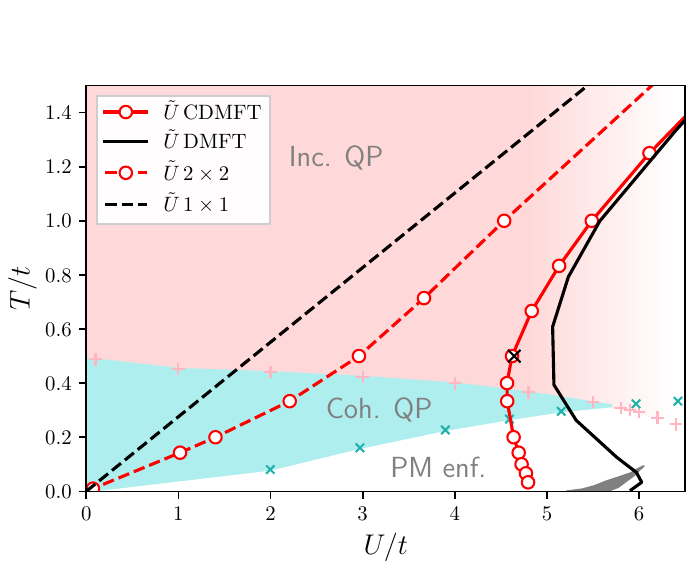}
      \hfill
      {\color{black}\vrule width 0.4pt height 0.36\linewidth}%
      \hfill
    \includegraphics[width=0.28895\linewidth]{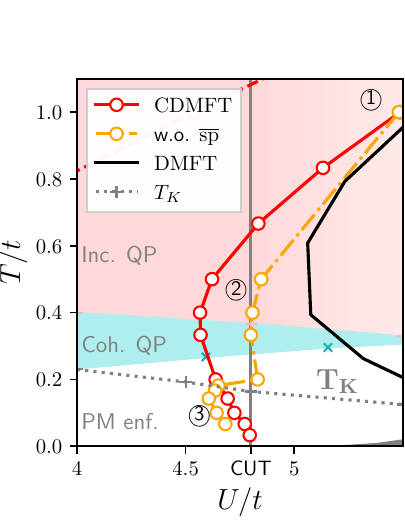}
    \includegraphics[width=0.19305\linewidth]{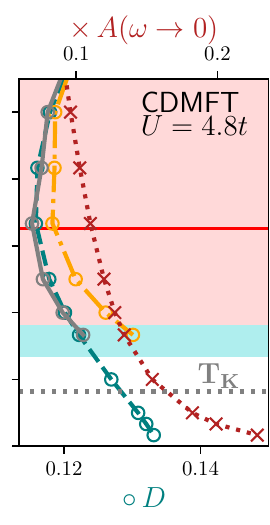}
    \caption{Left: Location of the first vertex divergence line in the $T-U$ phase-diagram of the 2D Hubbard model, computed in CDMFT with a $N_c=2\times 2$ cluster (red continuous line), and DMFT \cite{Schaefer2013} (black continuous line), compared to the corresponding results for a $2\times 2$ isolated Hubbard cluster (red dashed line) and the $1\times 1$ Hubbard atom \cite{Schaefer2013} (black dashed line). Red shading indicates the incoherent quasiparticle regime (Inc. QP), turquoise the coherent quasiparticle (Coh.~QP) one, and the white area the paramagnetically enforced (PM enf.) region, below the N\'eel-temperature of the cluster \cite{KlettPhd,Klett2020,Fratino2017,Meixner2024}. The gray shaded area gives the CDMFT metal-insulator coexistence region \cite{Park2008}. As comparison, the black $\times$-marker gives the divergence for CDMFT given for a single temperature in \cite{Vucicevic2018}. Middle: Zoom on the divergence line of $\Gamma_c$ with (red bold line) and without the non-local spin-transverse diagrams (orange dashed line). The black line indicates the corresponding DMFT divergence line \cite{Schaefer2013}. The vertical gray line indicates a temperature cut at $U=4.8t$, for which thermodynamic quantities are given in the right-hand side plot. To the right of the divergence lines, we present the local spectral weight (bordeaux-colored $\times$ marker) and the double occupancy $D$ ($\circ$ marker) on the horizontal axis, while the vertical axis gives the temperature. The dashed green line for $D$ results form the Migdal formula Eq.~(\ref{Eq:Migdal}), the grey, bold line from the generalized response function and the orange dash-dotted line excludes the non-local spin-transverse diagrams. This is contrasted to an area of incoherent quasiparticles in red shading, the red horizontal line the location of the first divergence line of $\Gamma_c$ and the gray, dotted line in the plot indicates the Kondo temperature $T_K$, see App.~\ref{App:Kondo}.}
    \label{fig:1st_div_line}
\end{figure*}

We begin our analysis of two-particle quantities in CDMFT by studying the two-particle irreducible vertex in the charge channel $\Gamma_\mathrm{ch}$. At the level of the cluster-local approximation \cite{Georges1996,Jarrell1992,Musshoff2021} $\Gamma_\mathrm{ch}$ of the converged cluster impurity problem corresponds to the lattice vertex of CDMFT.
 As it has been reported in the literature, for systems with strong electronic correlations $\Gamma_\mathrm{ch}$ can display several divergences (corresponding to vanishing eigenvalues of the generalized susceptibility of the impurity cluster appearing in Eq.~(\ref{Eq:Gamma})), also when the system does \textit{not} exhibit a thermodynamic phase transition \cite{Schaefer2013,Schaefer2016c}. In this case \cite{Kozik2015,Gunnarsson2017,Vucicevic2018,Essl2025}, techniques relying on the self-consistent resummation of a perturbative series of diagrams start to fail and iterative schemes might become unstable \cite{Essl2025}. In particular, the breakdown of self-consistent perturbation theory can be directly ascribed to the occurrence of the crossing of a physical and an unphysical solution~\cite{Kozik2015,Stan2015,Tarantino2018,Vucicevic2018,Kim2020} of the Luttinger-Ward functional $\Phi$, which implies~\cite{Gunnarsson2017} the divergence of the irreducible vertex functions~\cite{Schaefer2013,Schaefer2016c,Springer2020,Essl2024}. Hence, the ``first'' divergence of $\Gamma_\mathrm{ch}$, i.e., the one occurring at the lowest interaction values, can be regarded as a reliable indicator marking the regime of strong correlations~\cite{Chalupa2021,Reitner2025}, where self-consistent perturbation theory ceases to be applicable.

\subsection{Vertex divergence lines in the phase-diagram}

In Fig.~\ref{fig:1st_div_line}, we report the $T-U$ phase diagram of the 2D Hubbard model indicating the first divergence of the (matrix-valued) irreducible vertex \[\Gamma_\mathrm{ch}^{\overline{\alpha},\overline{\beta}} \rightarrow \infty\] encountered by increasing $U$, starting from weak coupling, comparing our CDMFT results with those obtained with other techniques. We can immediately note that the first divergence in the charge channel $\Gamma_\mathrm{ch}$ in CDMFT is occurring at \emph{systematically} lower interaction values $\tilde{U}^{\mathrm{CDMFT}}(T)$ (red solid line) than the corresponding (single-site) DMFT results $\tilde{U}^{\mathrm{DMFT}}(T)$ (black solid line). The respective $\tilde{U}^{2\times2}(T)$ and $\tilde{U}^{1\times1}(T)$ for isolated Hubbard cluster/atom results are also reported, for comparison, by red and black dashed lines.

While the observed systematic leftward shift of  $\tilde{U}^{\mathrm{CDMFT}}(T)$ w.r.t.~$\tilde{U}^{\mathrm{DMFT}}$ 
is consistent with sporadic results (for specific datasets only) obtained in previous DCA \cite{Gunnarsson2016} and CDMFT \cite{Vucicevic2018} studies, the  determination of  $\tilde{U}^{\mathrm{CDMFT}}(T)$ on the whole phase-diagram eventually paves the way for a thorough investigation of the underlying physical mechanisms controlling the vertex divergences in the presence of spatial correlations. To this aim, we proceed now with a systematic inspection of our finding in the different temperature regimes.

At high temperatures $T/t>1$, we find that the non-local part of $\Gamma_\mathrm{ch}$ (as well as, in general, the relative contribution of non-local correlations) is vanishingly small, so that our cluster impurity reproduces essentially the local physics of DMFT.
Hence, at these temperatures, the associated breakdown of the self-consistent perturbative expansion, indicated by the divergences of $\Gamma_\mathrm{ch}$, occurs almost at the same interaction values in CDMFT and DMFT.
The situation evidently changes by reducing the temperature: While the overall shape of the first divergence lines in DMFT and CDMFT appears to be rather similar, the non-perturbative breakdowns of DMFT and CDMFT do not coincide anymore for intermediate and low temperatures (i.e. for $T/t<1$). Here, the first vertex divergence in CDMFT already occurs at significantly \textit{lower} interaction values w.r.t.~the DMFT case $\tilde{U}^{\mathrm{CDMFT}}<\tilde{U}^{\mathrm{DMFT}}$. Evidently, the systematic shift of the convex shape of the first vertex divergence line in the phase diagram towards lower interactions when increasing the impurity size from DMFT to CDMFT indicates that non-local contributions play now a significant role. These contributions will be analyzed in detail in the following subsection. 

By further reducing the temperature, the convex shape of the divergence line displays a reentrance towards higher interaction values, qualitatively resembling the DMFT results \cite{Schaefer2013}, where this behavior has been unambiguously ascribed \cite{Chalupa2021,Adler2024} to the onset of Kondo screening in the auxiliary AIM of DMFT.

From a more quantitative point of view, we note that at the lowest temperature available ($T/t=1/30$), we find the interaction value at which the first vertex divergence  occurs $\tilde{U}^{\mathrm{CDMFT}}$ to be considerably lower than the interaction value $U_{c1}^{\mathrm{CDMFT}}$, which marks the onset of the coexistence region associated to the Mott MIT in CDMFT (gray shaded area). This trend is qualitatively similar, \emph{mutatis mutandis}, to the corresponding results of DMFT, in spite of the opposite slope of the DMFT coexistence region. We also remark that the smallest interaction value $\underset{T}{\mathrm{min}}\left[\tilde{U}^{\mathrm{CDMFT}}(T)\right]$ for which an irreducible vertex divergence occurs is at $T/t\approx 0.4$, the peak of the ``belly''-shape. Remarkably, at about this temperature, the single-particle description of the CDMFT impurity cluster changes from incoherent quasiparticles (light red shading) to coherent quasiparticles (turquoise shading)\footnote{This has been diagnosed by a change of sign of the difference between the first two Matsubara on-site self-energy data points \cite{Schaefer2021}.}. 

Further insight on to the breakdown of the perturbation expansions beyond the short-range description of CDMFT, can be gained by comparing our CDMFT/DMFT findings with the corresponding results for the isolated $2\times 2$ plaquette-cluster/Hubbard atom. The location of the first vertex divergences in the two latter cases has been thus reported in our phase-diagram of Fig.~\ref{fig:1st_div_line} with dashed lines\footnote{In the case of the isolated Hubbard atom, where no hopping is defined, the location of the vertex divergences is defined in terms of the ratio $T/U$}.

Interestingly, the comparison of the results for the embedded (CDMFT/DMFT) vs. isolated (plaquette/HA) shows a \textit{reversed} trend for the shift of the first vertex divergence lines w.r.t. an increasing cluster size.
In particular, while the vertex divergences of embedded theories occur systematically at lower interactions for an increased cluster impurity size  $\tilde{U}^{\mathrm{CDMFT}}\leq\tilde{U}^{\mathrm{DMFT}}$, the interaction values of the vertex divergences in the isolated Hubbard cluster\footnote{These calculations were performed via the exact diagonalization solver Pomerol due to high numerical performance \cite{Antipov2015a}.} are always at \emph{larger} interactions (red dashed line) w.r.t.~those of the atomic limit (black dashed line)\cite{Schaefer2013} (i.e., $1\times1$ Hubbard cluster, $\tilde{U}^{2\times 2}\geq\tilde{U}^{1\times 1}$).

Though apparently surprising, the opposite trend of the embedded and isolated systems can be directly rationalized in terms of fundamental physical considerations, on the basis of which interesting insight can be gained even on the (still unknown!) exact solution of the 2D Hubbard model.

In particular, the observed (opposite) trends directly match the corresponding 
charge-localization degree of the different cases. In fact, increasing the cluster size in the embedding methods reduces the delocalizing/itinerant effects associated to the auxiliary electronic bath, and, at the same time, enhances the electronic scattering, due to the inclusion of (progressively longer-ranged) non-local correlations. This determines, on the whole, a systematic reduction of the electronic mobility by larger cluster sizes. 

At the same time it is evident how, in the isolated systems, the most localized situation is realized in the Hubbard atom, and that, by increasing the system size, such a localization gets progressively mitigated by the hopping processes within the isolated clusters. This  features a \emph{reversed} localization trend w.r.t.~increasing cluster sizes. 
As the vertex divergences are a direct manifestation of the charge localization driven by the electronic scattering \cite{Gunnarsson2017,Chalupa2021,Adler2024}, the opposite localization trends of embedded/isolated calculations provide a clear explanation for the different locations of the vertex divergence lines in Fig.~\ref{fig:1st_div_line}.

Based on the above considerations, we can now outline some speculative, but rather plausible conclusions on how the perturbative breakdown might appear in the \textit{exact} solution of the 2D Hubbard model. Indeed, the exact solution does represents -per construction- the infinite cluster-size limit of both embedded and isolated calculations. Hence, as the (opposite) localization trends are essentially\footnote{While some oscillation may occur for small clusters due to finite size/geometrical effects, these  will become negligible in the large cluster size limit.} monotonous in the two classes of approaches, we can infer that the first divergence lines in the exact solution will be located in the area comprised between the red-dashed ($2\!\times \! 2$ isolated plaquette) and the red continuous (CDMFT) lines. Evidently, by increasing the cluster size of both embedded and isolated cluster calculations, the possible location area for the first divergence line of the exact solution would progressively shrink. Performing two-particle calculations for large clusters is certainly numerically highly demanding, but, even without that, we are now in a position to qualitatively infer how the exact first divergence vertex line should appear. While at high-T, where local correlations will prevail even in the exact solution, the line will approach the DMFT one, at low-T, the disappearance of the embedding bath (and of the associated Kondo effect) will eventually obliterate the re-entrance observed in (C)DMFT. Thus, at low-T the exact vertex divergence line will qualitatively resemble the isolated cluster results, only being gradually pushed down towards low temperatures for increasing cluster sizes, especially at small $U$. In fact, due to discrete nature of the isolated cluster solution at half-filling (entailing always some form of a spectral gap for perfect particle-hole symmetry and $T \rightarrow 0$),
we expect that all the isolated cluster solutions will feature divergence lines starting from $U \!= \! 0$. We can plausibly speculate, then, that also the first divergence line of the exact solution will start from $U \!= \!T \!= \! 0$, displaying, for small $U$ values, a slow exponential behavior and, then, interpolating, for intermediate $U$, towards the high-$T$ DMFT result. 

Such a heuristic picture of the location of the perturbative breakdown in the exact solution of the 2D Hubbard model highlights, firstly, the profoundly non-perturbative nature of the paramagnetic phase of the half-filled Hubbard model in 2D, and, secondly, its link to the (quasi-)long-ranged AF fluctuations associated to the Mermin-Wagner theorem \cite{Mermin1966,Hohenberg1967}. These fluctuations dominate the low-T regime of the model, due to its AF ordered ground state.

We finally remark that our considerations on the qualitative shape of the first vertex divergence line in the exact solution are also supported by the recent study \cite{Reitner2025} of the vertex function in the AF-ordered phase of DMFT calculations, see App.~\ref{App:AF_DMFT}.

Before analyzing the thermodynamic behavior of the CDMFT MIT in Sec.~\ref{sec:resultsMIT}, we now discuss the diagrammatic reason for the occurrence of divergences of $\Gamma_\mathrm{ch}$ at lower interactions in CDMFT than in the DMFT case.

\subsection{The role of non-local spin-fluctuations}
\label{Sec:Vertex_spin}
Let us now investigate in more detail the influence of short-ranged non-local correlations on the breakdown of the perturbation theory. In this respect, physical insight can be gained by taking a closer look at spectral and thermodynamic quantities in the vicinity of the first vertex divergence line.

In the rightmost panel of Fig.~\ref{fig:1st_div_line} the double occupancy $D$ is shown for a temperature cut for $U=4.8t$, i.e., in the close proximity to the location of the first divergence line at intermediate $T$. $D$ is calculated via the Galitskii-Migdal formula \cite{Galitskii1958}:

\begin{equation} \label{Eq:Migdal}\begin{split}
E_\text{pot}&=\frac{T}{N_c}\sum_{\nu_n}\sum_{\mathrm{ij}}\Sigma^{\mathrm{ij}}(\nu)G^{\mathrm{ji}}(\nu)\\
    D&=\frac{E_{pot}}{U},
\end{split}
\end{equation}
 where $E_\text{pot}$ is the respective potential energy.  
This study is enriched by the comparison with the on-site spectral function\footnote{using a third-order polynomial fitting procedure to extrapolate the fermionic Matsubara frequency to 0} $A(\omega=0)\approx - \mathrm{Im}\,G(\mathrm{i}\nu_n\rightarrow 0^+)/\pi$. 

Starting from high temperatures and progressively lowering $T$, we first observe a reduction of the double occupancy $D$ down to $T\approx 0.6t$---approximately at the temperature where the first vertex divergence is encountered, reflecting an initial suppression of the on-site charge fluctuations due to local moment formation in the isolated Hubbard cluster.
 
Lowering the temperature further results, however, in an enhancement of $D$, signaling a corresponding increase in metallicity of the system. Similarly as in DMFT, this can be ascribed to the screening of the local moment by cluster degrees of freedom on the impurity and by the bath. 
In fact, at temperatures below $T_K\approx0.2t$ (grey dotted line), we find that the static magnetic on-site response $\chi^{\mathrm{ii}}_\mathrm{sp}(\omega_m=0)$ for CDMFT exhibits a Kondo-like temperature dependence, similar to the Anderson model \cite{Krishna-murthy1980} and to DMFT \cite{Chalupa2021} (see App.~\ref{App:Kondo}). This indicates the progressive onset of screening of the cluster impurity by bath electrons.
This picture is coherent with the observed behavior of the spectral function: The initial, weakly linear $T$ dependence of $A(\omega=0)$ changes quantitatively by entering the coherent QP regime $T \leq 0.4 t$, consistent with a marked increase of metallic coherence at low $T$.
Hence, our analysis of the basic thermodynamic/spectral behavior of the impurity cluster indicates that the overall temperature behavior of the metallic coherence around the interaction values of the first vertex divergence $\tilde{U}^\text{CDMFT}$ is similar to the one in DMFT. This explains the qualitatively similar shape of the first divergence line in CDMFT and DMFT, which is in contrast to the difference in shape of the Mott MIT in these techniques.
However, the qualitatively remarkable shift towards lower $U$ values of the first vertex divergence line at intermediate-to-low temperatures must evidently be ascribed to the inclusion of short-ranged correlations within the CDMFT cluster considered.

In order to quantify this effect, we extend our non-local CDMFT case to the approach of Ref.~\cite{Adler2024}, where the effects of the local moment formation (and, especially, of the associated on-site spin transverse dynamical fluctuations) on the charge sector were investigated through a systematic SBE decomposition of the purely local two-particle correlation functions.

More specifically, we note that, within CDMFT, non-local correlations enter the corresponding cluster BSE in two possibly ways, (a) implicitly, in internal diagrams of the two-particle irreducible vertex ending on the same cluster site, $\Gamma^{\mathrm{iiii}}_\mathrm{ch}$, and (b) explicitly from the non-local components $\Gamma^{\mathrm{ijhl}}_\mathrm{ch}$. The former, $\Gamma^{\mathrm{iiii}}_\mathrm{ch}$, might be regarded as an implicit renormalization of the DMFT result, where only the on-site components are considered, while the latter is an explicit step beyond the DMFT approximation in the BSE equation. It is thus instructive to ask, to what degree these non-local correlations are contributing to the correlated system, and specifically, to the shifted shape of the cluster divergence line. Here, the real-space SBE decomposition of the generalized susceptibility \cite{Krien2019c} presents a valuable tool, which recent literature \cite{Adler2024,Meixner2025a} employed to underline the considerable importance of spin-diagrams in (C)DMFT.

In this context, we can compute the shift of the first vertex divergence line location, which one would observe, when \emph{manually switching off} the contribution of non-local spin-transverse SBE diagrams from the two-particle irreducible vertex $\Gamma^{ijhl}_\mathrm{ch}$, as follows:
\begin{equation}
    \label{eq:mod_gamma}
    \overline{\Gamma}_{\mathrm{ch}}^{\mathrm{ijhl}}=\left\{\begin{matrix}\begin{array}{l} \Gamma_\mathrm{ch}^{\mathrm{ijhl}}, \quad \mathrm{for}\quad \mathrm{i\!=\!j\!=\!h\!=\!l}\\
        \beta^2\left[\left(\chi_{\mathrm{ch}}^{\alpha\delta}-\chi_0^{\alpha\beta}\overline{\nabla}_{\overline{\mathrm{sp}}}^{\beta\gamma}\chi_0^{\gamma\delta}\right)^{-1}-\chi_0^{-1}\right]^{\mathrm{ijhl}}  \mathrm{else},\end{array}
       \end{matrix}\right. 
\end{equation}
where 
\begin{equation}
\begin{split}
    \overline{\nabla}_{\overline{\mathrm{sp}}}^{\beta\gamma}& =\\ -&\frac{3}{2} \Big(\sum_{\mathrm{mn}}\lambda^{\mathrm{mij}}_{\mathrm{sp}}(\nu'-\nu,\nu) w^{\mathrm{mn}}_{\mathrm{sp}} (\nu'-\nu)\lambda^{\mathrm{nhl}}_{\mathrm{sp}}(\nu'-\nu,\nu')\\
    &\phantom{\frac{3}{2}\quad}+U \Big),
\end{split}
\end{equation}
and $\lambda^{\mathrm{mij}}_{\mathrm{sp}}$ is the Hedin vertex, $w^{\mathrm{mn}}_{\mathrm{sp}}$ represents the bosonic propagator in the spin channel [for details see App.~\ref{App:SBE_def}] and  $\overline{\nabla}_{\overline{\mathrm{sp}}}$ all single spin-transverse boson irreducible diagrams in the charge channel \cite{Krien2019c} (except for the bare interaction). This corresponds to a subtraction of the non-local spin-transverse SBE diagrams of the generalized susceptibility before performing the BSE to obtain the charge vertex $\Gamma_{\mathrm{ch}}$.

The resulting ``shift'' of the divergence line is shown in  Fig.~\ref{fig:1st_div_line} as a yellow dot-dashed line. Two properties are noteworthy: First, in the absence of non-local transversal spin fluctuations, the first vertex divergence occurs at considerably higher interactions than the divergence of the ``full'' CDMFT vertex. Second, the prominent, ``belly shape'' characteristic of the vertex divergence line gets considerably flattened out in the parameter region $\textcircled{2}$ where the system's metallic coherence is mostly suppressed. Hence, this analysis clearly demonstrates the significant role of short-ranged transverse spin fluctuations in driving the vertex divergence line at intermediate temperatures. The remaining discrepancy of the divergence line location relative to the DMFT results can be attributed to non-local correlations of type (a), i.e., non-local correlations most likely consisting of spin transverse fluctuations, contributing to the on-site vertex $\Gamma^{\mathrm{iiii}}_\mathrm{ch}$.

To summarize, in our more physical analysis of the CDMFT vertex divergences ($\tilde{U}^{\mathrm{CDMFT}}$ in Fig.~\ref{fig:1st_div_line}), we can identify three regimes: $\textcircled{1}$ at high temperatures, $\tilde{U}^{\mathrm{CDMFT}}$ almost coincides with the one of DMFT due to the predominance of purely on-site correlations. At low temperatures $T\!<\!T_K$ $\textcircled{3}$, only the on-site vertex is relevant to CDMFT, as the manipulated divergence line essentially agrees with $\tilde{U}^{\mathrm{CDMFT}}$. This implies, that (non-local) anti-ferromagnetic spin fluctuations of type (b) do not play a relevant role here\footnote{The jump of the re-calculated vertex divergence line around the Kondo temperature can be ascribed to a crossing of the first singular eigenvalue of the generalized charge susceptibility with another eigenvalue, occurring when reducing the temperature in the modified calculations.} as the vertex is dominated by its on-site part, yielding a ``renormalized DMFT'' result. This, together with the Kondo-like spin response, further supports the statement that, in the lower temperature regime $\textcircled{3}$, our impurity cluster is effectively Kondo screened in the bath for $T<T_K$.
The most important effect of spin transverse correlations is found, thus, at intermediate temperatures $\textcircled{2}$, where a considerable shift of the divergence line is observed when turning off the non-local spin transverse correlations. Finally, the significance of these non-local short range spin fluctuations of type (b) becomes manifest when neglecting the former in our analysis of the thermodynamics. This leads to an increase of the double occupancy in the intermediate temperature regime (orange dash-dotted line), providing additional support of our interpretation.

After having investigated how the breakdown of the many-body perturbation theory is affected by short-ranged fluctuations, we now perform our two-particle studies for larger interaction values, namely in highly non-perturbative \cite{Pelz2023} regime, where the Mott MIT occurs in CDMFT.

\section{Mott transition at the two-particle level: CDMFT results}

\begin{figure*}
    \centering
    \includegraphics[width=0.9\linewidth]{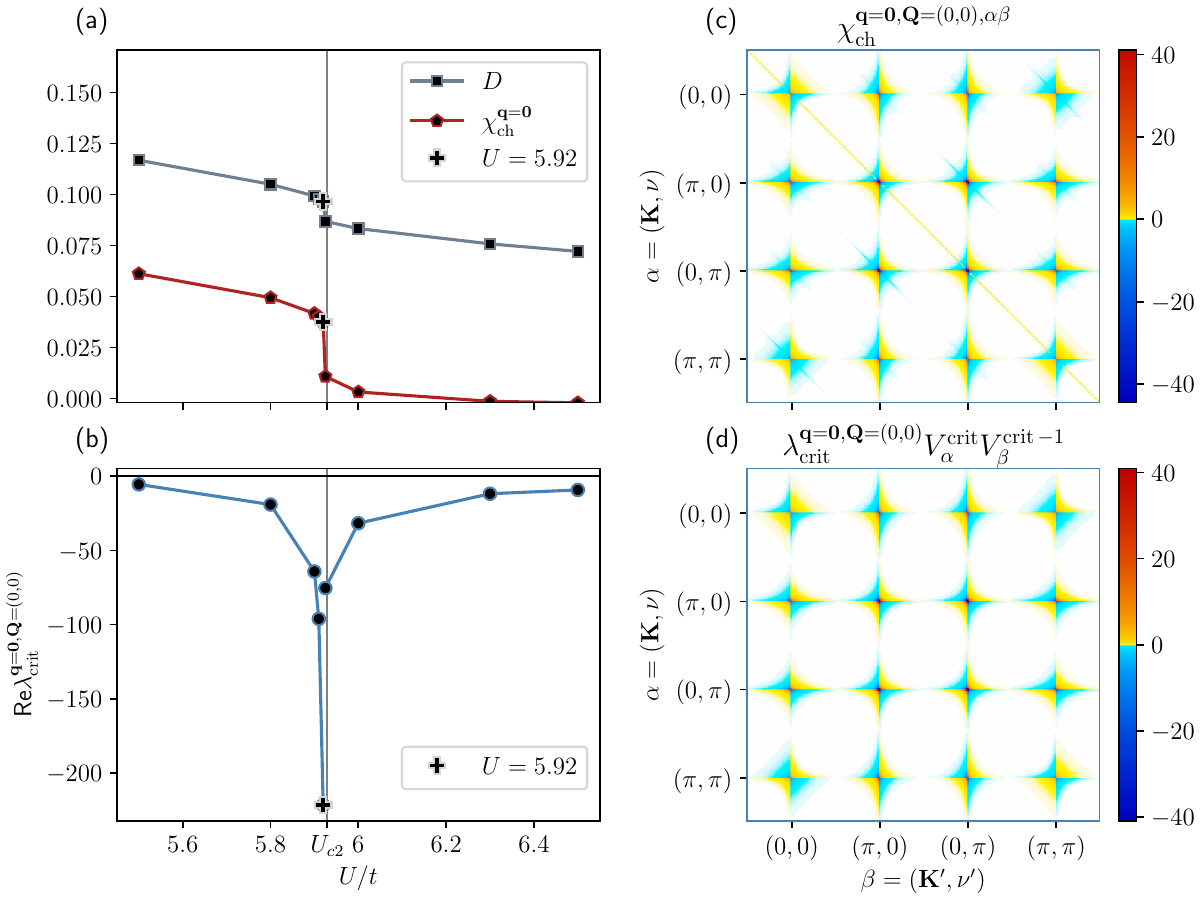}
    \caption{a) Interaction scan over the MIT of the double occupancy $D$ and the uniform charge response $\chi_c^{\vq=0}$ for $T=0.066t$. The gray vertical line indicates the critical interaction $U_{c2}$ of the MIT. b) The leading eigenvalues for the same interaction scan for $\vq=(0,0)$.
    The analysis of the eigenvectors corresponding to the eigenvalues in b) are given in c) for $\vQ=(0,0)$ for the data point of $U=5.92t$, short of $U_{c2}$.
    The corresponding charge responses on the reciprocal lattice is shown in d).}
    \label{fig:ch_inst}
\end{figure*}

\label{sec:resultsMIT}
The consistent description of the first order Mott metal-insulator-transition (MIT), ending with a second order critical endpoint, is notoriously one of the main DMFT \cite{Georges1992a,georges:1996} successes. In its paramagnetically restricted solution, for temperatures below a critical temperature $T<T_{c}$, a first order phase transition from a metal to an insulator occurs at a critical local Coulomb interaction value $U^{\mathrm{DMFT}}_{c2}$ due to purely local correlations. When including non-local correlations via CDMFT \cite{Kotliar2001,Parcollet2004,Lichtenstein2000}, the MIT is, somewhat similar to the vertex divergences, shifted to lower interaction values $U^{\mathrm{CDMFT}}_{c2}<U^{\mathrm{DMFT}}_{c2}$ \cite{Park2008}. Let us note that, different from the vertex divergence lines, the low-$T$ curvature of the MIT line in CDMFT gets reversed w.r.t.~DMFT.

\subsection{Leading instability at the CDMFT Mott transition}

The thermodynamic first order transition of the MIT can be studied via the double occupancy $D$ Eq.~(\ref{Eq:Migdal}), which corresponds to the first derivative $D=(1/V)(\partial \Omega/\partial U)$ of the Landau free energy $\Omega$ \cite{Sordi2011,Strand2011,Walsh2019,Kowalski2024}. In Fig.~\ref{fig:ch_inst}a), we follow $D$ (square markers) from the metallic solution, over the coexistence region across the MIT, into the insulating regime via an interaction scan for $T=1/15t<T_c\approx1/11t$ \cite{Park2008}. $D$ displays a discontinuous drop at the critical interaction $U_{c2}$ \cite{Walsh2019,Reymbaut2020}, which is highlighted by a gray vertical line. This discontinuity indicates the presence of the characteristic hysteresis region associated to a first order transition \cite{georges:1996}, which is delimited by the two $T$-dependent values $U_{c1}$ and $U_{c2}$. By increasing the temperature, this coexistence region gradually shrinks, terminating in a second order critical endpoint, where $-\partial D/\partial U$ diverges. Alternatively, the MIT can be identified from the associated abrupt suppression of the static charge response function $\chi_{\mathrm{ch}}^{\vq=0}(\omega\!=\!0)$ at $U_{c2}$, which is given for the half-filled case in Fig.~\ref{fig:ch_inst}a) as pentagon markers\footnote{We find good agreement of the critical
eigenvalue with $\lambda_\text{crit}\propto 1/(U-U_{c2})$ in the proximity of the MIT, which allows for determining  $U_{c2}$ by fitting the former
in Fig.~\ref{fig:ch_inst}b).}. While the double occupancy $D$ is computed from the potential energy via the Migdal-Galitskii formula, the lattice charge response is obtained by integrating out the fermionic degrees of freedom of the generalized charge reciprocal lattice susceptibility:
\begin{equation}
\chi_{\mathrm{ch}}^{\vq=0}=\frac{T^2}{N_c}\sum_{\vK,\vK'}\sum_{\nu,\nu'}\chi_{\mathrm{ch},\vK\vK'}^{\vq=0}(\omega=0,\nu,\nu'),
\end{equation} where a tail treatment for the fermionic Matsubara sum has been performed\footnote{\begin{minipage}[t]{\linewidth} Regarding the fermionic Matsubara sum, we first sum the connected part of the generalized susceptibility on its frequency grid independent of the summation of the bubble terms. Second, we sum the bubble contribution to $\chi_{\mathrm{ch},\vK\vK'}^{\vq=0}(\omega=0,\nu,\nu')$ on a large frequency grid of $\pm n$ available frequencies. Third, the part of the bubble not entailed in the larger frequency grid can be estimated analytically as follows \cite{Reitner2020DA}: \begin{equation}
    \chi_{0,\mathrm{ch}/\mathrm{sp}}^\vq \approx\frac{2}{\beta}\left(\sum_{\nu,\nu'}^{\pm n}\chi_{0,\mathrm{ch}/\mathrm{sp}}^\vq(\nu,\nu')+2\sum_{n}^{\infty}\delta_{\nu,\nu'}\frac{\beta}{\nu^2}\right)
\end{equation} with
\begin{equation}
    \frac{4}{\beta^2}\sum_n^{\infty}\delta_{\nu,\nu'}\frac{\beta}{\nu^2}=\frac{4\beta}{\pi^2}\left(\frac{\pi^2}{8}-\sum_{\tilde{n}=0}^{n}\frac{1}{(2\tilde{n}+1)^2}\right).
\end{equation}\end{minipage}}.
For finite doping, the latter expression shows, similar as $-\partial D/\partial U$, an actual divergence at the critical endpoint of the transition, as is the case in DMFT \cite{Kotliar2002,Walsh2019,Reitner2020}. However, for the unfrustrated model at half-filling, the respective contribution to $\chi_{\mathrm{ch}}^{\vq=0}(\omega\!=\!0)$ vanishes due to the perfect particle-hole symmetry \cite{vanLoon2020,Reitner2020,Kowalski2024}, which we explicitly verified for CDMFT and discuss in the following.

 Our two-particle analysis extends the results of DMFT \cite{Reitner2020,Kowalski2024}, by showing that also in CDMFT the divergences of these response functions can be directly traced to the divergence of a \emph{single} eigenvalue of the generalized uniform charge susceptibility. To identify such a critical eigenvalue $\lambda_{\mathrm{crit}}$ for the transition, we express the generalized lattice charge susceptibility in its orthonormal spectral representation: 
\begin{equation}
\label{Eq:Spectral_representation}
    \chi_{\mathrm{ch}}^{\vq=0,\vQ=0,\alpha\beta}=\sum_n V^n_{\alpha} \lambda_n V^n_{\beta},
\end{equation} where $V_{\alpha}^{n}$ is the respective eigenvector to the corresponding eigenvalue $\lambda_n$, and $\alpha,\beta$ represent now multi-indices with respect to fermionic frequencies and momenta $\vK,\vK'$. For the generalized charge susceptibility at $\vQ=(0,0)$, the leading eigenvalue is displayed in Fig.~\ref{fig:ch_inst}b) in blue circle markers for the interaction scan across the MIT. It is negative but very small for interactions $U\ll U_{c2}$, however it increases significantly close to the MIT, displaying an almost divergent behavior at $U_{c2}$.

Fig.~\ref{fig:ch_inst}c) displays the generalized charge response for the reciprocal lattice at $U=5.92t$, closely before the MIT, which displays a predominant butterfly-like structure similar for each momentum patch and which is anti-symmetric with respect to $\nu \rightarrow -\nu$ and $\nu \rightarrow -\nu'$. When comparing to the span of the eigenvectors corresponding to the leading eigenvalue $\lambda_{\mathrm{crit}}$ in Fig.~\ref{fig:ch_inst}d), we observe that this frequency structure is associated with the instability. At the same time, as anticipated from the vanishing physical charge response, the anti-symmetry of the frequency structure in the generalized charge response leads to a cancellation of the divergent contribution when conducting the fermionic Matsubara sums, similar as for the MIT in the DMFT \cite{Reitner2020}. We verified, that this does not hold anymore out of half-filling, where upon doping the eigenvalue generates a divergent contribution to the physical charge response. This cancellation does not hold for the quantity $-\partial D/\partial U$, even at half-filling, allowing its systematic divergence at the critical MIT endpoint.

From a more algorithmic perspective, it is worth stressing the following: The divergent eigenvalue of the generalized charge susceptibility as $\delta G/\delta\mathcal{G}^{-1}$ demonstrates that the system is instable towards an infinitesimal variation in the inverse retarded fermionic propagator $\delta G_0^{-1}$ at the MIT. Hence, in (C)DMFT, in contrast to other iterative many-electron approaches \cite{Essl2025}, the (in)stability condition of the corresponding self-consistent algorithm can only originate from thermodynamic (in)stability conditions \cite{Kotliar2000,Strand2011,vanLoon2020,vanLoon2022,vanLoon2024,Kowalski2024,Essl2025}.

\subsection{Effective instability threshold}
\begin{figure}
    \centering
    \includegraphics[width=0.37\linewidth]{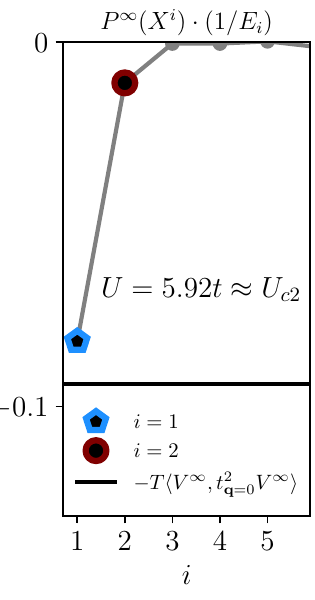}
    \includegraphics[width=0.617\linewidth]{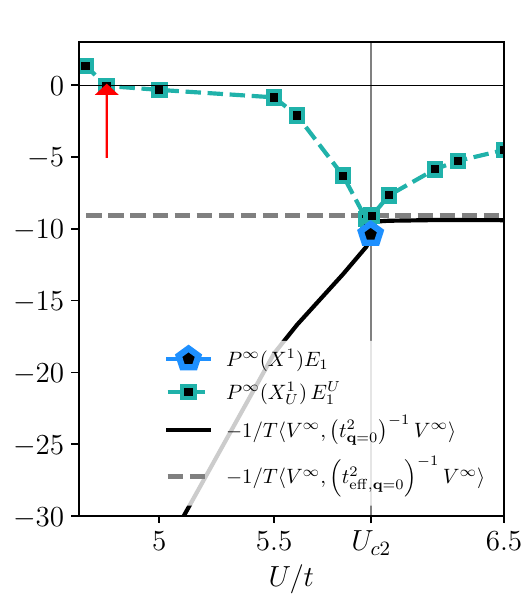}
    \caption{Left panel: Projections of eigenvalues of the impurity charge response function onto the leading eigenvector of the lattice charge response (gray) at $U_{c2}$. The largest contributing eigenvalue $E_1$ of the impurity susceptibility is highlighted in blue, to which we can associate the eigenvector $X^1$. The black line encodes the effects of the inverse bubble difference $t^2_{\vq=0}$, which serves as a threshold for the divergence of the lattice charge response with the respective symmetry. Right panel: the contribution in the direction of the dominant eigenvector $X^1$ to the cluster-impurity charge response (blue dot) at $U_{c2}$. We follow the closest eigenvector to $X^1$ of the impurity charge response over the interaction $X^1_U$. The red arrow indicates the location of the charge vertex divergence at this temperature $T=1/15t$. The black line gives the projected inverse bubble difference as a threshold for the MIT and the gray dashed line is its strong coupling approximation.}
    \label{fig:Eigenvalues_to_inverse_bubble_U=5.92}
\end{figure}

Beyond the general eigenvalue analysis as an indication of a (potential) divergence of a thermodynamic response of CDMFT, as discussed above, it is also of interest how such divergences might be driven by the information encoded in the two-particle properties of the underlying cluster impurity problem, extending the studies hitherto made a the purely single-site DMFT level only. In CDMFT, strong correlations which render the thermodynamic lattice quantities instable, originate from the impurity response. They are connected via the \emph{real-space} superlattice Eqs.~(\ref{Eq:Gamma}-\ref{Eq:Chi_c_q}):

\begin{align}
\chi_{\mathrm{ch}}^{\vq,\alpha,\beta}&=\left( \left[\chi^{-1}_{\mathrm{ch}}\right]^{\alpha,\beta}+\left[\chi^{\vq,-1}_{0}\right]^{\alpha,\beta}-\left[\chi^{-1}_{0}\right]^{\alpha,\beta}\right)^{-1}\\
&=\left( \left[\chi^{-1}_{\mathrm{ch}}\right]^{\alpha,\beta}+T t_{\vq}^2\,^{\alpha,\beta}\right)^{-1}\label{Eq:impurity_to_lattice}\\
    &\approx\left( \left[\chi^{-1}_{\mathrm{ch}}\right]^{\alpha,\beta}+T t_{\mathrm{eff},\vq}^2\,^{\alpha,\beta}\right)^{-1}.
\end{align} 
The effect of the lattice onto the two-particle quantity can thus be analyzed as the difference of the inverse lattice bubble and the inverse impurity bubble $\mathfrak{t}_{\mathbf{q}}^{2\,\alpha,\beta}$ \cite{Kowalski2024}, for which, similar to DMFT \cite{Moghadas2025}, a strong coupling approximation $\mathfrak{t}_{\mathrm{eff},\mathbf{q}}^{2\,\alpha,\beta}$ can be found. Specifically, we can approximate the difference of the inverse lattice bubble and the inverse impurity bubble as a second order expansion of the respective Green function in the hopping matrix Eq.~(\ref{eq:hopping}):
\begin{equation}
\begin{split}   
    \mathfrak{t}_{\mathrm{eff},\mathbf{q}}^{2\,\alpha,\beta} \coloneqq \delta_{\nu\nu'} &\left[ \frac{1}{N_\mathbf{k}}\sum_\mathbf{k} \varepsilon^{\mathrm{il}}_\mathbf{k}\, \varepsilon^{\mathrm{hj}}_\mathbf{k+q}\right. \\
    &\left.-\left(\frac{1}{N_\mathbf{k}}\sum_\mathbf{k} \varepsilon^{\mathrm{il}}_\mathbf{k}\right)\left(\frac{1}{N_\mathbf{k}}\sum_\mathbf{k} \varepsilon^{\mathrm{hj}}_\mathbf{k}\right)\right].
\end{split}
\end{equation}
As mentioned above, such an approximation is valid for the strong coupling case, where the self-energy contribution is large in comparison to the hopping term in the Green function. For details and derivations see App.~\ref{App:Strong-coupling-BSE}.\\
Remarkably, while the quantity $t_{\vq}^2\,^{\alpha,\beta}$ is, in general,  a frequency-dependent diagonal matrix in Matsubara frequency space, its strong coupling approximation  $\mathfrak{t}_{\mathrm{eff},\mathbf{q}}^{2\,\alpha,\beta}$ does no longer depend on frequency. Evidently, this greatly simplifies the investigation of the link between the eigenspectral properties of the lattice and of the impurity system at the phase-instability.
In particular, as the lattice generalized susceptibility in the proximity of the the critical endpoint of the MIT is fully dominated by the eigenvalue
$\lambda_{\mathrm{crit}}$, we can approximate its spectral representation in Eq.~(\ref{Eq:Spectral_representation}) as follows:
\begin{equation}
    \label{Eq:eigenvector_charge_diverence}
\left(\chi_{\mathrm{ch}}^{-1}+Tt_{\vq=0}^2\right)^{-1}=\chi_{\mathrm{ch}}^{\vq=0}\approx V^{\text{crit}} \lambda_{\text{crit}} V^{\text{crit},-1},
\end{equation} where we omitted the multi-indices $\alpha,\beta$ for brevity. For an orthonormal basis we study the effect of the impurity quantities in the direction of the leading superlattice eigenvector. Specifically, for a diverging $ \lambda_{\infty}\coloneq \lambda_{\text{crit}}(U_{c2})$ and using the notation of the scalar product $\langle.,.\rangle$ \footnote{We use the right- and left hand side eigenvectors for the respective sides of the scalar product.}, we can write Eq.~(\ref{Eq:eigenvector_charge_diverence}):
\begin{equation}
\label{Eq:leading_Eigenvector_charge}
    0=\frac{1}{\lambda_\infty}=\langle V^{\infty},\left(\chi_{\mathrm{ch}}^{-1}+Tt_{\vq=0}^2\right)V^{\infty}\rangle,
\end{equation}
which gives a necessary condition for the divergence of the lattice charge response. Further, we can exploit the spectral decomposition of the impurity susceptibility
\begin{equation}
\chi_{\mathrm{ch}}=\sum_i X^{i}E_i X^{i,-1}
\end{equation} and find
\begin{equation}
\begin{split}
\label{Eq:overlapp_impurity_charge_respone}
    -T\langle V^{\infty}, t_{\vq=0}^2 V^{\infty}\rangle&=\langle V^{\infty},\chi_{\mathrm{ch}}^{-1}V^{\infty}\rangle\\
    &=\sum_i\langle V^{\infty},X^{i}\rangle\frac{1}{E_i}\langle X^{i},V^{\infty}\rangle\\
    &=\sum_iP^\infty(X^i)\frac{1}{E_i},
\end{split}
\end{equation} where we introduced the projection $P^\infty(X^i)$ of the eigenvector $X^i$ of the cluster-impurity charge susceptibility onto the leading eigenvector of the lattice charge response $V^\infty$.
The contributions of the right- and left hand side of this equation, respectively, are displayed in the left plot of Fig.~\ref{fig:Eigenvalues_to_inverse_bubble_U=5.92}: The gray line is a guide to the eye for the summands of the right hand side of Eq.~(\ref{Eq:overlapp_impurity_charge_respone}), which are the eigenvalues of the cluster-impurity charge response $E_i$ projected to the leading lattice eigenvector $V^\infty$, in descending order. Their sign is determined by the sign of the eigenvalues $E_i$ of the cluster-impurity charge response, and is negative. As it can be readily seen, only one eigenvalue ($i=1$) dominates, whereas the second contribution ($i=2$) is of marginal significance and the remaining summands are basically vanishing. The additional, crucial piece of information is provided, however, by the projection of the inverse bubble difference $t_{\vq=0}^2$ onto the leading lattice eigenvector (black line). As stated in Eq.~(\ref{Eq:eigenvector_charge_diverence}), the lattice charge response diverges if both terms are equal, which implies that at least one $E_i$ must be negative at the MIT.

In order to quantitatively analyze the role played by the relevant eigenvalue of the impurity charge response, and its evolution as a function of interaction, we can invert  Eq.~(\ref{Eq:leading_Eigenvector_charge}) and perform the spectral decomposition for the inverted hopping and charge-response matrices. This amounts to focusing on the contribution of $E_1$:
\begin{equation}
\begin{split}
    -\frac{1}{T}\langle V^{\infty}, \left(t_{\vq=0}^2\right)^{-1} V^{\infty}\rangle&\approx P^\infty(X^1) E_1,
\end{split}
\end{equation} displayed in Fig.~\ref{fig:Eigenvalues_to_inverse_bubble_U=5.92}. Here, the black line defines the threshold for the MIT from the inverse bubble difference $t_{\vq=0}^2$ and the inverse dominant eigenvalue is again depicted as a blue pentagon marker at $U_{c2}$. We follow the contribution of the impurity charge eigenvectors projected onto $V^\infty$ for different interaction values, by singling out for each interaction value the respective impurity eigenvector $X^1_U$  (green square marker line), offering the largest overlap with the dominant impurity eigenvector $X^1_{U_{c2}}$ at $U_{c2}$. We can observe that such a contribution starts  at positive values for low interactions, displaying then a change in sign of its eigenvalue, indicated by the red arrow at low interactions $\tilde{U}\approx4.77t$. This change of sign is evidently equivalent to the divergence of the irreducible charge vertex discussed in Sec.~\ref{sec:resultsVertex}. When approaching the MIT by further increasing the interaction value ($U \rightarrow U_{c2}$), the associated contribution decreases further towards the negative instability threshold (black line) at the critical interaction $U_{c2}$. Here, the strong coupling approximation of the Bethe-Salpeter equation (gray dashed line) yields a fairly good description of the instability threshold. Hence, also in CDMFT, the evolution of the green line from positive values at weak coupling to the negative values at the instability threshold via a vertex divergence is a \textit{necessary} condition for the MIT instability. It is worth noticing that this finding might have relevant implications for the non-perturbative correlation-driven enhancement of the electron-phonon scattering in two-dimensional strongly correlated systems \cite{Moghadas2025}.

\begin{figure*}
    \centering
    \includegraphics[width=\linewidth]{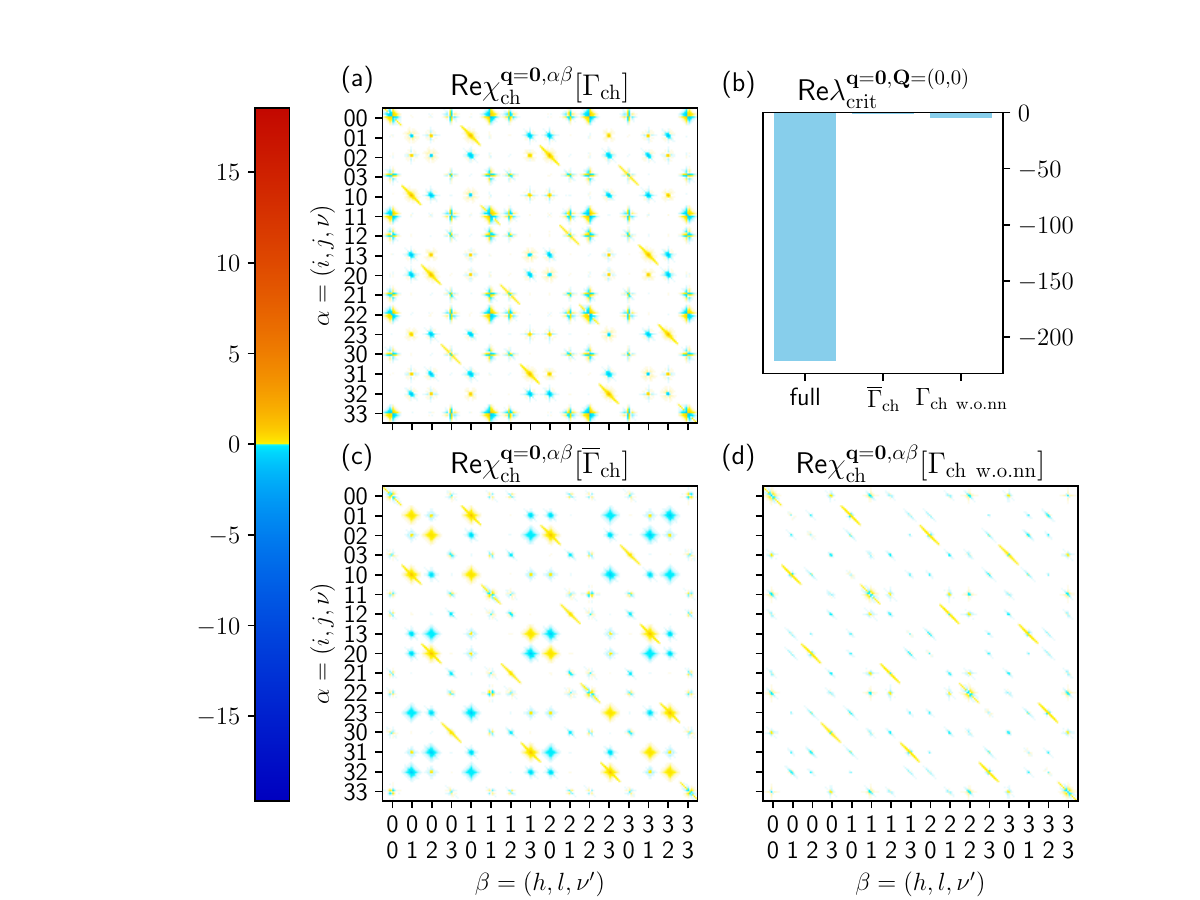}
    \caption{Analysis of the real-space charge response on the reciprocal superlattice for the full $\Gamma_c$ for the data point closest to the MIT on the metallic side for $T=0.066t$ in a). b) The leading eigenvalues for the full vertex and the vertex modified in two ways: $\overline{\Gamma}_{\mathrm{ch}}$ with all SBE spin diagrams in the charge channel subtracted as in Eq.~\ref{eq:mod_gamma} and $\overline{\Gamma}_{\mathrm{ch}}$ with missing SBE spin diagrams for nearest-neighbor distances, subtracted on the level of the two-particle irreducible charge vertex.
    c) and d) give the corresponding charge response on the reciprocal superlattice. }
    \label{fig:ch_inst_SBE}
\end{figure*}
\subsection{Importance of spin-diagrams to MIT}
While we discussed the role played by the short-range antiferromagnetic fluctuations for the non-perturbative vertex divergence in Sec.~\ref{Sec:Vertex_spin}, a recent real-space analysis for CDMFT \cite{Meixner2025a} argued that the nearest-neighboring spin-boson diagrams are of crucial relevance for the one-particle spectral properties of the MIT in the CDMFT.
\begin{figure}
    \centering
    \includegraphics[width=0.9\linewidth]{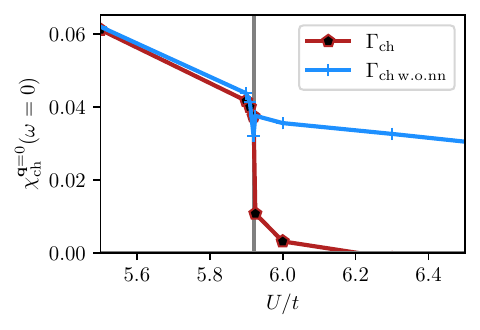}
    \caption{Suppression of the uniform charge response at the MIT for $T=0/15t$ for the full BSE scheme (red pentagon marker) and without the nearest-neighbor SBE spin-transverse diagram in $\Gamma_c$ (blue $+$-marker).}
    \label{fig:chq0_c_manip}
\end{figure}
Motivated by this result, we study here the impact of non-local spin fluctuations onto the two-particle properties relevant for the Mott MIT, aiming at an analysis of the real-space symmetry of the associated instabilities on the superlattice cluster level. In particular, we will now focus on the generalized charge susceptibility for the superlattice at the MIT, before Fourier-transforming it to the reciprocal lattice via Eq.~(\ref{Eq:ToReciprocal_ph}). This quantity, shown in Fig.~\ref{fig:ch_inst_SBE}a), contains all thermodynamic information discussed in the previous section, but on the real-space level. In particular, we observe the following points:
\begin{itemize}
    \item First, we focus on the real-space on-site index combination $i,j=(0,0)$ and $h,l=(0,0)$ in the top left hand part of the plot. Here, the butterfly-like frequency structure, which was found to diverge at the MIT in $\chi_{\mathrm{ch}}^{\vQ=0}$ is clearly visible. This frequency structure shows up for all on-site contributions $i=j$ and $h=l$.
 \item Second, the yellow features at, e.g., $i,j=(0,1)$ and $h,l=(1,0)$ and the corresponding blue features in the real-space patch below $i,j=(0,2)$ and $h,l=(1,0)$ display nearest-neighbor anti-symmetry. 
\end{itemize} 
The leading eigenvalue of the corresponding full reciprocal lattice generalized charge susceptibility is given in Fig.~\ref{fig:ch_inst_SBE}b) by the bar with the label ``full''. In order to single out the specific impact of non-local spin correlations, we now present the leading eigenvalues of the lattice charge susceptibility which we again manipulate in two different ways: i) we calculate via the modified vertex $\overline{\Gamma}_{\mathrm{ch}}$ in Eq.~(\ref{eq:mod_gamma}) which we used to analyze the non-local contributions to the vertex divergence a modified $\chi^{\vq=\mathbf{0}}_{\mathrm{ch}}[\overline{\Gamma}_{\mathrm{ch}}]$ via the BSE, and, ii) we subtract from the full $\Gamma_{\mathrm{ch}}$ the nearest-neighbor $\overline{\mathrm{sp}}$-SBE diagrams to obtain $\Gamma_{\mathrm{ch w.o.nn}}$, and \emph{then} compute the lattice generalized susceptibility $\chi^{\vq=\mathbf{0}}_{\mathrm{ch}}[\Gamma_{\mathrm{ch w.o.nn}}]$.

Evidently, in both cases, we do \emph{not} find any longer a critical (diverging) eigenvalue for $\vQ=(0,0)$, as indicated by the vanishing blue bars in b). The corresponding response functions are shown in c) and d).

In addition, when performing the fermionic Matsubara sum as discussed earlier, the physical charge response for this manipulated charge vertex  Fig.~\ref{fig:chq0_c_manip} displays only a minor suppression in contrast to the full one at the MIT. This further underlines the importance of nearest-neighbor spin diagrams to the MIT in CDMFT.

Hence, our data unequivocally demonstrate the pivotal role of non-local spin single-boson diagrams, and specifically of the non-local nearest-neighbor diagrams, associated with short-range anti-ferromagnetic fluctuations, in driving the charge sector in the proximity of the Mott MIT in CDMFT. In fact, their removal directly suppress the two-particle seed for the charge instabilities associated to the Mott MIT.

\section{Conclusion and Outlook}
\label{sec:conclusion}

In our paper, we have presented a systematic two-particle CDMFT investigation of how short-range correlations drive the breakdown of self-consistent perturbation theory in two-dimensional correlated systems, overcoming the purely local picture of most preceding works.

To this aim a general, self-contained derivation of the Bethe-Salpeter formalism for CDMFT, necessary to compute thermodynamic (and, in perspective) Kubo response-functions in all scattering sectors, has been explicitly presented. In this context, we have been also able to rigorously illustrate important properties related to the Ward identities linking the CDMFT quantities to the one- and two-particle level.

Our numerical results for the half-filled square lattice Hubbard model allowed for a thorough discussion of the first line where the irreducible vertex diverges, marking the breakdown of perturbation theory. The CDMFT results have been put in relation to those obtained in the purely local picture of DMFT and have been interpreted in terms of a temperature-dependent interplay between short-ranged spin fluctuations (at intermediate $T$) and on-site Kondo screening effects (at low $T$).
Further, the comparison with two-particle calculations of isolated clusters has made possible a well-grounded speculation on how the perturbative breakdown should occur in the thermodynamic limit, i.e., for the exact solution of the unfrustrated Hubbard model in two dimensions.

Our two-particle investigation has then been extended to higher interaction values, where the Mott MIT occurs in CDMFT. This has allowed us to highlight the specific role played by the short-range transversal spin fluctuations in driving the thermodynamic instabilities associated to the MIT in the charge sector, as well as to demonstrate the intrinsic link of the Mott MIT to the preceding breakdown of the perturbation theory, beyond the purely local description of DMFT \cite{Reitner2020,vanLoon2020,Kowalski2024}. 

Our findings call for future studies of the impact of nonlocal correlations over increasing spatial distances on the perturbative breakdown in two dimensions. Perspectively, this might further allow for a comparison with the foreseen trend of non-perturbative effects towards the thermodynamic/exact limit. Let us finally note that ($2\times 2$) CDMFT has been employed to study instabilities of the Hubbard model for a more realistic parameter set, i.e., out of particle-hole symmetry, to make connection to the phase diagram of high temperature superconductors such as cuprates \cite{Lichtenstein2000,Sordi2011,Sordi2019,Harland2019,Walsh2023}. Our method provides the instabilities in an unbiased way and for all possible momentum symmetries, and therefore can help to further unveil, on a fundamental level, the multi-faceted interplay of nonlocal correlations and support the understanding of complex materials, which are a manifestation of the non-perturbative many-electron physics in two-dimensions.

\acknowledgments 
We would like to thank M\'ario Malcolms de Oliveira, Sabine Andergassen, Marcel Kr\"amer, Emin Moghadas, Herbert E{\ss}l, and Dominik Kiese for fruitful discussions. We further thank Demetrio Vilardi for critical reading of our manuscript. The authors acknowledge the members of the computer service facility of the MPI-FKF for their help. We acknowledge financial support from the Austrian Science Fund (FWF) through the projects with Grant DOIs 10.55776/I5487 (M.R.) and 10.55776/I5868 (A.T.), which is the P1 project part of the QUAST research unit of the German Research Foundation (DFG), for 5249, as well as through P4 of the research unit QUAST, for 5249 ID No. 449872909 (T.S.). M.R.~further acknowledges support as a recipient of a DOC fellowship of the Austrian Academy of Sciences.‌ Further, we appreciate the cooperation with Nils Wentzell of the Center for Computational Quantum Physics at the Flatiron Institute (New York City) of the Simons Foundation.
\appendix
\section*{Appendices}
\section{Cluster Kondo temperature} \label{App:Kondo}
\begin{figure}
    \centering
    \includegraphics[width=0.85\linewidth]{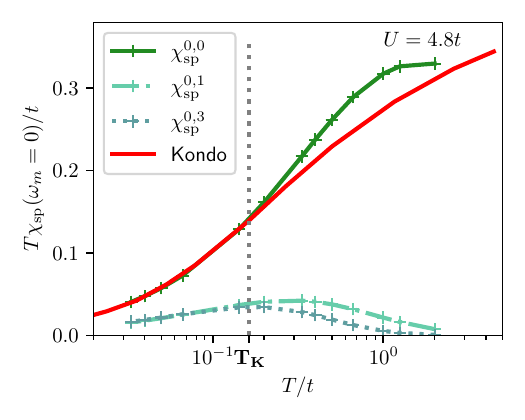}
    \caption{Analysis of the low temperature magnetic response. The green lines give the on-site $\chi_\mathrm{sp}^{0,0}$ the next neighbor $\chi_\mathrm{sp}^{0,1}$ and the second next neighbor $\chi_\mathrm{sp}^{0,2}$ magnetic response, while the red bold line gives the universal Kondo behavior. The gray dotted, vertical line indicates the Kondo temperature $T_K=0.14t$.}
    \label{fig:T_K}
\end{figure}

One established procedure to determine the Kondo temperature (see e.g. \cite{Krishna-murthy1975,Krishna-murthy1980,Chalupa2021}), displayed in Fig.~\ref{fig:T_K}, is to fit the universal magnetic response of the Kondo model (red bold line), which displays a suppression of the magnetic response, to the on-site impurity magnetic response function $T\chi^{0,0}_\mathrm{sp}(\omega_m=0)$. The necessary shift of the Kondo magnetic response \cite{Krishna-murthy1980} on the temperature axis gives the Kondo temperature $T_K$, indicated by the gray dotted line. Below the Kondo temperature, we find excellent agreement of the magnetic response of our system to the Kondo magnetic response. A detailed explanation of the procedure may be found in the Supplemental Material of \cite{Chalupa2021}. For the Hubbard model considered, we estaimated $T_K$ for each point in the $(U,T)$ plance separately, since the respective self-consistently determined impurity model changes with those parameters.
\section{Definitions and derivations}\label{App:Def}
\subsection{Derivation of the particle-particle Bethe-Salpeter equation}
For the numeric computation of the real-space BSE, a matrix form is necessary. To obtain the particle-particle BSE in this form as given in Sec.~\ref{sec:modelandmethods}, frequency- and index shifts are necessary, which we discuss here in detail. We define the generalized susceptibility in the particle-particle (pp) channel as (cv. Eq.~(7) of \cite{Rohringer2012}):
\begin{equation}
\begin{split}
\chi^{\alpha|\beta}_{\mathrm{pp},\sigma\sigma'}&(\omega)
\coloneq\chi^{\mathrm{ih|jl}}_{\mathrm{pp},\sigma\sigma'|\sigma\sigma'}(\omega,\nu,\nu')\\=&\int_0^\beta \mathrm{d}\tau_1 \mathrm{d}\tau_2 \mathrm{d}\tau_3 \mathrm{e}^{-\i\nu(\tau_1)} \mathrm{e}^{\i(\omega-\nu')\tau_2}\mathrm{e}^{-\i(\omega-\nu)\tau_3} \\&\left[\langle T_{\tau}c_{\mathrm{i},\sigma}^\dagger (\tau_1)c_{\mathrm{j},\sigma}(\tau_2)c^\dagger_{\mathrm{h},\sigma'}(\tau_3)c_{l,\sigma'}(0)\rangle \right]\\&-\beta\delta(\omega-\nu'-\nu)G^{\mathrm{ji}}_{\sigma}(\nu)G^{\mathrm{lh}}_{\sigma'}(\nu'),
\end{split}
\end{equation} where we use the notation ``$\text{outgoing}|\text{incoming}$", and place the indices in the order of their appearance. 
The generalized susceptibility can be expressed as Feynman diagrams:
\begin{equation}
    \begin{split}
&\chi^{\mathrm{ih|jl}}_{\mathrm{pp},\sigma\sigma'|\sigma\sigma'}(\omega,\nu,\nu') \\&=
    \begin{tikzpicture}[baseline=0.cm]
        \begin{feynhand}
            \node at (-1.4,0.7) {$\mathrm{h}$};
            \node at (-1.4,-0.5) {$\mathrm{i}$};
            \node at (1.4,-0.5) {$\mathrm{l}$};
            \node at (1.4,0.7) {$\mathrm{j}$};
            \node at (0,-0.8) {$\beta \delta(\omega-\nu'-\nu)$};
            \vertex (a) at (-1.2,0.7);
            \vertex (u) at (0,0.1);
            \vertex (b) at (-1.2,-0.5); 
            \vertex (c) at (1.2,-0.5); 
            \vertex (d) at (1.2,0.7); 
            \propag [plain] (d) to (u);
            \propag [plain] (c) to (u);
            \propag [fer] (u) to (b);
            \propag [fer] (u) to (a);
        \end{feynhand}
    \end{tikzpicture}
    -
    \begin{tikzpicture}[baseline=0.cm]
        \begin{feynhand}
            \node at (-1.4,1) {$\mathrm{h},\sigma',\omega-\nu$};
            \node at (-1.4,-0.8) {$\mathrm{i},\sigma,\nu$};
            \node at (1.4,-0.8) {$\mathrm{l},\sigma',\nu'$};
            \node at (1.4,1) {$\mathrm{j},\sigma,\omega-\nu'$};
            \vertex (a) at (-1.2,0.7);
            \vertex (b) at (-1.2,-0.5); 
            \vertex (c) at (1.2,-0.5); 
            \vertex (d) at (1.2,0.7); 
            \vertex (u) at (-0.6,0.7);
            \vertex (v) at (-0.6,-0.5); 
            \vertex (w) at (0.6,-0.5); 
            \vertex (x) at (0.6,0.7); 
            \fill[TUgreen] (v) rectangle (x);
            \node at (0,0.1) {$G^{(2)}$};
            \propag [antfer] (a) to (u);
            \propag [antfer] (x) to (d);
            \propag [antfer] (b) to (v);
            \propag [antfer] (w) to (c);
            \propag (u) to (x);
            \propag (v) to (w);
        \end{feynhand}
    \end{tikzpicture}, \\
\end{split}
\end{equation}
where we switched the legs $\mathrm{h}$ and $\mathrm{j}$ relative to the ph-notation (in contrast to Rohringer \cite{Rohringer2013}), resulting in a global minus sign and similar we define the particle-particle transverse channel:
\begin{equation}
\begin{split}
\chi^{\alpha|\beta}_{\mathrm{pp},\overline{\sigma\sigma'}}&(\omega)
\coloneq\chi^{\mathrm{ih|jl}}_{\mathrm{pp},\sigma\sigma'|\sigma'\sigma}(\omega,\nu,\nu')\\=&\int_0^\beta \mathrm{d}\tau_1 \mathrm{d}\tau_2 \mathrm{d}\tau_3 \mathrm{e}^{-\i\nu(\tau_1)} \mathrm{e}^{\i(\omega-\nu')\tau_2}\mathrm{e}^{-\i(\omega-\nu)\tau_3} \\&\left[\langle T_{\tau}c_{\mathrm{i},\sigma}^\dagger (\tau_1)c_{\mathrm{j},\sigma'}(\tau_2)c^\dagger_{\mathrm{h},\sigma'}(\tau_3)c_{l,\sigma}(0)\rangle \right]\\&-\beta\delta_{\sigma,\sigma'}\delta(\omega-\nu'-\nu)G^{\mathrm{jl}}_{\sigma}(\nu)G^{\mathrm{lh}}_{\sigma'}(\nu'),
\end{split}
\end{equation} which results in the real-space crossing relation
\begin{equation}
\begin{split}
    &\chi_{\mathrm{pp},\overline{\sigma\sigma'}}^{\mathrm{ih|jl}}(\omega,\nu,\nu')-\chi_{0,\mathrm{pp},\overline{\sigma\sigma'}}^{\mathrm{ih|jl}}(\omega,\nu,\nu')\\&\quad=-\chi_{\mathrm{pp},\sigma\sigma'}^{\mathrm{ih|lj}}(\omega,\nu,\omega-\nu')+\delta_{\sigma\sigma'}\chi_{0,\mathrm{pp},\sigma\sigma'}^{\mathrm{ih|lj}}(\omega,\nu,\omega-\nu')\\
    &\Gamma_{\mathrm{pp},\overline{\sigma\sigma'}}^{\mathrm{ih|jl}}(\omega,\nu,\nu')=-\Gamma_{\mathrm{pp},\sigma\sigma'}^{\mathrm{hi|jl}}(\omega,\omega-\nu,\nu') 
\end{split}
\end{equation}
The bubble part contained in the generalized susceptibility reads
\begin{equation}
\begin{split}
    \chi^{\mathrm{ih|jl}}_{0,\mathrm{pp},\sigma\sigma'}(\omega,\nu,\nu')&=-\beta\delta(\nu-\nu') G^{\mathrm{li}}_\sigma (\nu)G^{\mathrm{jh}}_{\sigma'}(\omega-\nu'),\\
\end{split}
\end{equation}which is independent of $\sigma,\sigma'$ for the SU(2) symmetric case.

Following Rohringer \cite{Rohringer2013a} Eq.~(B.22), the Bethe-Salpeter equation can then be written as
\begin{equation}
\begin{split}
    &F_{\mathrm{pp},\uparrow\downarrow}^{\mathrm{ih|jl}}(\omega,\nu,\nu')= 
    \Gamma_{\mathrm{pp},\uparrow\downarrow}^{\mathrm{ih|jl}}(\omega,\nu,\nu')\\ &-\frac{1}{\beta}\sum_{\nu_1, \mathrm{abmn}}\left[F_{\mathrm{pp}}^{\mathrm{ih|mn}}(\omega,\nu,\omega-\nu_1)\,\chi_{0,\mathrm{pp}}^{\mathrm{mn|ab}}\,\Gamma^{\mathrm{ba|jl}}_{\mathrm{pp}}(\omega,\nu_1,\nu')\right].
\end{split}
\end{equation}
Employing the real-space crossing relation for $F$ we have
\begin{equation}
\chi_{\mathrm{pp},\sigma\sigma'}^{\mathrm{go|ba}}=-\chi_{0,\mathrm{pp}}^{\mathrm{go|hi}}F_{\mathrm{pp},\sigma\sigma'}^{\mathrm{ih|mn}}\chi_{0,\mathrm{pp}}^{\mathrm{nm|ba}}+\delta_{\sigma\sigma'}\chi_{0,\mathrm{pp}}^{\mathrm{go|ba}}
\end{equation}
and relabeling the indices leads to the real-space equivalent of [\cite{Rohringer2013a} Eq.~(B.24)]:
\begin{equation}
\begin{split}
    \chi_{\mathrm{pp},\uparrow\downarrow}^{\mathrm{ih|lj}}(\omega,\nu,\omega-\nu')&=\\-\frac{1}{\beta^2}\sum_{\nu_1,\nu_2,abcd}&\left[\chi^{\mathrm{ih|ba}}_{0,\mathrm{pp}}(\omega,\nu,\nu_1)-\chi_{\mathrm{pp},\uparrow\downarrow}^{\mathrm{ih|ab}}(\omega,\nu,\omega-\nu_1)\right]\\&\cdot\Gamma_{\mathrm{pp}}^{\mathrm{ab|dc}}(\omega,\nu_1,\omega-\nu_2)\chi_{0,\mathrm{pp}}^{\mathrm{dc|jl}}(\omega,\nu_2,\nu')
\end{split}
\end{equation}
where the indices $\mathrm{l,j}$ have been exchanged relative to the definition Eq.~(\ref{Eq:Def_pp}) due to the exploitation of crossing relations Eq.~(\ref{Eq:crossing}). This equation can be brought to matrix form by switching the incoming legs $\tilde{\chi}_{\mathrm{pp},\uparrow\downarrow}^{\mathrm{ih|jl}}(\omega,\nu,\nu')=\chi_{\mathrm{pp},\uparrow\downarrow}^{\mathrm{ih|lj}}(\omega,\nu,\omega-\nu')$. 
This results in the real-space equivalent to Eq.~(B.25) of \cite{Rohringer2013a}:
\begin{equation}
\begin{split}
\tilde{\chi}_{\mathrm{pp},\uparrow\downarrow}^{\mathrm{ih|jl}}(\omega,\nu,\nu')&=\\-\frac{1}{\beta^2}\sum_{\nu_1,\nu_2,\mathrm{abcd}}&\left[\chi^{\mathrm{ih|ba}}_{0,\mathrm{pp}}(\omega,\nu,\nu_1)-\tilde{\chi}_{\mathrm{pp}}^{\mathrm{ih|ba}}(\omega,\nu,\nu_1)\right]\\&\cdot\tilde{\Gamma}_{\mathrm{pp}}^{\mathrm{ab|cd}}(\omega,\nu_1,\nu_2)\chi_{0,\mathrm{pp}}^{\mathrm{dc|jl}}(\omega,\nu_2,\nu').
\end{split}
\end{equation}
We now switch to multi-index notation, gathering:
\begin{equation}
\begin{split}
    \alpha&=\mathrm{i,h},\nu\\
    \beta&=\mathrm{a,b},\nu_1\\
    \gamma&=\mathrm{d,c},\nu_2\\
    \delta&=\mathrm{l,j},\nu',
\end{split}
\end{equation} which results in
\begin{equation}
\begin{split}
    \tilde{\chi}_{\mathrm{pp},\uparrow\downarrow}^{\alpha|\delta}=-\frac{1}{\beta^2}\left[\chi_{0,\mathrm{pp}}-\tilde{\chi}_{\mathrm{pp}}^{\alpha|\beta}\right]\cdot\tilde{\Gamma}_{\mathrm{pp}}^{\overline{\beta}|\overline{\gamma}}\,\chi_{0,\mathrm{pp}}^{\gamma|\delta},
\end{split}
\end{equation} where repeated indices in this matrix equation are to be summed over. A feynman-diagrammatic illustration of this equation can be found in Fig.~\ref{fig:BSE_pp}.
This matrix equation can then be inverted similar to Eq.~(B.26) in \cite{Rohringer2013a}:
\begin{equation}
\tilde{\Gamma}_{\mathrm{pp}}^{\overline{\beta}|\overline{\gamma}}=\beta^2\left[\left(\tilde{\chi}_{\mathrm{pp}}-\chi_{0,\mathrm{pp}}\right)^{-1}+\left(\chi_{0,\mathrm{pp}}\right)^{-1}\right]^{\beta|\gamma}.
\end{equation}

The momentum dependent pp-response $\tilde{\chi}_{\mathrm{pp},\uparrow\downarrow}^{\vq,\alpha|\delta}$ in CDMFT can then be obtained from
\begin{equation}
\label{Eq:Chi_pp_q_app}
\begin{split}
     \tilde{\chi}&_{\mathrm{pp},\uparrow\downarrow}^{\vq,\alpha|\delta}=\chi^{\vq,\alpha,\beta}_{0,pp}+\left( T^2\tilde{\Gamma}_{pp}^{\overline{\alpha},\overline{\delta}}-\left[\chi^{\vq,\alpha,\beta}_{0,pp}\right]^{-1}\right)^{-1}.
\end{split}
\end{equation}
\subsection{Strong coupling expression for the cluster Bethe-Salpeter equation}
\label{App:Strong-coupling-BSE}
At strong coupling, when the dispersion relation $\varepsilon^{ij}_{\mathbf{k}}$ in the denominator of the Green's function $G^{ij}_\mathbf{k}(\nu)$ becomes small in respect to its self-energy $\Sigma^{ij}(\nu)$ (or when $\i\nu$ becomes large for $\nu\to\pm \infty$), the BSE can be brought into an approximate form (cf.~the Supplemental of Ref.~\cite{Moghadas2025} for the corresponding single site DMFT expression):
\begin{equation}
\label{eq:strong_coup_bse}
    \begin{split}
\chi_{\mathrm{ch}/\mathrm{sp}}^{\vq,\alpha,\beta}
&=\left( T^2\Gamma_{\mathrm{ch}/\mathrm{sp}}^{\overline{\alpha},\overline{\beta}}+\left[\chi^{\vq,\alpha,\beta}_{0}\right]^{-1}\right)^{-1}\\
&=\left( \left[\chi^{\alpha,\beta}_{\mathrm{ch}/\mathrm{sp}}\right]^{-1}-\left[\chi^{\alpha,\beta}_{\mathrm{ch}/\mathrm{sp},0}\right]^{-1}+\left[\chi^{\vq,\alpha,\beta}_{0}\right]^{-1}\right)^{-1}\\
&\approx \left( \left[\chi^{\alpha,\beta}_{\mathrm{ch}/\mathrm{sp}}\right]^{-1} + T\, \mathfrak{t}_{\mathrm{eff},\mathbf{q}}^{2\,\alpha,\beta}\right)^{-1}.
    \end{split}
\end{equation}
Here, in the second line of Eq.~(\ref{eq:strong_coup_bse}), we first inserted the BSE for $\Gamma_{\mathrm{ch}/\mathrm{sp}}$ on the cluster  and then, in the third line, for the strong coupling approximation we introduced
\begin{equation}
\begin{split}   
    \mathfrak{t}_{\mathrm{eff},\mathbf{q}}^{2\,\alpha,\beta} \coloneqq \delta_{\nu\nu'} &\left[ \frac{1}{N_\mathbf{k}}\sum_\mathbf{k} \varepsilon^{\mathrm{il}}_\mathbf{k}\, \varepsilon^{\mathrm{hj}}_\mathbf{k+q}\right. \\
    &\left.-\left(\frac{1}{N_\mathbf{k}}\sum_\mathbf{k} \varepsilon^{\mathrm{il}}_\mathbf{k}\right)\left(\frac{1}{N_\mathbf{k}}\sum_\mathbf{k} \varepsilon^{\mathrm{hj}}_\mathbf{k}\right)\right].
\end{split}
\end{equation}
 This expression can be obtained in analog to the derivation in the Supplemental of Ref.~\cite{Moghadas2025}.

To see this, let us first introduce a compact matrix notation on the cluster for convenience, for which
the definitions of the cluster and lattice bubble read [here denoted with the tensor product $\otimes$, cf.~Eqs.~(\ref{eq:loc_bubble}) and (\ref{eq:latt_bubble})]:
\begin{align}
    \bm{\chi}_0 &= - \frac{\delta_{\nu\nu'}}{T}\left(\frac{1}{N_\mathbf{k}}\sum_\mathbf{k} \bm{G}_\mathbf{k}(\nu)\right) \otimes \left(\frac{1}{N_\mathbf{k}}\sum_\mathbf{k} \bm{G}_\mathbf{k}(\nu)\right), \\
    \bm{\chi}^{\mathbf{q}}_0 &= - \frac{\delta_{\nu\nu'}}{T}\frac{1}{N_\mathbf{k}}\sum_\mathbf{k} \bm{G}_\mathbf{k}(\nu)\otimes  \bm{G}_\mathbf{k+q}(\nu),
\end{align}
where 
\begin{equation}
    \bm{G}_\mathbf{k}(\nu) = [ \bm{\zeta}_\nu-\bm{\varepsilon}_\mathbf{k}]^{-1},
\end{equation}
 with $\bm{\zeta}_\nu = (\i\nu +\mu) \mathbb{1} - \bm{\Sigma}(\nu)$. For strong coupling, we can expand $\bm{G}_\mathbf{k}(\nu)$ in powers of $\bm{\varepsilon}_\mathbf{k}$ up to second order:
 \begin{equation}
     \bm{G}_\mathbf{k}(\nu) \approx \bm{\zeta}^{-1}_\nu + \bm{\zeta}^{-1}_\nu \bm{\varepsilon}_\mathbf{k} \bm{\zeta}^{-1}_\nu + \bm{\zeta}^{-1}_\nu \bm{\varepsilon}_\mathbf{k} \bm{\zeta}^{-1}_\nu \bm{\varepsilon}_\mathbf{k} \bm{\zeta}^{-1}_\nu + \dots
 \end{equation}
 for which the cluster and lattice bubble become
\begin{equation}
\label{eq:strong_coupl_chi0}
\begin{split}
    \bm{\chi}_0 &\approx - \frac{\delta_{\nu\nu'}}{T} \left[ \bm{\zeta}^{-1}_\nu \otimes \bm{\zeta}^{-1}_\nu + \bm{\zeta}^{-1}_\nu \otimes \left( \frac{1}{N_\mathbf{k}}\sum_\mathbf{k}\bm{\zeta}^{-1}_\nu \bm{\varepsilon}_\mathbf{k} \bm{\zeta}^{-1}_\nu\right)\right.\\
    &+ \left( \frac{1}{N_\mathbf{k}}\sum_\mathbf{k}\bm{\zeta}^{-1}_\nu \bm{\varepsilon}_\mathbf{k} \bm{\zeta}^{-1}_\nu\right) \otimes \bm{\zeta}^{-1}_\nu \\
    &+ \left( \frac{1}{N_\mathbf{k}}\sum_\mathbf{k}\bm{\zeta}^{-1}_\nu \bm{\varepsilon}_\mathbf{k} \bm{\zeta}^{-1}_\nu\right) \otimes \left( \frac{1}{N_\mathbf{k}}\sum_\mathbf{k}\bm{\zeta}^{-1}_\nu \bm{\varepsilon}_\mathbf{k} \bm{\zeta}^{-1}_\nu\right)\\
    &+\bm{\zeta}^{-1}_\nu\otimes \left(\frac{1}{N_\mathbf{k}}\sum_\mathbf{k} \bm{\zeta}^{-1}_\nu \bm{\varepsilon}_\mathbf{k} \bm{\zeta}^{-1}_\nu \bm{\varepsilon}_\mathbf{k} \bm{\zeta}^{-1}_\nu\right)\\
    &\left.+\left(\frac{1}{N_\mathbf{k}}\sum_\mathbf{k} \bm{\zeta}^{-1}_\nu \bm{\varepsilon}_\mathbf{k} \bm{\zeta}^{-1}_\nu \bm{\varepsilon}_\mathbf{k} \bm{\zeta}^{-1}_\nu\right) \otimes \bm{\zeta}^{-1}_\nu + \dots \right]
\end{split}
\end{equation}
and
\begin{equation}
\label{eq:strong_coupl_chi0q}
\begin{split}
    \bm{\chi}^\mathbf{q}_0 &\approx - \frac{\delta_{\nu\nu'}}{T}\left[ \bm{\zeta}^{-1}_\nu \otimes \bm{\zeta}^{-1}_\nu + \bm{\zeta}^{-1}_\nu \otimes \left( \frac{1}{N_\mathbf{k}}\sum_\mathbf{k}\bm{\zeta}^{-1}_\nu \bm{\varepsilon}_\mathbf{k+q} \bm{\zeta}^{-1}_\nu\right)\right.\\
    &+ \left( \frac{1}{N_\mathbf{k}}\sum_\mathbf{k}\bm{\zeta}^{-1}_\nu \bm{\varepsilon}_\mathbf{k} \bm{\zeta}^{-1}_\nu\right) \otimes \bm{\zeta}^{-1}_\nu \\
    &+ \frac{1}{N_\mathbf{k}}\sum_\mathbf{k}\left( \bm{\zeta}^{-1}_\nu \bm{\varepsilon}_\mathbf{k} \bm{\zeta}^{-1}_\nu\right) \otimes \left( \bm{\zeta}^{-1}_\nu \bm{\varepsilon}_\mathbf{k+q} \bm{\zeta}^{-1}_\nu\right)\\
    &+\bm{\zeta}^{-1}_\nu\otimes \left(\frac{1}{N_\mathbf{k}}\sum_\mathbf{k} \bm{\zeta}^{-1}_\nu \bm{\varepsilon}_\mathbf{k+q} \bm{\zeta}^{-1}_\nu \bm{\varepsilon}_\mathbf{k+q} \bm{\zeta}^{-1}_\nu\right)\\
    &\left.+\left(\frac{1}{N_\mathbf{k}}\sum_\mathbf{k} \bm{\zeta}^{-1}_\nu \bm{\varepsilon}_\mathbf{k} \bm{\zeta}^{-1}_\nu \bm{\varepsilon}_\mathbf{k} \bm{\zeta}^{-1}_\nu\right) \otimes \bm{\zeta}^{-1}_\nu + \dots \right].
\end{split}
\end{equation} 
In $\bm{\chi}^\mathbf{q}_0$, due to the periodicity of $\bm{\varepsilon}_\mathbf{k}$ in the RBZ, in the first, and fourth line of Eq.~(\ref{eq:strong_coupl_chi0q}) we can shift the summation index by  $\mathbf{k}\to \mathbf{k-q}$, since the corresponding terms depend on $\bm{\varepsilon}_\mathbf{k+q}$ only. Hence, between Eq.~(\ref{eq:strong_coupl_chi0}) and Eq.~(\ref{eq:strong_coupl_chi0q}) only the third line differs. Further expanding the inverses of the two bubble susceptibilities in powers of $\bm{\varepsilon}_\mathbf{k}$ (up to second order) leads to 
\begin{equation}
\begin{split}    
    \left[\bm{\chi}^\mathbf{q}_0\right]^{-1}& - \left[\bm{\chi}_0\right]^{-1} 
    \approx T \delta_{\nu\nu'} \left[ \frac{1}{N_\mathbf{k}}\sum_\mathbf{k}  \bm{\varepsilon}_\mathbf{k}  \otimes  \bm{\varepsilon}_\mathbf{k+q} \right.\\
    &\left.- \left( \frac{1}{N_\mathbf{k}}\sum_\mathbf{k}\bm{\varepsilon}_\mathbf{k} \right) \otimes \left( \frac{1}{N_\mathbf{k}}\sum_\mathbf{k} \bm{\varepsilon}_\mathbf{k} \right)\right]
    \eqqcolon T\bm{\mathfrak{t}}_{\mathrm{eff},\mathbf{q}}^{\bm{2}} .
\end{split}
\end{equation}
Note that in difference to the single site DMFT result~\cite{Moghadas2025}, the sum over the RBZ $(1/N_\mathbf{k})\sum_\mathbf{k}\bm{\varepsilon}_\mathbf{k}$ in general does not vanish. For the hopping matrix in Eq.~(\ref{eq:hopping}) for $(1/N_\mathbf{k})\sum_\mathbf{k}\to \int^{\pi/2}_{-\pi/2} d^2k/\pi^2$ the expression reads:
\begin{widetext}
\setcounter{MaxMatrixCols}{16}
\begin{equation}
    \bm{\mathfrak{t}}_{\mathrm{eff},\mathbf{q}}^{\bm{2}} =\delta_{\nu\nu'} t^2
    \begin{pmatrix}
0&0&0&0&0&e^{\i2\mathbf{q}_x}&0&0&0&0&e^{\i2\mathbf{q}_y}&0&0&0&0&0\\
0&0&0&0&0&0&0&0&0&0&0&0&0&0&e^{\i2\mathbf{q}_y}&0\\
0&0&0&0&0&0&0&0&0&0&0&0&0&e^{\i2\mathbf{q}_x}&0&0\\
0&0&0&0&0&0&0&0&0&0&0&0&0&0&0&0\\
0&0&0&0&0&0&0&0&0&0&0&e^{\i2\mathbf{q}_y}&0&0&0&0\\
e^{-\i2\mathbf{q}_x}&0&0&0&0&0&0&0&0&0&0&0&0&0&0&e^{\i2\mathbf{q}_y}\\
0&0&0&0&0&0&0&0&0&0&0&0&0&0&0&0\\
0&0&0&0&0&0&0&0&e^{-\i2\mathbf{q}_x}&0&0&0&0&0&0&0\\
0&0&0&0&0&0&0&e^{\i2\mathbf{q}_x}&0&0&0&0&0&0&0&0\\
0&0&0&0&0&0&0&0&0&0&0&0&0&0&0&0\\
e^{-\i2\mathbf{q}_y}&0&0&0&0&0&0&0&0&0&0&0&0&0&0&e^{\i2\mathbf{q}_x}\\
0&0&0&0&e^{-\i2\mathbf{q}_y}&0&0&0&0&0&0&0&0&0&0&0\\
0&0&0&0&0&0&0&0&0&0&0&0&0&0&0&0\\
0&0&e^{-\i2\mathbf{q}_x}&0&0&0&0&0&0&0&0&0&0&0&0&0\\
0&e^{-\i2\mathbf{q}_y}&0&0&0&0&0&0&0&0&0&0&0&0&0&0\\
0&0&0&0&0&e^{-\i2\mathbf{q}_y}&0&0&0&0&e^{-\i2\mathbf{q}_x}&0&0&0&0&0
    \end{pmatrix}.
\end{equation}
\end{widetext}

\subsection{real-space single boson exchange decomposition}\label{App:SBE_def} To analyze the fluctuations on the impurity level, we employ the single-boson exchange decomposition for the full vertex $F$ \cite{Krien2019c} and the two-particle irreducible vertex \cite{Krien2020b}, to the real-space cluster quantities for the $\mathrm{SU}(2)$-symmetric case. To that end, we define the bosonic propagators as\footnote{This results in the same $w_{r}$ as in \cite{Krien2019c}. In comparison to \cite{Krien2019c}, we removed the hidden prefactors and minus-signs of the effective interaction values and the response functions $\chi_r$ on the right hand side, which are defined negative in \cite{Krien2019c}.}
\begin{equation}
    \begin{split}
        w^{\mathrm{mn}}_{\text{ch}}&=U\delta_{\mathrm{mn}}-U^2\chi^{\mathrm{mn}}_{\text{ch}}\\
        w^{\mathrm{mn}}_{\text{sp}}&=-U\delta_{\mathrm{mn}}-U^2\chi^{\mathrm{mn}}_{\text{sp}}\\
        w^{\mathrm{m|n}}_{\text{sing}}&=2U\left(\delta_{\mathrm{mn}}-U\chi^{\mathrm{m|n}}_{\text{sing}}\right),
    \end{split}
\end{equation} where $\chi^{\mathrm{m|n}}_r$ represents the connected physical response function in the respective channel $r$, which can be obtained from the generalized susceptibilities by integrating out all fermionic frequencies and $\chi^{\mathrm{m|n}}_{\text{sing}}=\chi^{\mathrm{m|n}}_{\text{pp},\uparrow\uparrow}-\chi^{\mathrm{m|n}}_{\text{pp},\uparrow\downarrow}$. Further, $\lambda_r$ represents the Hedin vertex \cite{Hedin1965} in the respective channel, which can be obtained from the corresponding fermion-boson response function \begin{equation}
    \chi^{3,\mathrm{bfe}}_{\text{r}}(\omega,\nu)=\sum_{\nu'}\chi_\text{r}^{\mathrm{febb}}(\omega,\nu,\nu')
\end{equation} by removing the fermionic- and bosonic legs \cite{Meixner2025a} \begin{equation}
    \begin{split}
        \lambda^{\mathrm{kij}}_{\mathrm{sp/ch}}(\omega,\nu)&=\pm \sum_{\mathrm{mno}}\frac{U\chi^{3,\mathrm{omn}}_{\mathrm{sp/ch}}(\omega,\nu)}{G^{\mathrm{mi}}(\nu)G^{\mathrm{jn}}(\nu+\omega)w^{\mathrm{ok}}_{\mathrm{sp/ch}}(\omega)}\\
        \lambda^{\mathrm{kij}}_{\mathrm{sing}}(\omega,\nu)&= \sum_{\mathrm{mno}}\frac{2U\chi^{3,\mathrm{omn}}_{\mathrm{sing}}(\omega,\nu)}{G^{\mathrm{mi}}(\nu)G^{\mathrm{jn}}(\omega-\nu)w^{\mathrm{ok}}_{\mathrm{sing}}(\omega)},\\
    \end{split}
\end{equation} which have a high-frequency tail towards the value 1.
From the decomposition of the two-particle irreducible vertex $\Gamma$ in momentum space \cite{Krien2020b}, we can deduce the decomposition:
\begin{equation}
\label{Eq:SBE_decomposition}
\begin{split}
\Gamma^{\mathrm{ijhl}}_{\text{ch}}&=-\frac{1}{2}\overline{\Delta}_{\text{ch}}^{\mathrm{ijhl}}-\frac{3}{2}\overline{\Delta}_{\text{sp}}^{\mathrm{ijhl}}+\frac{1}{2}\Delta_{\text{sing}}^{ijhl}+\Box_{\text{ch}}-U\\
\Gamma^{\mathrm{ijhl}}_{\text{sp}}&=-\frac{1}{2}\overline{\Delta}_{\text{ch}}^{\mathrm{ijhl}}+\frac{1}{2}\overline{\Delta}_{\text{sp}}^{\mathrm{ijhl}}-\frac{1}{2}\Delta_{\text{sing}}^{\mathrm{ijhl}}+\Box_{\text{sp}}+U,\\
\Gamma^{\mathrm{ih|lj}}_{\text{pp},\uparrow\downarrow}&=\frac{1}{2}\Delta_{\text{ch}}^{\mathrm{ih|lj}}-\frac{1}{2}\Delta_{\text{sp}}^{\mathrm{ih|lj}}-\overline{\Delta}_{\text{sp}}^{\mathrm{ih|lj}}+\Box_{\text{pp},\uparrow\downarrow}-U\\
\Gamma_{\text{pp},\uparrow\downarrow}&=\frac{1}{2}\left(\Gamma_{\text{pp},s}+\Gamma_{\text{pp},t}\right)
\end{split}
\end{equation}
with \begin{equation}
\begin{split}
\overline{\Delta}_{\text{r}}^{\mathrm{ijhl}}=\sum_{\mathrm{mn}}&\lambda^{\mathrm{mli}}_{\text{r}}\left(\nu'-\nu,\nu\right)\,w^{\mathrm{m|n}}_{\text{r}}\left(\nu'-\nu\right)\\&\cdot\,\lambda^{\mathrm{njh}}_{\text{r}}\left(\nu-\nu',\omega+\nu'\right),\\
\Delta_{\text{sing}}^{\mathrm{ijhl}}=\sum_{\mathrm{mn}}&\lambda^{\mathrm{nhi}}_{\text{sing}}\left(\omega+\nu'+\nu,\nu\right)\,w^{\mathrm{m|n}}_{\text{sing}}\left(\nu'+\nu+\omega\right)\\&\cdot\,\lambda^{\mathrm{mjl}}_{\text{sing}}\left(-\nu-\nu'-\omega,-\nu'\right),\\
\Delta_{\text{r}}^{\mathrm{ih|lj}}=\sum_{\mathrm{mn}}&\lambda^{\mathrm{mij}}_{\text{r}}\left(\omega-\nu-\nu',\nu\right)\,w^{\mathrm{m|n}}_{\text{r}}\left(\omega-\nu-\nu'\right)\\
&\cdot\,\lambda^{\mathrm{nlh}}_{\text{r}}\left(\nu+\nu'-\omega,\omega-\nu\right)\\
\overline{\Delta}_{\text{r}}^{\mathrm{ih|lj}}=\sum_{\mathrm{mn}}&\lambda^{\mathrm{mli}}_{\text{r}}\left(\nu'-\nu,\nu\right)\,w^{\mathrm{m|n}}_{\text{r}}\left(\nu'-\nu\right)\\&\cdot\,\lambda^{\mathrm{njh}}_{\text{r}}\left(\nu-\nu',\omega-\nu\right),\\
\end{split}
\end{equation} where $\text{r}\in\left\{\text{sp},\text{ch}\right\}$ and $\Box$ contains all multi-boson diagrams and all the $U$-irreducible diagrams in the respective channel. For $\overline{\Delta}$, the bar indicates that the bosonic propagator is transversal (vertical) in our convention. A comparison of the vertex SBE decomposition to the BSE scheme can be found in the benchmarking section App.~\ref{App:SBE}. Let us finally note, that $\lambda$ can in the high-frequency limit for each channel be approximated to $\lambda\approx 1$ which for DMFT results in a decomposition employed in \cite{DelRe2019b} for the purpose of fluctuation diagnostics.
\section{Benchmarking, Ward identities and consistency checks}
\subsection{Benchmarking of the lattice response to applied field calculations} \label{App:AFM} We benchmark the BSE particle-hole calculations setting the static anti-ferromagnetic (AFM) correlator \begin{equation}\chi_\mathrm{sp}^{\vq=0,\vQ=(\pi,\pi)}\left(\omega_m\right)=T^2\sum_{\nu_n,\nu'_n}\chi_\mathrm{sp}^{\vq=0,\vQ=(\pi,\pi)}\left(\nu_n,\nu'_n,\omega_m\right)\end{equation} into contrast to calculations with the linear response to an applied AFM field for different cluster sizes, namely the $2\times2$-CDMFT, performed by us and DMFT and $8\times8$-CDMFT from \cite{Schaefer2021}. Both methods should yield the same result  \cite{Hafermann2014a}. We find excellent agreement in Fig.~\ref{fig:BM} for both methods\footnote{As described in the main text, a tail treatment has been performed. Otherwise, the qualitative picture would remain the same, although the response would be slightly underestimated by the BSE.} for the $2\times 2$-CDMFT, and at high temperatures also for the other cluster sizes. When approaching the lowest temperatures $T$ depicted $0.1t>T>T_{\text{N\'eel}}$, our $2\times2$-CDMFT calculations display a slightly lower AFM response than DMFT and a slightly higher AFM response than $8\times8$-CDMFT. This can be explained by the reduction of $T_{\text{N\'eel}}$ with increased cluster size in the particle-hole symmetric square lattice Hubbard model \cite{Klett2020,Meixner2024} as the Mermin-Wagner theorem \cite{Mermin1966}, which is violated by mean-field theories such as (C)DMFT \cite{Georges1996}, is gradually reconstituted when increasing the impurity size towards the thermodynamic limit.

\subsection{Benchmarking of the two-particle irreducible vertex $\Gamma$ to the SBE}\label{App:SBE}
\begin{figure}
    \centering
    \includegraphics[width=0.8\linewidth]{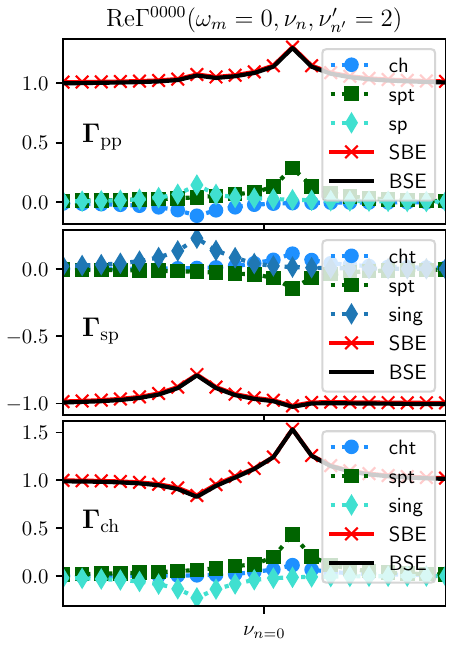}
    \caption{Comparison between the two-particle irreducible vertex and its SBE components Eq.~(\ref{Eq:SBE_decomposition}) for $U=1t,\,\beta t=5,\,t'=-0.25t$ and $n=0.95$ in each channel, where the multi-boson diagrams are neglected.}
    \label{fig:Benchmark_SBE}
\end{figure}

In the following, benchmarking plots of the Bethe-Salpeter equations to the SBE decomposition of $\Gamma$ (see App.~\ref{App:SBE_def}) in the three channels 
Fig.~\ref{fig:Benchmark_SBE} are presented in the weak coupling $U=1t,\,\beta t=5t$, particle-hole asymmetric case $t'=-0.25t,\, n=0.95$. Here, all three SBE contributions are significant, while the multi-boson diagrams can be neglected. Hence, good agreement to $\Gamma$ obtained from the BSE scheme is found.
\subsection{Ward identities for real-space}
\label{App:Ward}
 Ward identities relate lower point correlation functions with higher order correlation functions by an equation of motion approach \cite{Ward1950} and are therefore a formidable consistency check. In the Hubbard model, spin- and charge conservation can be used to relate the self-energy to the full vertex $F$ \cite{Behn1978}. A Ward identity has been recently computed for the single impurity Anderson model \cite{Ritz2025} and the DMFT \cite{Krien2018} via the operator formalism. Obtaining Ward identities for continuous symmetries via the action principle is an established procedure, see e.g. \cite{Kopietz2010,Bonetti2022b}.
 We employ the time inversion invariance of the Hubbard model action $S_{Hubbard}$ to obtain real-space ward identities for all three physical channels for the correlation function and the irreducible quantities, without the use of the Bethe-Salpeter equation. To that end, we define $S_{Hubbard}$ and the partition function $\mathcal{Z}$ \cite{Kotliar2001} on arbitrary real-space geometries
 \begin{equation}\begin{split}
 S_{Hubbard}\left[\overline{c},c\right]=&-\int_{1',2'}\sum_{\sigma}\overline{c}_{1'\sigma}G^{0^{-1}}_{1'2',\sigma}c_{2'\sigma}+\int_{1'} Un_{1'\uparrow}n_{1'\uparrow}\\ \mathcal{Z}=&\int\mathcal{D}\left[\overline{c},c\right]\e^{-S}, 
 \end{split}\end{equation} where $\overline{c}_{1'},c_{1'}$ represent independent, anit-commuting Grassmann numbers $\overline{c}_{1'}c_{2'}=-c_{2'}\overline{c}_{1'}$, 
 where $1'=(\tau_{1'},r_{1'} )$ is a tuple of imaginary time $\tau_{1'}$ and space $r_{1'}$ variables and we employ the short hand notation $\int\d\tau_{1'}\d r_{1'}\rightarrow \int_{1'}$ for readability. In the course of this derivation, primed indices indicate internal variables to be summed over, while non-primed variables will later become external indices. We define the grand potential
 \begin{equation}
     \Omega=-T\ln{\mathcal{Z}}.
 \end{equation}
 \begin{widetext}
 We define the Green function $G_{12,\sigma}$ and the generalized susceptibility $\chi_{1234,\sigma\sigma'}$ in the particle-hole channel \cite{Rohringer2012}
 \begin{equation}
     \begin{split}
         G_{12,\sigma}=&-\frac{\delta \Omega}{\delta  G^{0^{-1}}_{21,\sigma}}=\frac{1}{\mathcal{Z}}\int\mathcal{D}\left[\overline{c},c\right]\overline{c}_{2\sigma} c_{1\sigma }\e^{-S}=-\langle \T_\tau c_{1\sigma }c_{2\sigma }^{\dagger}\rangle\\
         \chi_{1234,\sigma\sigma'}=&\frac{\delta  G_{21,\sigma}}{\delta  G^{0^{-1}}_{34,\sigma'}}=-\frac{\delta ^2 \Omega}{\delta  G^{0^{-1}}_{34,\sigma'} \delta  G^{0^{-1}}_{12,\sigma}}=\frac{1}{\mathcal{Z}}\int\mathcal{D}\left[\overline{c},c\right]\overline{c}_{1\sigma} c_{2\sigma }\overline{c}_{3\sigma'} c_{4\sigma' }\e^{-S}-G_{21,\sigma} \frac{1}{\part} \frac{\delta  \part}{\delta  G_{34,\sigma'}^{0^{-1}}}\\=&\langle \T_\tau c_{1\sigma }^{\dagger}c_{2\sigma } c_{3\sigma' }^{\dagger}c_{4\sigma' }\rangle-G_{21,\sigma}G_{43,\sigma'},\\         
     \end{split}
 \end{equation} where the disconnected contributions stem from the derivative of the partition function and we introduce the time-ordering operator $\T_\tau$ for the correlation functions. Let us consider
 \begin{equation}
     \begin{split}
         \frac{\delta (S-S_j)}{\delta \overline{c}_{3\uparrow}}&=-\int_{2'} \Gn_{32'\uparrow}c_{2'\uparrow}+Uc_{3\uparrow}n_{3\downarrow}+j_{3\uparrow}\\
         \frac{\delta (S-S_j)}{\delta c_{3\uparrow}}&=\int_{2'} \overline{c}_{2'\uparrow}\Gn_{2'3\uparrow}-U\overline{c}_{3\uparrow}n_{3\downarrow}-\overline{j}_{3\uparrow},
     \end{split}
 \end{equation} where fermionic sources $j_{1},\overline{j}_1$ have been introduced as
 \begin{equation}
     S_j\left[\overline{c},c\right]=\sum_\sigma\int_{1'} \overline{j}_{1'\sigma}c_{1'\sigma}+\overline{c}_{1'\sigma}j_{1'\sigma}.
 \end{equation}
As Grassmann integrals are invariant under a linear shift of the variables \cite{Kopietz2010}, we may again write
\begin{equation}
    0=\int\mathcal{D}\left[\overline{c},c \right]\frac{\delta}{\delta \overline{c}_{3\uparrow}}\e^{-S+S_j}
\end{equation} Conducting a derivative with respect to the fermionic source $j_{3\uparrow}$ and setting the sources to zero yields
\begin{equation}
    \begin{split}
        0&=\int\mathcal{D}\left[\overline{c},c \right]\frac{\delta^2}{\delta j_{1\uparrow}\delta \overline{c}_{1\uparrow}}\e^{-S+S_j}\\
        0&\overset{j\rightarrow 0}{=}\int\mathcal{D}\left[\overline{c},c \right]\left(-\int_{2'}\Gn_{32'\uparrow}c_{2'\uparrow}\overline{c}_{3\uparrow}+Uc_{3\uparrow}\overline{c}_{3\uparrow}n_{3\downarrow}-\mathbb{1} \right)\e^{-S}
    \end{split}
\end{equation} from which we can obtain an identity\footnote{Note, that this is equivalent to the Migdal-Galitskii formula \cite{Galitskii1958} when eliminating $\Gn$ via the Dyson equation.} for the potential energy
\begin{equation}
    \int\mathcal{D}\left[\overline{c},c \right]Un_{3\uparrow}n_{3\downarrow}\e^{-S}=\int\mathcal{D}\left[\overline{c},c \right]\left(-\int_{2'}\Gn_{32'\uparrow}c_{2'\uparrow}\overline{c}_{3\uparrow}-\mathbb{1} \right)\e^{-S},
\end{equation} and dividing by $\mathcal{Z}$ and switching to operator formalism results in
\begin{equation}\label{Eq:Ward-Action_1}
    \langle H_{pot}\rangle=\langle Un_{3\uparrow} n_{3\downarrow}\rangle=\int_{2'}\Gn_{32'\uparrow}G_{2'3\uparrow}-\mathbb{1}.
\end{equation} Similar, from \begin{equation}
    0=\int\mathcal{D}\left[\overline{c},c \right]\frac{\delta^2}{\delta \overline{j}_{3\uparrow}\delta c_{3\uparrow}}\e^{-S+S_j}
\end{equation} one may obtain
\begin{equation}\label{Eq:Ward-Action_2}
        \langle H_{pot}\rangle=\langle Un_{3\uparrow} n_{3\downarrow}\rangle=\int_{2'}G_{32'\uparrow}\Gn_{2'3\uparrow}-\mathbb{1}.
\end{equation} By combining the two identities for the potential energy Eq.~(\ref{Eq:Ward-Action_1}),(\ref{Eq:Ward-Action_2})
\begin{equation}
    0=\left(\int_{2'}\Gn_{32'\uparrow}G_{2'3\uparrow}-\int_{2'}G_{32'\uparrow}\Gn_{2'3\uparrow}\right),
\end{equation} 
and conducting the derivative $\delta\Gn_{12,\uparrow}$ via 
\begin{equation}
    \frac{\delta\Gn_{2'3,\uparrow}}{\delta\Gn_{12,\sigma}}=\delta_{2'1}\delta_{32}\delta_{\uparrow\sigma}
\end{equation} yields a Ward identity on the imaginary space-time contour
\begin{equation} \label{Eq:Ward_Action_Imaginary_Time}
    G_{21,\uparrow}\delta_{32}\delta_{\uparrow\sigma}-G_{21,\uparrow}\delta_{31}\delta_{\uparrow\sigma}=\int_{2'}\left(\chi_{1232'\uparrow\uparrow} \Gn_{32',\uparrow}-\chi_{122'3\uparrow\uparrow} \Gn_{2'3,\uparrow}\right).
\end{equation}
\paragraph{A notion of time inversion symmetry} Before concluding the discussion by conveying the Ward identity Eq.~(\ref{Eq:Ward_Action_Imaginary_Time}) to the Matsubara formalism, we want to stress, that it is a manifestation of the time-inversion invariance of the system. Consider, that fermionic fields transform as $c_{i\sigma}\rightarrow-\overline{c}_{-i\sigma}$ and $\overline{c}_{i\sigma}\rightarrow c_{-i\sigma}$ under the time inversion operator $T_\text{inversion}$ \cite{Kopietz2010}. While the action is invariant under time inversion, the derivative transforms as
\begin{equation}
    \frac{\delta^2 S}{\delta j_{1\uparrow} \delta \overline{c}_{1\uparrow}}\rightarrow -\frac{\delta^2 S}{ \delta \overline{j}_{-1\uparrow} \delta c_{-1\uparrow}}.
\end{equation} These derivatives were used to obtain the identities for the potential energy Eq.~(\ref{Eq:Ward-Action_1}),(\ref{Eq:Ward-Action_2}), only with positive time arguments on the right hand side. As the derivatives result in the potential energy of the system, which is invariant under the time inversion symmetry, we can essentially find above Ward identity by means of applying time inversion to the functional derivatives.
\paragraph{A Ward identity in the Matsubara formalism}
The inverse Fourier transforms are defined as
\begin{equation}
\begin{split}
G_{\mathrm{ijlh},\sigma\sigma'}\left(\tau_1,\tau_2,\tau_3,\tau_4\right)&=\frac{1}{\beta^3}\sum_{\omega\nu_1\nu_2} G_{\mathrm{ijlh}}(\omega,\nu_1,\nu_2)\e^{\i\nu_1\tau_1}\e^{-\i(\nu_1+\omega)\tau_2}\e^{\i(\nu_2+\omega)\tau_3}\e^{-\i\nu_2\tau_4}\\
    G_{\mathrm{lh},\sigma}\left(\tau_3,\tau_4\right)&=\frac{1}{\beta}\sum_{\nu_3} G_{\mathrm{lh}}(\nu_3)\e^{\i\nu_3(\tau_4-\tau_3)},
\end{split}
\end{equation} and hence, by employing $\int \d\tau e^{-\i\omega \tau}=\beta\delta(\omega)$ for bosonic Matsubara frequencies $\omega$, one finds
\begin{equation}\label{Eq:Fouriertransforms}\begin{split}
\int\d\tau_3\sum_\mathrm{l}G_{\mathrm{ijlh},\sigma\sigma'}\left(\tau_1,\tau_2,\tau_3,\tau_4\right)\Gn_{\mathrm{lh},\sigma'}\left(\tau_3,\tau_4\right)&=\frac{1}{\beta^3}\sum_{\mathrm{l},\omega,\nu_1,\nu_2}G_{\mathrm{ijlh},\sigma\sigma'}(\omega,\nu_1,\nu_2)\Gn_{\mathrm{lh},\sigma'}(\nu_2+\omega)\e^{\i\nu_1\tau_1}\e^{-\i(\nu_1+\omega)\tau_2}\e^{\i(\nu_2+\omega)\tau_4}\\
\int\d\tau_4\sum_\mathrm{l}G_{\mathrm{ijhl},\sigma\sigma'}\left(\tau_1,\tau_2,\tau_3,\tau_4\right)\Gn_{\mathrm{hl},\sigma'}\left(\tau_3,\tau_4\right)&=\frac{1}{\beta^3}\sum_{\mathrm{l},\omega,\nu_1,\nu_2}G_{\mathrm{ijhl},\sigma\sigma'}(\omega,\nu_1,\nu_2)\Gn_{\mathrm{hl},\sigma'}(\nu_2)\e^{\i\nu_1\tau_1}\e^{-\i(\nu_1+\omega)\tau_2}\e^{\i(\nu_2+\omega)\tau_3},\\
\end{split}
\end{equation} where we split imaginary time arguments $\tau$, and real-space orbitals $\mathrm{i,j,h,l}$, resulting in the Ward identity 
\begin{equation} 
    G_{\mathrm{ji}\uparrow}(\nu)\delta_{\mathrm{jh}}\delta_{\uparrow\sigma}-G_{\mathrm{ji}\uparrow}(\nu+\omega)\delta_{\mathrm{ih}}\delta_{\uparrow\sigma}=\frac{1}{\beta}\sum_{l}\sum_{\nu'}\left(\chi_{\mathrm{ijhl}\sigma\uparrow}(\omega,\nu,\nu') \Gn_{\mathrm{hl},\uparrow}(\nu')-\chi_{\mathrm{ijlh}\sigma\uparrow}(\omega,\nu,\nu') \Gn_{\mathrm{lh},\uparrow}(\nu'+\omega)\right).
\end{equation} As the left-hand part of the Ward identity vanishes for the spin combination $\sigma=\downarrow$ and it is trivially fulfilled. Hence, it holds for $\chi_{\mathrm{ch}/\mathrm{sp}}$.
For the particle-particle channel the inverse four-point Fourier transform reads
\begin{equation}
\begin{split}
G_{\mathrm{ijlh},\sigma\sigma'}\left(\tau_1,\tau_2,\tau_3,\tau_4\right)&=\frac{1}{\beta^3}\sum_{\omega\nu_1\nu_2} \tilde{G}_{\mathrm{il|hj}}(\omega,\nu_1,\nu_2)\e^{\i\nu_1\tau_1}\e^{-\i\nu_2\tau_2}\e^{\i(\omega-\nu)\tau_3}\e^{-\i(\omega-\nu_2)\tau_4}
\end{split}
\end{equation}
resulting in
\begin{equation}
\begin{split}
\int\d\tau_3\sum_\mathrm{l}G_{\mathrm{ijlh},\sigma\sigma'}\left(\tau_1,\tau_2,\tau_3,\tau_4\right)\Gn_{\mathrm{lh},\sigma'}\left(\tau_3,\tau_4\right)&=\frac{1}{\beta^3}\sum_{\mathrm{l},\omega,\nu_1,\nu_2}\tilde{G}_{\mathrm{il|hj},\sigma\sigma'}(\omega,\nu_1,\nu_2)\Gn_{\mathrm{lh},\sigma'}(\omega-\nu)\e^{\i\nu_1\tau_1}\e^{-\i\nu_2\tau_2}\e^{\i(\nu-\nu_2)\tau_4}\\
\int\d\tau_4\sum_\mathrm{l}G_{\mathrm{ijhl},\sigma\sigma'}\left(\tau_1,\tau_2,\tau_3,\tau_4\right)\Gn_{\mathrm{hl},\sigma'}\left(\tau_3,\tau_4\right)&=\frac{1}{\beta^3}\sum_{\mathrm{l},\omega,\nu_1,\nu_2}\tilde{G}_{\mathrm{ih|lj},\sigma\sigma'}(\omega,\nu_1,\nu_2)\Gn_{\mathrm{hl},\sigma'}(\omega-\nu_2)\e^{\i\nu_1\tau_1}\e^{-\i\nu_2\tau_2}\e^{\i(\nu-\nu_2)\tau_4},\\
\end{split}
\end{equation} where we observe, that the bosonic frequency is not Fourier transformed but rather eliminated by summation. By replacing the full four-point function with the generalized susceptibility in the particle-particle channel as the disconnected contribution does not contribute to the sum, the identity reads
\begin{equation} \label{Eq:Ward_Action_Imaginary_Time_pp}
    G_{\mathrm{ji}\uparrow}(\nu)\delta_{\mathrm{jh}}\delta_{\uparrow\sigma}-G_{\mathrm{ji}\uparrow}(\nu')\delta_{\mathrm{ih}}\delta_{\uparrow\sigma}=\frac{1}{\beta}\sum_{\mathrm{l}}\sum_{\omega}\left(\tilde{\chi}_{\mathrm{ih|lj},\sigma\uparrow}(\omega,\nu,\nu') \Gn_{\mathrm{hl},\uparrow}(\omega-\nu')-\tilde{\chi}_{\mathrm{il|hj},\sigma\uparrow}(\omega,\nu,\nu') \Gn_{\mathrm{lh},\uparrow}(\omega-\nu)\right),
\end{equation} which we checked to hold numerically for the strong coupling case.
\paragraph{A ward identity for irreducible quantities from the action formalism} 
Employing the Dyson equation $\Gn_{2'3}=\Sigma_{2'3}+G^{-1}_{2'3}$ yields for Eq.~(\ref{Eq:Ward-Action_1}) and(\ref{Eq:Ward-Action_2}) the Migdal-Galitskii formula \cite{Galitskii1958}:
\begin{equation}
    U\langle n_{3\uparrow}n_{3\downarrow}\rangle=\int_{2'}\Sigma_{32'}G_{2'3}=\int_{2'}G_{32'}\Sigma_{2'3}.
\end{equation}
In analogy to the previous derivation, we perform the derivative $\delta G_{12,\sigma}$ and employ the product rule to obtain
 \begin{equation}
 \begin{split}
     \Sigma_{23}\delta_{31}\delta_{\uparrow\sigma}-\Sigma_{31}\delta_{32}\delta_{\uparrow\sigma}&=\int_{2'}\Gamma_{2'312}G_{2'3}-\int_{2'}\Gamma_{32'12}G_{32'}.\\
     \end{split}
 \end{equation} Using real-space particle-hole crossing relations $\Gamma_{3412}=\Gamma_{1234}$ yields
  \begin{equation}
 \begin{split}
     \Sigma_{23\uparrow}\delta_{31}\delta_{\uparrow\sigma}-\Sigma_{31\uparrow}\delta_{32}\delta_{\uparrow\sigma}&=\int_{2'}\Gamma_{122'3\uparrow\sigma}G_{2'3}-\int_{2'}\Gamma_{1232'\uparrow\sigma}G_{32'},
     \end{split}
 \end{equation}
a Ward identity for the irreducible quantities $\Sigma$ and $\delta\Sigma_{2'3}/\delta G_{12}=\Gamma_{32'12}$ with the Fourier transform 
\begin{equation} \label{Eq:Ward_Irreducible_Action_Imaginary_Time}
    \Sigma_{\mathrm{ji},\uparrow}(\nu+\omega)\delta_{\mathrm{ih}}\delta_{\uparrow\sigma}-\Sigma_{\mathrm{ji},\uparrow}(\nu)\delta_{\mathrm{jh}}\delta_{\uparrow\sigma}=\frac{1}{\beta}\sum_{l}\sum_{\nu'}\left(\Gamma_{\mathrm{ijlh}\sigma\uparrow}(\omega,\nu,\nu') G_{\mathrm{lh},\uparrow}(\nu'+\omega)-\Gamma_{\mathrm{ijhl}\sigma\uparrow}(\omega,\nu,\nu') G_{\mathrm{hl},\uparrow}(\nu')\right).
\end{equation} 
\end{widetext}
 We finalize by remarking that we were able to obtain the same Ward identities by a real-space cavity construction for CDMFT \cite{Kotliar2001,Bolech03}, similar to the procedure of \cite{Krien2018} but for CDMFT, however, the derivation is  less compact. Further, the Ward identities are perfectly fulfilled for the impurity generalized susceptibilities in all three channels and also for the irreducible vertex in the ch/m channel in the strong coupling case by our implementation. 
\section{Definition of the single particle Green function}\label{App:Gf} 
Our definition of the single-particle imaginary time Green function $G_\sigma^{ij}(\tau_1,\tau_2)\!=\!-\langle T_\tau c_{i\sigma}(\tau_1)c^\dagger_{j\sigma}(\tau_2)\rangle$ matches the literature of fundamental derivations of CDMFT, such as \cite{Kotliar2001,Maier2005} and the one of the TRIQS and TPRF libraries \cite{TRIQS,tprf}, together with the Fourier transform Eq.~(\ref{Eq:Fouriertransforms}). However, the imaginary time definition differs from the one of Rohringer \cite{Rohringer2013a} who defines:
\begin{equation}
\begin{split}
    G_{\sigma,\mathrm{Rohr.}}^{\mathrm{ji}}(\tau_2,\tau_1)=&\langle T_\tau c^\dagger_{\mathrm{j}\sigma}(\tau_2)c_{\mathrm{i}\sigma}(\tau_1)\rangle\\=&-\langle T_\tau c_{i\sigma}(\tau_1)c^\dagger_{\mathrm{j}\sigma}(\tau_2)\rangle=G_\sigma^{\mathrm{ij}}(\tau_1,\tau_2).
\end{split}
\end{equation}
The definition of the Fourier transform for creation and annihilation operators, respectively, is equivalent in both cases. Hence, since,
\begin{equation}
\begin{split}
 \frac{1}{\beta}&\sum_{\nu} G^{\mathrm{ji}}_{\sigma,\mathrm{Rohr.}}(\nu)\e^{\i\nu(\tau_2-\tau_1)}
    =G_{\sigma,\mathrm{Rohr.}}^{\mathrm{ji}}(\tau_2,\tau_1)\\&=G_\sigma^{\mathrm{ij}}(\tau_1,\tau_2) =\frac{1}{\beta}\sum_{\nu} G^{\mathrm{ij}}(\nu)\e^{\i\nu(\tau_2-\tau_1)},
\end{split}
\end{equation} where the sign of the time arguments is consistent for both sides of the equation in relation to creation- or annihilation operators. We hence formally obtain an index flip, which is equivalent to a time inversion operation, cf.~\cite{Rohringer2013a} Eq.~(2.94).
\section{Vertex divergences in the thermodynamic limit and anti-ferromagnetic DMFT} \label{App:AF_DMFT}
\begin{figure}
    \centering
    \includegraphics[width=\linewidth]{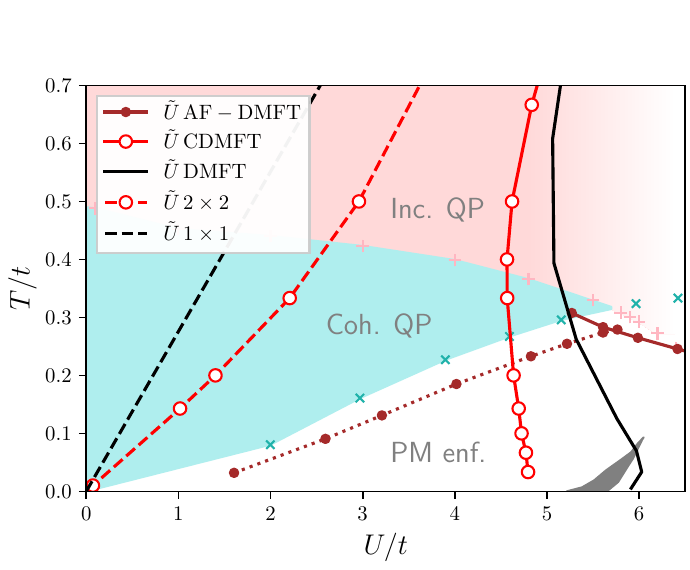}
    \caption{Phase diagram of the vertex divergences for the $1\times 1$ DMFT (black solid line) and $2\times 2$ CDMFT (red solid line) and  their respective isolated clusters (dashed lines). This is set into contrast to the anti-ferromagnetic solution of DMFT \cite{Reitner2025}, blow the Ne\'el temperature $T_{\text{N\'eel}}$, where (C)DMFT is paramagnetically enforced (PM enf.). Note, that $T_{\text{N\'eel}}$ (turquoise crosses) does only differ slightly in the two embedding theories. Below $T_{\text{N\'eel}}$, AF-DMFT (brown $\circ$-markers) displays both, vertex divergences (solid line) and pseudo-divergences (dotted line). Sources of other lines as in Fig.~\ref{fig:1st_div_line}.}
    \label{fig:Phasediagram_AF}
\end{figure}
Here we discuss the link between our speculative considerations about the vertex divergences in the thermodynamic limit of the 2D Hubbard model at half-filling presented in Sec.~\ref{sec:resultsVertex} to the recent literature results on the vertex divergences in the antiferromagnetically ordered phase of DMFT \cite{Reitner2025}, see Fig.~\ref{fig:Phasediagram_AF} (brown $\circ$-marker), i.e., below its N\'eel temperature. In contrast to (C)DMFT solutions, where the paramagnetic phase is enforced (PM enf.), breaking the  SU(2)-symmetry allows \cite{Essl2024,Reitner2025} for complex conjugate eigenvalues of the on-site generalized susceptibilities even for perfect-particle hole symmetry. As a result, one does not observe true vertex divergences in the Slater-AF phase at weak-coupling, but rather exponentially shaped lines starting from $U \!= \!T \!= \! 0$, where the real part of one eigenvalue of the generalized particle-hole (charge/longitudinal spin)  susceptibility vanishes in the presence of a \emph{finite imaginary part} (brown dotted line). On the other hand, the absence of an actual SU(2)-symmetry breaking in the quasi AF-ordered regime of the \emph{exact} solution of the 2D Hubbard model will render such eigenvalues real. Hence the (dotted brown) lines in the phase space will be transformed back into true vertex divergence lines of the charge sector. These, in the thermodynamic limit at lower interactions, will display a similar shape as the brown dotted lines of Fig.~\ref{fig:Phasediagram_AF}, yielding a qualitative picture of the perturbative breakdown in 2D, fully consistent with the considerations made in Sec.~\ref{sec:resultsVertex}.
\bibliography{main}

@Article{Meixner2024,
	title={{Mott transition and pseudogap of the square-lattice Hubbard model: Results from center-focused cellular dynamical mean-field theory}},
	author={Michael Meixner and Henri Menke and Marcel Klett and Sarah Heinzelmann and Sabine Andergassen and Philipp Hansmann and Thomas Schäfer},
	journal={SciPost Phys.},
	volume={16},
	pages={059},
	year={2024},
	publisher={SciPost},
	doi={10.21468/SciPostPhys.16.2.059},
	url={https://scipost.org/10.21468/SciPostPhys.16.2.059},
}

@article{TRIQS,
title = "TRIQS: A toolbox for research on interacting quantum systems",
journal = "Computer Physics Communications",
volume = "196",
pages = "398 - 415",
year = "2015",
issn = "0010-4655",
doi = "https://doi.org/10.1016/j.cpc.2015.04.023",
url = "http://www.sciencedirect.com/science/article/pii/S0010465515001666",
author = "Olivier Parcollet and Michel Ferrero and Thomas Ayral and Hartmut Hafermann and Igor Krivenko and Laura Messio and Priyanka Seth"
}

@misc{Meixner2025a,
      title={{Disentangling real space fluctuations: the diagnostics of metal-insulator transitions beyond single-particle spectral functions}}, 
      author={Meixner, Michael and Kr\"amer, Marcel and Wentzell, Nils and Bonetti, Pietro Maria and Andergassen, Sabine and Toschi, Alessandro and Sch\"afer, Thomas},
      year={2025},
      eprint={2501.18325v1},
      archivePrefix={arXiv},
      primaryClass={cond-mat.str-el}
}

@article{Rossi2015,
	doi = {10.1088/1751-8113/48/48/485202},
	url = {https://doi.org/10.1088%2F1751-8113%2F48%2F48%2F485202},
	year = 2015,
	month = {oct},
	publisher = {{IOP} Publishing},
	volume = {48},
	number = {48},
	pages = {485202},
	author = {Riccardo Rossi and F{\'{e}}lix Werner},
	title = {{Skeleton series and multivaluedness of the self-energy functional in zero space-time dimensions}},
	journal = {Journal of Physics A: Mathematical and Theoretical}
}

@article{Parcollet2004,
  title = {{Cluster Dynamical Mean Field Analysis of the Mott Transition}},
  author = {Parcollet, O. and Biroli, G. and Kotliar, G.},
  journal = {Phys. Rev. Lett.},
  volume = {92},
  issue = {22},
  pages = {226402},
  numpages = {4},
  year = {2004},
  month = {Jun},
  publisher = {American Physical Society},
  doi = {10.1103/PhysRevLett.92.226402},
  url = {https://link.aps.org/doi/10.1103/PhysRevLett.92.226402}
}

@online{tprf,
  author = {Strand, H. U. R. and Wentzell, N. and Parcollet, O.},
  title = {{tprf - two-particle response function tools based on the TRIQS library}},
  year = 2019,
  url = {https://triqs.github.io/tprf/2.1.x/index.html},
  urldate = {2020-01-06}
}

@INCOLLECTION{Senechal2011,
  author = {S\'en\'echal, David},
  title = {{Cluster Perturbation Theory}},
  booktitle = {Strongly Correlated Systems: Theoretical Methods},
  publisher = {Springer series},
  year = {2011},
  editor = {Mancini, F. and Avella, A.},
  chapter = {8}
}

@article{Jarrell1992,
  title = {{Hubbard model in infinite dimensions: A quantum Monte Carlo study}},
  author = {Jarrell, M.},
  journal = {Phys. Rev. Lett.},
  volume = {69},
  issue = {1},
  pages = {168--171},
  numpages = {0},
  year = {1992},
  month = {Jul},
  publisher = {American Physical Society},
  doi = {10.1103/PhysRevLett.69.168},
  url = {https://link.aps.org/doi/10.1103/PhysRevLett.69.168}
}

@article{georges:1996,
  title = {Dynamical mean-field theory of strongly correlated fermion systems and the limit of infinite dimensions},
  author = {Georges, Antoine and Kotliar, Gabriel and Krauth, Werner and Rozenberg, Marcelo J.},
  journal = {Rev. Mod. Phys.},
  volume = {68},
  issue = {1},
  pages = {13--125},
  numpages = {0},
  year = {1996},
  month = {Jan},
  publisher = {American Physical Society},
  doi = {10.1103/RevModPhys.68.13},
  url = {https://link.aps.org/doi/10.1103/RevModPhys.68.13}
}

@article{Kim2020b,
  title = {{Multivaluedness of the Luttinger-Ward functional in the fermionic and bosonic system with replicas}},
  author = {Kim, Aaram J. and Sacksteder, Vincent},
  journal = {Phys. Rev. B},
  volume = {101},
  issue = {11},
  pages = {115146},
  numpages = {8},
  year = {2020},
  month = {Mar},
  publisher = {American Physical Society},
  doi = {10.1103/PhysRevB.101.115146},
  url = {https://link.aps.org/doi/10.1103/PhysRevB.101.115146}
}

@article{Kim2020,
  title = {{Spin and Charge Correlations across the Metal-to-Insulator Crossover in the Half-Filled 2D Hubbard Model}},
  author = {Kim, Aaram J. and Simkovic, Fedor and Kozik, Evgeny},
  journal = {Phys. Rev. Lett.},
  volume = {124},
  issue = {11},
  pages = {117602},
  numpages = {6},
  year = {2020},
  month = {Mar},
  publisher = {American Physical Society},
  doi = {10.1103/PhysRevLett.124.117602},
  url = {https://link.aps.org/doi/10.1103/PhysRevLett.124.117602}
}

@article{Kotliar2000,
  title     = {Landau Theory of the Finite Temperature Mott Transition},
  author    = {Kotliar, G. and Lange, E. and Rozenberg, M. J.},
  journal   = {Phys. Rev. Lett.},
  volume    = {84},
  issue     = {22},
  pages     = {5180--5183},
  numpages  = {0},
  year      = {2000},
  month     = {May},
  publisher = {American Physical Society},
  doi       = {10.1103/PhysRevLett.84.5180},
  url       = {https://link.aps.org/doi/10.1103/PhysRevLett.84.5180}
}

@incollection{Bickers04,
  author = "Bickers, N. E.",
  title = "Self-Consistent Many-Body Theory for Condensed Matter Systems",
  editor = {S\'en\'echal, D. and Tremblay, A.-M. and Bourbonnais, C.},
  booktitle = {Theoretical Methods for Strongly Correlated Electrons. CRM Series in Mathematical Physics},
  year = 2004,
  publisher = "Springer",
  adress = "New York",
  url = {https://link.springer.com/chapter/10.1007/0-387-21717-7_6},
}

@article{Schaefer2021,
  title = {{Tracking the Footprints of Spin Fluctuations: A MultiMethod, MultiMessenger Study of the Two-Dimensional Hubbard Model}},
  author = {Sch\"afer, Thomas and Wentzell, Nils and \ifmmode \check{S}\else \v{S}\fi{}imkovic, Fedor and He, Yuan-Yao and Hille, Cornelia and Klett, Marcel and Eckhardt, Christian J. and Arzhang, Behnam and Harkov, Viktor and Le R\'egent, {Fran\ifmmode \mbox{\c{c}}\else \c{c}\fi{}ois-Marie} and Kirsch, Alfred and Wang, Yan and Kim, Aaram J. and Kozik, Evgeny and Stepanov, Evgeny A. and Kauch, Anna and Andergassen, Sabine and Hansmann, Philipp and Rohe, Daniel and Vilk, Yuri M. and LeBlanc, James P. F. and Zhang, Shiwei and Tremblay, A.-M. S. and Ferrero, Michel and Parcollet, Olivier and Georges, Antoine},
  journal = {Phys. Rev. X},
  volume = {11},
  issue = {1},
  pages = {011058},
  numpages = {53},
  year = {2021},
  month = {Mar},
  publisher = {American Physical Society},
  doi = {10.1103/PhysRevX.11.011058},
  url = {https://link.aps.org/doi/10.1103/PhysRevX.11.011058}
}

@article{Pelz2023,
  title = {{Highly nonperturbative nature of the Mott metal-insulator transition: Two-particle vertex divergences in the coexistence region}},
  author = {Pelz, M. and Adler, S. and Reitner, M. and Toschi, A.},
  journal = {Phys. Rev. B},
  volume = {108},
  issue = {15},
  pages = {155101},
  numpages = {14},
  year = {2023},
  month = {Oct},
  publisher = {American Physical Society},
  doi = {10.1103/PhysRevB.108.155101},
  url = {https://link.aps.org/doi/10.1103/PhysRevB.108.155101}
}

@Article{Adler2024,
	title={{Non-perturbative intertwining between spin and charge correlations: A ``smoking gun'' single-boson-exchange result}},
	author={Severino Adler and Friedrich Krien and Patrick Chalupa-Gantner and Giorgio Sangiovanni and Alessandro Toschi},
	journal={SciPost Phys.},
	volume={16},
	pages={054},
	year={2024},
	publisher={SciPost},
	doi={10.21468/SciPostPhys.16.2.054},
	url={https://scipost.org/10.21468/SciPostPhys.16.2.054},
}

@article{Klett2020,
  title = {{Real-space cluster dynamical mean-field theory: Center-focused extrapolation on the one- and two particle-levels}},
  author = {Klett, Marcel and Wentzell, Nils and Sch\"afer, Thomas and Simkovic, Fedor and Parcollet, Olivier and Andergassen, Sabine and Hansmann, Philipp},
  journal = {Phys. Rev. Research},
  volume = {2},
  issue = {3},
  pages = {033476},
  numpages = {11},
  year = {2020},
  month = {Sep},
  publisher = {American Physical Society},
  doi = {10.1103/PhysRevResearch.2.033476},
  url = {https://link.aps.org/doi/10.1103/PhysRevResearch.2.033476}
}

@article{Schaefer2013,
  title = {{Divergent Precursors of the Mott-Hubbard Transition at the Two-Particle Level}},
  author = {Sch\"afer, T. and Rohringer, G. and Gunnarsson, O. and Ciuchi, S. and Sangiovanni, G. and Toschi, A.},
  journal = {Phys. Rev. Lett.},
  volume = {110},
  issue = {24},
  pages = {246405},
  numpages = {5},
  year = {2013},
  month = {Jun},
  publisher = {American Physical Society},
  doi = {10.1103/PhysRevLett.110.246405},
  url = {https://link.aps.org/doi/10.1103/PhysRevLett.110.246405}
}

@article{DelRe2019b,
  title = {{Fluctuations analysis of spin susceptibility: N\'eel ordering revisited in dynamical mean field theory}},
  author = {Del Re, Lorenzo and Rohringer, Georg},
  journal = {Phys. Rev. B},
  volume = {104},
  issue = {23},
  pages = {235128},
  numpages = {14},
  year = {2021},
  month = {Dec},
  publisher = {American Physical Society},
  doi = {10.1103/PhysRevB.104.235128},
  url = {https://link.aps.org/doi/10.1103/PhysRevB.104.235128}
}

@article{Musshoff2021,
  title = {Magnetic response trends in cuprates and the $t\ensuremath{-}{t}^{\ensuremath{'}}$ Hubbard model},
  author = {Mu\ss{}hoff, Julian and Kiani, Amin and Pavarini, Eva},
  journal = {Phys. Rev. B},
  volume = {103},
  issue = {7},
  pages = {075136},
  numpages = {18},
  year = {2021},
  month = {Feb},
  publisher = {American Physical Society},
  doi = {10.1103/PhysRevB.103.075136},
  url = {https://link.aps.org/doi/10.1103/PhysRevB.103.075136}
}

@Misc{Antipov2015a,
  Title                    = {pomerol: 1.1},

  Author                   = {Andrey E. Antipov and
 Igor Krivenko},
  Year                     = {2015},

  Doi                      = {10.5281/zenodo.17900},
  Url                      = {http://dx.doi.org/10.5281/zenodo.17900}
}

@article{Kowalski2024,
  title = {Thermodynamic Stability at the Two-Particle Level},
  author = {Kowalski, A. and Reitner, M. and Del Re, L. and Chatzieleftheriou, M. and Amaricci, A. and Toschi, A. and de' Medici, L. and Sangiovanni, G. and Sch\"afer, T.},
  journal = {Phys. Rev. Lett.},
  volume = {133},
  issue = {6},
  pages = {066502},
  numpages = {7},
  year = {2024},
  month = {Aug},
  publisher = {American Physical Society},
  doi = {10.1103/PhysRevLett.133.066502},
  url = {https://link.aps.org/doi/10.1103/PhysRevLett.133.066502}
}

@article{Krien2019c,
  title = {{Single-boson exchange decomposition of the vertex function}},
  author = {Krien, Friedrich and Valli, Angelo and Capone, Massimo},
  journal = {Phys. Rev. B},
  volume = {100},
  issue = {15},
  pages = {155149},
  numpages = {15},
  year = {2019},
  month = {Oct},
  publisher = {American Physical Society},
  doi = {10.1103/PhysRevB.100.155149},
  url = {https://link.aps.org/doi/10.1103/PhysRevB.100.155149}
}

@Article{Ayral2017,
  Title                    = {{Fierz Convergence Criterion: A Controlled Approach to Strongly Interacting Systems with Small Embedded Clusters}},
  Author                   = {Ayral, Thomas and Vu\ifmmode \check{c}\else \v{c}\fi{}i\ifmmode \check{c}\else \v{c}\fi{}evi\ifmmode \acute{c}\else \'{c}\fi{}, Jaksa and Parcollet, Olivier},
  Journal                  = {Phys. Rev. Lett.},
  Year                     = {2017},

  Month                    = {Oct},
  Pages                    = {166401},
  Volume                   = {119},

  Doi                      = {10.1103/PhysRevLett.119.166401},
  Issue                    = {16},
  Numpages                 = {7},
  Publisher                = {American Physical Society},
  Url                      = {https://link.aps.org/doi/10.1103/PhysRevLett.119.166401}
}

@Article{Baym1962,
  Title                    = {Self-Consistent Approximations in Many-Body Systems},
  Author                   = {Baym, Gordon},
  Journal                  = {Phys. Rev.},
  Year                     = {1962},

  Month                    = {Aug},
  Pages                    = {1391--1401},
  Volume                   = {127},

  Doi                      = {10.1103/PhysRev.127.1391},
  Issue                    = {4},
  Numpages                 = {0},
  Publisher                = {American Physical Society},
  Url                      = {http://link.aps.org/doi/10.1103/PhysRev.127.1391}
}

@InBook{Bickers2004,
  Title                    = {Theoretical Methods for Strongly Correlated Electrons},
  Author                   = {N. E. Bickers},
  Chapter                  = {6},
  Editor                   = {D. S\'en\'echal and A.-M. Trembly and C. Bourbonnais},
  Pages                    = {237-296},
  Publisher                = {Springer-Verlag New York Berlin Heidelbert},
  Year                     = {2004}
}

@Article{Bolech03,
  Title                    = {Cellular dynamical mean-field theory for the one-dimensional extended Hubbard model},
  Author                   = {Bolech, C. J. and Kancharla, S. S. and Kotliar, G.},
  Journal                  = {Phys. Rev. B},
  Year                     = {2003},

  Month                    = {Feb},
  Pages                    = {075110},
  Volume                   = {67},

  Doi                      = {10.1103/PhysRevB.67.075110},
  Issue                    = {7},
  Numpages                 = {9},
  Publisher                = {American Physical Society},
  Url                      = {http://link.aps.org/doi/10.1103/PhysRevB.67.075110}
}

@Article{Galitskii1958,
  Author                   = {V. M. Galitskii and A. B. Migdal},
  title = {{Application of quantum fiel theoretical methods to the many-body problem}},
  Journal                  = {Sov. Phys. JETP},
  Year                     = {1958},
  Pages                    = {139},
  Volume                   = {34},
  url = {http://www.jetp.ac.ru/cgi-bin/e/index/e/7/1/p96?a=list}
}

@article{Bonetti2022b,
  title = {{Local Ward identities for collective excitations in fermionic systems with spontaneously broken symmetries}},
  author = {Bonetti, Pietro M.},
  journal = {Phys. Rev. B},
  volume = {106},
  issue = {15},
  pages = {155105},
  numpages = {17},
  year = {2022},
  month = {Oct},
  publisher = {American Physical Society},
  doi = {10.1103/PhysRevB.106.155105},
  url = {https://link.aps.org/doi/10.1103/PhysRevB.106.155105}
}

@article{Chalupa2017,
  title = {{Divergences of the irreducible vertex functions in correlated metallic systems: Insights from the Anderson impurity model}},
  author = {Chalupa, P. and Gunacker, P. and Sch\"afer, T. and Held, K. and Toschi, A.},
  journal = {Phys. Rev. B},
  volume = {97},
  issue = {24},
  pages = {245136},
  numpages = {15},
  year = {2018},
  month = {Jun},
  publisher = {American Physical Society},
  doi = {10.1103/PhysRevB.97.245136},
  url = {https://link.aps.org/doi/10.1103/PhysRevB.97.245136}
}

@misc{Moghadas2025,
  Title                    = {{Effective enhancement of the electron-phonon coupling driven by nonperturbative electronic density fluctuations}},
  Author                   = {Emin Moghadas and Matthias Reitner and Tim Wehling and Giorgio Sangiovanni and Sergio Ciuchi and Alessandro Toschi},
  Year                     = {2025},
  Archiveprefix            = {arXiv},
  Eprint                   = {2503.12113}
}

@article{Harland2019,
  title = {Josephson lattice model for phase fluctuations of local pairs in copper oxide superconductors},
  author = {Harland, Malte and Brener, Sergey and Lichtenstein, Alexander I. and Katsnelson, Mikhail I.},
  journal = {Phys. Rev. B},
  volume = {100},
  issue = {2},
  pages = {024510},
  numpages = {12},
  year = {2019},
  month = {Jul},
  publisher = {American Physical Society},
  doi = {10.1103/PhysRevB.100.024510},
  url = {https://link.aps.org/doi/10.1103/PhysRevB.100.024510}
}

@Article{Fratino2017,
  Title                    = {{Signatures of the Mott transition in the antiferromagnetic state of the two-dimensional Hubbard model}},
  Author                   = {Fratino, L. and S\'emon, P. and Charlebois, M. and Sordi, G. and Tremblay, A.-M. S.},
  Journal                  = {Phys. Rev. B},
  Year                     = {2017},

  Month                    = {Jun},
  Pages                    = {235109},
  Volume                   = {95},

  Doi                      = {10.1103/PhysRevB.95.235109},
  Issue                    = {23},
  Numpages                 = {11},
  Publisher                = {American Physical Society},
  Url                      = {https://link.aps.org/doi/10.1103/PhysRevB.95.235109}
}

@Article{Georges1992a,
  Title                    = {{Hubbard model in infinite dimensions}},
  Author                   = {Georges, Antoine and Kotliar, Gabriel},
  Journal                  = {Phys. Rev. B},
  Year                     = {1992},

  Month                    = {Mar},
  Pages                    = {6479--6483},
  Volume                   = {45},

  Doi                      = {10.1103/PhysRevB.45.6479},
  Numpages                 = {4},
  Publisher                = {American Physical Society},
  Url                      = {http://link.aps.org/doi/10.1103/PhysRevB.45.6479}
}

@Article{Georges1996,
  Title                    = {{Dynamical mean-field theory of strongly correlated fermion systems and the limit of infinite dimensions}},
  Author                   = {Georges, Antoine and Kotliar, Gabriel and Krauth, Werner and Rozenberg, Marcelo J.},
  Journal                  = {Rev. Mod. Phys.},
  Year                     = {1996},

  Month                    = {Jan},
  Number                   = {1},
  Pages                    = {13},
  Volume                   = {68},

  Doi                      = {10.1103/RevModPhys.68.13},
  Publisher                = {American Physical Society},
  Url                      = {http://dx.doi.org/10.1103/RevModPhys.68.13}
}

@article{Melnick2020,
  title = {{Fermi-liquid theory and divergences of the two-particle irreducible vertex in the periodic Anderson lattice}},
  author = {Melnick, Corey and Kotliar, Gabriel},
  journal = {Phys. Rev. B},
  volume = {101},
  issue = {16},
  pages = {165105},
  numpages = {16},
  year = {2020},
  month = {Apr},
  publisher = {American Physical Society},
  doi = {10.1103/PhysRevB.101.165105},
  url = {https://link.aps.org/doi/10.1103/PhysRevB.101.165105}
}

@article{Mazitov2022b,
  title = {{Effect of local magnetic moments on spectral properties and resistivity near interaction- and doping-induced Mott transitions}},
  author = {Mazitov, T. B. and Katanin, A. A.},
  journal = {Phys. Rev. B},
  volume = {106},
  issue = {20},
  pages = {205148},
  numpages = {10},
  year = {2022},
  month = {Nov},
  publisher = {American Physical Society},
  doi = {10.1103/PhysRevB.106.205148},
  url = {https://link.aps.org/doi/10.1103/PhysRevB.106.205148}
}

@article{Mazitov2022,
  title = {{Local magnetic moment formation and Kondo screening in the half-filled single-band Hubbard model}},
  author = {Mazitov, T. B. and Katanin, A. A.},
  journal = {Phys. Rev. B},
  volume = {105},
  issue = {8},
  pages = {L081111},
  numpages = {6},
  year = {2022},
  month = {Feb},
  publisher = {American Physical Society},
  doi = {10.1103/PhysRevB.105.L081111},
  url = {https://link.aps.org/doi/10.1103/PhysRevB.105.L081111}
}

@Article{Gull2008a,
  Title                    = {{Continuous-time auxiliary-field Monte Carlo for quantum impurity models}},
  Author                   = {E. Gull and P. Werner and O. Parcollet and M. Troyer},
  Journal                  = {EPL (Europhysics Letters)},
  Year                     = {2008},
  Number                   = {5},
  Pages                    = {57003},
  Volume                   = {82},
  Url                      = {http://stacks.iop.org/0295-5075/82/i=5/a=57003}
}

@Article{Gunnarsson2017,
  Title                    = {{Breakdown of Traditional Many-Body Theories for Correlated Electrons}},
  Author                   = {Gunnarsson, O. and Rohringer, G. and Sch\"afer, T. and Sangiovanni, G. and Toschi, A.},
  Journal                  = {Phys. Rev. Lett.},
  Year                     = {2017},

  Month                    = {Aug},
  Pages                    = {056402},
  Volume                   = {119},

  Doi                      = {10.1103/PhysRevLett.119.056402},
  Issue                    = {5},
  Numpages                 = {5},
  Publisher                = {American Physical Society},
  Url                      = {https://link.aps.org/do, (iii) check the Introduction.i/10.1103/PhysRevLett.119.056402}
}

@article{Qin2022,
author = {Qin, Mingpu and Schäfer, Thomas and Andergassen, Sabine and Corboz, Philippe and Gull, Emanuel},
title = {{The Hubbard Model: A Computational Perspective}},
journal = {{Annual Review of Condensed Matter Physics}},
volume = {13},
year = {2022},
doi = {10.1146/annurev-conmatphys-090921-033948},

URL = { 
        https://doi.org/10.1146/annurev-conmatphys-090921-033948
    
},
eprint = { 
        https://doi.org/10.1146/annurev-conmatphys-090921-033948
    
}
}

@Article{Gunnarsson2016,
  Title                    = {{Parquet decomposition calculations of the electronic self-energy}},
  Author                   = {Gunnarsson, O. and Sch\"afer, T. and LeBlanc, J. P. F. and Merino, J. and Sangiovanni, G. and Rohringer, G. and Toschi, A.},
  Journal                  = {Phys. Rev. B},
  Year                     = {2016},

  Month                    = {Jun},
  Pages                    = {245102},
  Volume                   = {93},

  Doi                      = {10.1103/PhysRevB.93.245102},
  Issue                    = {24},
  Numpages                 = {17},
  Publisher                = {American Physical Society},
  Url                      = {http://link.aps.org/doi/10.1103/PhysRevB.93.245102}
}

@Article{Gutzwiller1963,
  Title                    = {{Effect of Correlation on the Ferromagnetism of Transition Metals}},
  Author                   = {Gutzwiller, Martin C.},
  Journal                  = {Phys. Rev. Lett.},
  Year                     = {1963},

  Month                    = {Mar},
  Pages                    = {159--162},
  Volume                   = {10},

  Doi                      = {10.1103/PhysRevLett.10.159},
  Issue                    = {5},
  Numpages                 = {0},
  Publisher                = {American Physical Society},
  Url                      = {http://link.aps.org/doi/10.1103/PhysRevLett.10.159}
}

@Article{Hafermann2014a,
  Title                    = {Collective charge excitations of strongly correlated electrons, vertex corrections, and gauge invariance},
  Author                   = {Hafermann, Hartmut and van Loon, Erik G. C. P. and Katsnelson, Mikhail I. and Lichtenstein, Alexander I. and Parcollet, Olivier},
  Journal                  = {Phys. Rev. B},
  Year                     = {2014},

  Month                    = {Dec},
  Pages                    = {235105},
  Volume                   = {90},

  Doi                      = {10.1103/PhysRevB.90.235105},
  Issue                    = {23},
  Numpages                 = {19},
  Publisher                = {American Physical Society},
  Url                      = {http://link.aps.org/doi/10.1103/PhysRevB.90.235105}
}

@MastersThesis{Badr2024,
  Title                    = {Efficiently solving the parquet equations using a sparse representation},
  Author                   = {S. Badr},
  School                   = {TU Wien},
  Year                     = {2024},
}

@article{Reitner2024,
  title = {{Protection of correlation-induced phase instabilities by exceptional susceptibilities}},
  author = {Reitner, M. and Crippa, L. and Fus, D. R. and Budich, J. C. and Toschi, A. and Sangiovanni, G.},
  journal = {Phys. Rev. Res.},
  volume = {6},
  issue = {2},
  pages = {L022031},
  numpages = {7},
  year = {2024},
  month = {May},
  publisher = {American Physical Society},
  doi = {10.1103/PhysRevResearch.6.L022031},
  url = {https://link.aps.org/doi/10.1103/PhysRevResearch.6.L022031}
}

@Article{Hedin1965,
  Title                    = {New Method for Calculating the One-Particle Green's Function with Application to the Electron-Gas Problem},
  Author                   = {Hedin, Lars},
  Journal                  = {Phys. Rev.},
  Year                     = {1965},

  Month                    = {Aug},
  Pages                    = {A796--A823},
  Volume                   = {139},

  Doi                      = {10.1103/PhysRev.139.A796},
  Issue                    = {3A},
  Numpages                 = {0},
  Publisher                = {American Physical Society},
  Url                      = {http://link.aps.org/doi/10.1103/PhysRev.139.A796}
}

@Article{Hettler2000,
  Title                    = {Dynamical cluster approximation: Nonlocal dynamics of correlated electron systems},
  Author                   = {Hettler, M. H. and Mukherjee, M. and Jarrell, M. and Krishnamurthy, H. R.},
  Journal                  = {Phys. Rev. B},
  Year                     = {2000},

  Month                    = {May},
  Pages                    = {12739--12756},
  Volume                   = {61},

  Doi                      = {10.1103/PhysRevB.61.12739},
  Issue                    = {19},
  Numpages                 = {0},
  Publisher                = {American Physical Society},
  Url                      = {http://link.aps.org/doi/10.1103/PhysRevB.61.12739}
}

@article{Hohenberg1967,
  title = {{Existence of Long-Range Order in One and Two Dimensions}},
  author = {Hohenberg, P. C.},
  journal = {Phys. Rev.},
  volume = {158},
  issue = {2},
  pages = {383--386},
  numpages = {0},
  year = {1967},
  month = {Jun},
  publisher = {American Physical Society},
  doi = {10.1103/PhysRev.158.383},
  url = {https://link.aps.org/doi/10.1103/PhysRev.158.383}
}

@Article{Hubbard1964,
  Title                    = {{Electron Correlations in Narrow Energy Bands. III. An Improved Solution}},
  Author                   = {Hubbard, J.},
  Journal                  = {Proc R. Soc. London},
  Year                     = {1964},
  Number                   = {1386},
  Pages                    = {401-419},
  Volume                   = {281},
  Doi                      = {10.1098/rspa.1964.0190}
}

@Article{Hubbard1963,
  Title                    = {{Electron Correlations in Narrow Energy Bands}},
  Author                   = {Hubbard, J.},
  Journal                  = {Proceedings of the Royal Society of London. Series A, Mathematical and Physical Sciences},
  Year                     = {1963},
  Number                   = {1365},
  Pages                    = {238-257},
  Volume                   = {276},
  Doi                      = {10.1098/rspa.1963.0204},
  Url                      = {http://rspa.royalsocietypublishing.org/content/276/1365/238.abstract}
}

@article{Thunstrom2018,
  title = {{Analytical investigation of singularities in two-particle irreducible vertex functions of the Hubbard atom}},
  author = {Thunstr\"om, P. and Gunnarsson, O. and Ciuchi, Sergio and Rohringer, G.},
  journal = {Phys. Rev. B},
  volume = {98},
  issue = {23},
  pages = {235107},
  numpages = {16},
  year = {2018},
  month = {Dec},
  publisher = {American Physical Society},
  doi = {10.1103/PhysRevB.98.235107},
  url = {https://link.aps.org/doi/10.1103/PhysRevB.98.235107}
}

@Article{Janis2014,
  Title                    = {Critical metal-insulator transition and divergence in a two-particle irreducible vertex in disordered and interacting electron systems},
  Author                   = {V. Jani\v{s} and V. Pokorn\'y},
  Journal                  = {Phys. Rev. B},
  Year                     = {2014},

  Month                    = {Jul},
  Pages                    = {045143},
  Volume                   = {90},

  Doi                      = {10.1103/PhysRevB.90.045143},
  Issue                    = {4},
  Numpages                 = {11},
  Publisher                = {American Physical Society},
  Url                      = {http://link.aps.org/doi/10.1103/PhysRevB.90.045143}
}

@PhdThesis{KlettPhd,
  Title                    = {Cellular Dynamical Mean-field Theory in large Impurity Clusters},
  Author                   = {Klett, M.},
  School                   = {University of T\"ubingen},
  Year                     = {2020}
}

@article{Kanamori1963,
    author = {Kanamori, Junjiro},
    title = "{Electron Correlation and Ferromagnetism of Transition Metals}",
    journal = {Progress of Theoretical Physics},
    volume = {30},
    number = {3},
    pages = {275-289},
    year = {1963},
    month = {09},
    issn = {0033-068X},
    doi = {10.1143/PTP.30.275},
    url = {https://doi.org/10.1143/PTP.30.275},
    eprint = {https://academic.oup.com/ptp/article-pdf/30/3/275/5278869/30-3-275.pdf},
}

@article{Krien2020b,
 title = {{Tiling with triangles: parquet and $GW\ensuremath{\gamma}$ methods unified}},
 author = {Krien, Friedrich and Kauch, Anna and Held, Karsten},
 journal = {Phys. Rev. Research},
 volume = {3},
 issue = {1},
 pages = {013149},
 numpages = {12},
 year = {2021},
 month = {Feb},
 publisher = {American Physical Society},
 doi = {10.1103/PhysRevResearch.3.013149},
 url = {https://link.aps.org/doi/10.1103/PhysRevResearch.3.013149}
}

@article{Kotliar2002,
  title     = {Compressibility Divergence and the Finite Temperature Mott Transition},
  author    = {Kotliar, G. and Murthy, Sahana and Rozenberg, M. J.},
  journal   = {Phys. Rev. Lett.},
  volume    = {89},
  issue     = {4},
  pages     = {046401},
  numpages  = {4},
  year      = {2002},
  month     = {Jul},
  publisher = {American Physical Society},
  doi       = {10.1103/PhysRevLett.89.046401},
  url       = {https://link.aps.org/doi/10.1103/PhysRevLett.89.046401}
}

@Article{Kotliar2001,
  Title                    = {{Cellular Dynamical Mean Field Approach to Strongly Correlated Systems}},
  Author                   = {Kotliar, Gabriel and Savrasov, Sergej Y. and P\'alsson, Gunnar and Biroli, Giulio},
  Journal                  = {Phys. Rev. Lett.},
  Year                     = {2001},

  Month                    = {Oct},
  Pages                    = {186401},
  Volume                   = {87},

  Doi                      = {10.1103/PhysRevLett.87.186401},
  Issue                    = {18},
  Numpages                 = {4},
  Publisher                = {American Physical Society},
  Url                      = {http://link.aps.org/doi/10.1103/PhysRevLett.87.186401}
}

@Article{Kozik2015,
  Title                    = {{Nonexistence of the Luttinger-Ward Functional and Misleading Convergence of Skeleton Diagrammatic Series for Hubbard-Like Models}},
  Author                   = {Kozik, Evgeny and Ferrero, Michel and Georges, Antoine},
  Journal                  = {Phys. Rev. Lett.},
  Year                     = {2015},

  Month                    = {Apr},
  Pages                    = {156402},
  Volume                   = {114},

  Doi                      = {10.1103/PhysRevLett.114.156402},
  Issue                    = {15},
  Numpages                 = {5},
  Publisher                = {American Physical Society},
  Url                      = {http://link.aps.org/doi/10.1103/PhysRevLett.114.156402}
}

@Article{Krien2017,
  Title                    = {Conservation in two-particle self-consistent extensions of dynamical mean-field theory},
  Author                   = {Krien, Friedrich and van Loon, Erik G. C. P. and Hafermann, Hartmut and Otsuki, Junya and Katsnelson, Mikhail I. and Lichtenstein, Alexander I.},
  Journal                  = {Phys. Rev. B},
  Year                     = {2017},

  Month                    = {Aug},
  Pages                    = {075155},
  Volume                   = {96},

  Doi                      = {10.1103/PhysRevB.96.075155},
  Issue                    = {7},
  Numpages                 = {17},
  Publisher                = {American Physical Society},
  Url                      = {https://link.aps.org/doi/10.1103/PhysRevB.96.075155}
}

@book{Senechal2004Book,
    author ={David S\'en\'echal, Andr\'e-Marie Tremblay, Claude Bourbonnais},
    title = {Theoretical Methods for Strongly Correlated Electrons},
    publisher = {Springer, New York},
    year = {2004}
}

@misc{Essl2025,
  Title                    = {{How to stay on the physical branch in self-consistent many-electron approaches}},
  Author                   = {E\ss{}l, Herbert and Reitner, Matthias and Kozik, Evgeny and Toschi, Alessandro},
  Year                     = {2025},
  Archiveprefix            = {arXiv},
  Eprint                   = {2502.01420},
  Url                      = {https://arxiv.org/pdf/2502.01420}
}

@Article{Lichtenstein2000,
  Title                    = {{Antiferromagnetism and \textit{d}-wave superconductivity in cuprates: A cluster dynamical mean-field theory}},
  Author                   = {Lichtenstein, A. I. and Katsnelson, M. I.},
  Journal                  = {Phys. Rev. B},
  Year                     = {2000},

  Month                    = {Oct},
  Pages                    = {R9283--R9286},
  Volume                   = {62},

  Doi                      = {10.1103/PhysRevB.62.R9283},
  Issue                    = {14},
  Numpages                 = {0},
  Publisher                = {American Physical Society},
  Url                      = {http://link.aps.org/doi/10.1103/PhysRevB.62.R9283}
}

@Article{Lichtenstein2001,
  Title                    = {Finite-Temperature Magnetism of Transition Metals: An {\sl ab initio} Dynamical Mean-Field Theory},
  Author                   = {A. I. Lichtenstein and M. I. Katsnelson and G. Kotliar},
  Journal                  = {Phys. Rev. Lett.},
  Year                     = {2001},
  Pages                    = {067205},
  Volume                   = {87}
}

@article{Ritz2025,
   title={{Testing the parquet equations and the U(1) Ward identity for real-frequency correlation functions from the multipoint numerical renormalization group}},
   volume={7},
   ISSN={2643-1564},
   url={http://dx.doi.org/10.1103/3jtq-5wf5},
   DOI={10.1103/3jtq-5wf5},
   number={3},
   journal={Physical Review Research},
   publisher={American Physical Society (APS)},
   author={Ritz, Nepomuk and Ge, Anxiang and Frankenbach, Markus and Pelz, Mathias and von Delft, Jan and Kugler, Fabian B.},
   year={2025},
   month=aug
}

@article{vanLoon2024-2,
  title = {{Dual Bethe-Salpeter equation for the multiorbital lattice susceptibility within dynamical mean-field theory}},
  author = {van Loon, Erik G. C. P. and Strand, Hugo U. R.},
  journal = {Phys. Rev. B},
  volume = {109},
  issue = {15},
  pages = {155157},
  numpages = {15},
  year = {2024},
  month = {Apr},
  publisher = {American Physical Society},
  doi = {10.1103/PhysRevB.109.155157},
  url = {https://link.aps.org/doi/10.1103/PhysRevB.109.155157}
}

@article{vanLoon2020,
  title     = {Bethe-Salpeter Equation at the Critical End Point of the Mott Transition},
  author    = {van Loon, Erik G. C. P. and Krien, Friedrich and Katanin, Andrey A.},
  journal   = {Phys. Rev. Lett.},
  volume    = {125},
  issue     = {13},
  pages     = {136402},
  numpages  = {8},
  year      = {2020},
  month     = {Sep},
  publisher = {American Physical Society},
  doi       = {10.1103/PhysRevLett.125.136402},
  url       = {https://link.aps.org/doi/10.1103/PhysRevLett.125.136402}
}

@article{vanLoon2022,
  title     = {Two-particle correlations and the metal-insulator transition: Iterated perturbation theory revisited},
  author    = {van Loon, Erik G. C. P.},
  journal   = {Phys. Rev. B},
  volume    = {105},
  issue     = {24},
  pages     = {245104},
  numpages  = {14},
  year      = {2022},
  month     = {Jun},
  publisher = {American Physical Society},
  doi       = {10.1103/PhysRevB.105.245104},
  url       = {https://link.aps.org/doi/10.1103/PhysRevB.105.245104}
}

@article{vanLoon2024,
  title     = {Second-order phase transitions and divergent linear response in dynamical mean-field theory},
  author    = {van Loon, Erik G. C. P.},
  journal   = {Phys. Rev. B},
  volume    = {109},
  issue     = {24},
  pages     = {L241110},
  numpages  = {7},
  year      = {2024},
  month     = {Jun},
  publisher = {American Physical Society},
  doi       = {10.1103/PhysRevB.109.L241110},
  url       = {https://link.aps.org/doi/10.1103/PhysRevB.109.L241110}
}

@Article{Maier2005,
  Title                    = {{Quantum Cluster Theories}},
  Author                   = {T. A. Maier and M. Jarrell and T. Pruschke and M. Hettler},
  Journal                  = {Rev. Mod. Phys.},
  Year                     = {2005},

  Month                    = {Oct},
  Pages                    = {1027},
  Volume                   = {77},

  Doi                      = {10.1103/RevModPhys.77.1027},
  Issue                    = {3},
  Publisher                = {American Physical Society},
  Url                      = {http://link.aps.org/doi/10.1103/RevModPhys.77.1027}
}

@article{Arovas2022,
author = {Arovas, Daniel P. and Berg, Erez and Kivelson, Steven A. and Raghu, Srinivas},
title = {{The Hubbard Model}},
journal = {{Annual Review of Condensed Matter Physics}},
volume = {13},
number = {1},
year = {2022},
doi = {10.1146/annurev-conmatphys-031620-102024},

URL = { 
        https://doi.org/10.1146/annurev-conmatphys-031620-102024
    
},
eprint = { 
        https://doi.org/10.1146/annurev-conmatphys-031620-102024
    
}
}

@Article{Mermin1966,
  Title                    = {{Absence of Ferromagnetism or Antiferromagnetism in One- or Two-Dimensional Isotropic Heisenberg Models}},
  Author                   = {Mermin, N. D. and Wagner, H.},
  Journal                  = {Phys. Rev. Lett.},
  Year                     = {1966},

  Month                    = {Dec},
  Pages                    = {1307--1307},
  Volume                   = {17},

  Doi                      = {10.1103/PhysRevLett.17.1307},
  Issue                    = {26},
  Numpages                 = {0},
  Publisher                = {American Physical Society},
  Url                      = {http://link.aps.org/doi/10.1103/PhysRevLett.17.1307}
}

@Article{Metzner1989,
  Title                    = {{Correlated Lattice Fermions in $d=\infty$ Dimensions}},
  Author                   = {Metzner, Walter and Vollhardt, Dieter},
  Journal                  = {Phys. Rev. Lett.},
  Year                     = {1989},

  Month                    = {Jan},
  Pages                    = {324--327},
  Volume                   = {62},

  Doi                      = {10.1103/PhysRevLett.62.324},
  Numpages                 = {3},
  Publisher                = {American Physical Society},
 URL                   = {http://link.aps.org/doi/10.1103/PhysRevLett.62.324}
}

@Article{Park2008,
  Title                    = {{Cluster Dynamical Mean Field Theory of the Mott Transition}},
  Author                   = {Park, H. and Haule, K. and Kotliar, G.},
  Journal                  = {Phys. Rev. Lett.},
  Year                     = {2008},

  Month                    = {Oct},
  Pages                    = {186403},
  Volume                   = {101},

  Doi                      = {10.1103/PhysRevLett.101.186403},
  Issue                    = {18},
  Numpages                 = {4},
  Publisher                = {American Physical Society},
  Url                      = {http://link.aps.org/doi/10.1103/PhysRevLett.101.186403}
}

@Article{Potthoff2003,
  Title                    = {Self-energy-functional approach to systems of correlated electrons},
  Author                   = {Potthoff, Michael},
  Journal                  = {Eur. Phys. J. B},
  Year                     = {2003},
  Pages                    = {429},
  Volume                   = {32},
  Doi                      = {10.1140/epjb/e2003-00121-8},
  Numpages                 = {7}
}

@incollection{Potthoff2018,
  author    = {Potthoff, Michael},
  title     = {Cluster Extensions of Dynamical Mean-Field Theory},
  booktitle = {{DMFT}:{F}rom {I}nfinite {D}imensions to {R}eal {M}aterials},
  volume    = {8},
  publisher = {Forschungszentrum Jülich},
  isbn      = {978-3-95806-313-6},
  year      = {2018},
  chapter   = {5},
  url       = {http://hdl.handle.net/2128/19720}
}

@article{Walsh2023,
  title = {{Superconductivity in the two-dimensional Hubbard model with cellular dynamical mean-field theory: A quantum impurity model analysis}},
  author = {Walsh, C. and Charlebois, M. and S\'emon, P. and Tremblay, A.-M. S. and Sordi, G.},
  journal = {Phys. Rev. B},
  volume = {108},
  issue = {7},
  pages = {075163},
  numpages = {17},
  year = {2023},
  month = {Aug},
  publisher = {American Physical Society},
  doi = {10.1103/PhysRevB.108.075163},
  url = {https://link.aps.org/doi/10.1103/PhysRevB.108.075163}
}

@article{Sordi2019,
  title = {{Specific heat maximum as a signature of Mott physics in the two-dimensional Hubbard model}},
  author = {Sordi, G. and Walsh, C. and S\'emon, P. and Tremblay, A.-M. S.},
  journal = {Phys. Rev. B},
  volume = {100},
  issue = {12},
  pages = {121105},
  numpages = {6},
  year = {2019},
  month = {Sep},
  publisher = {American Physical Society},
  doi = {10.1103/PhysRevB.100.121105},
  url = {https://link.aps.org/doi/10.1103/PhysRevB.100.121105}
}

@PhdThesis{Rohringer2013a,
  Title                    = {New routes towards a theoretical treatment of nonlocal electronic correlations},
  Author                   = {Georg Rohringer},
  School                   = {Vienna University of Technology},
  Year                     = {2013},
  url                      = {http://digital.obvsg.at/download/pdf/1631831}
}

@Article{Rohringer2018,
  Title                    = {{LadderD$\Gamma$A code}},
  Author                   = {Rohringer, G. and Katanin, A. and Sch{\"a}fer, T. and Hausoel, A. and Held, K. and Toschi, A.},
  Year                     = {2018},
  Journal                  = {github.com/ladderDGA},
  Url                      = {https://github.com/ladderDGA/ladderDGA}
}

@article{Reitner2020,
   title={{Attractive Effect of a Strong Electronic Repulsion: The Physics of Vertex Divergences}},
   volume={125},
   ISSN={1079-7114},
   url={http://dx.doi.org/10.1103/PhysRevLett.125.196403},
   DOI={10.1103/physrevlett.125.196403},
   number={19},
   journal={Physical Review Letters},
   publisher={American Physical Society (APS)},
   author={Reitner, M. and Chalupa, P. and Del Re, L. and Springer, D. and Ciuchi, S. and Sangiovanni, G. and Toschi, A.},
   year={2020},
   month={Nov}
}

@misc{Reymbaut2020,
      title={{Mott transition and high-temperature crossovers at half-filling}}, 
      author={A. Reymbaut and M. Boulay and L. Fratino and P. Sémon and Wei Wu and G. Sordi and A. -M. S. Tremblay},
      year={2020},
      eprint={2004.02302},
      archivePrefix={arXiv},
      primaryClass={cond-mat.str-el},
      doi={https://doi.org/10.48550/arXiv.2004.02302}
}

@Article{Rohringer2013,
  Title                    = {One-particle irreducible functional approach: A route to diagrammatic extensions of the dynamical mean-field theory},
  Author                   = {G. Rohringer and A. Toschi and H. Hafermann and K. Held and V. I. Anisimov and A. A. Katanin},
  Journal                  = {Phys. Rev. B},
  Year                     = {2013},
  Pages                    = {115112},
  Volume                   = {88},
  Url                      = {http://link.aps.org/doi/10.1103/PhysRevB.88.115112}
}

@Article{Rohringer2012,
  Title                    = {Local electronic correlation at the two-particle level},
  Author                   = {Rohringer, G. and Valli, A. and Toschi, A.},
  Journal                  = {Phys. Rev. B},
  Year                     = {2012},

  Month                    = {Sep},
  Pages                    = {125114},
  Volume                   = {86},

  Doi                      = {10.1103/PhysRevB.86.125114},
  Issue                    = {12},
  Numpages                 = {26},
  Publisher                = {American Physical Society},
  Url                      = {http://link.aps.org/doi/10.1103/PhysRevB.86.125114}
}

@Article{Rossi2016,
  Title                    = {{Shifted-action expansion and applicability of dressed diagrammatic schemes}},
  Author                   = {Rossi, Riccardo and Werner, F\'elix and Prokof'ev, Nikolay and Svistunov, Boris},
  Journal                  = {Phys. Rev. B},
  Year                     = {2016},

  Month                    = {Apr},
  Pages                    = {161102(R)},
  Volume                   = {93},

  Doi                      = {10.1103/PhysRevB.93.161102},
  Issue                    = {16},
  Numpages                 = {5},
  Publisher                = {American Physical Society},
  Url                      = {https://link.aps.org/doi/10.1103/PhysRevB.93.161102}
}

@Article{Rubtsov2005,
  Title                    = {{Continuous-time quantum Monte Carlo method for fermions}},
  Author                   = {Rubtsov, A. N. and Savkin, V. V. and Lichtenstein, A. I.},
  Journal                  = {Phys. Rev. B},
  Year                     = {2005},

  Month                    = {Jul},
  Pages                    = {035122},
  Volume                   = {72},

  Doi                      = {10.1103/PhysRevB.72.035122},
  Issue                    = {3},
  Numpages                 = {9},
  Publisher                = {American Physical Society},
  Url                      = {http://link.aps.org/doi/10.1103/PhysRevB.72.035122}
}

@article{Krishna-murthy1975,
  title = {{Temperature-Dependent Susceptibility of the Symmetric Anderson Model: Connection to the Kondo Model}},
  author = {Krishna-murthy, H. R. and Wilson, K. G. and Wilkins, J. W.},
  journal = {Phys. Rev. Lett.},
  volume = {35},
  issue = {16},
  pages = {1101--1104},
  numpages = {0},
  year = {1975},
  month = {Oct},
  publisher = {American Physical Society},
  doi = {10.1103/PhysRevLett.35.1101},
  url = {https://link.aps.org/doi/10.1103/PhysRevLett.35.1101}
}

@article{Krishna-murthy1980,
  title = {{Renormalization-group approach to the Anderson model of dilute magnetic alloys. I. Static properties for the symmetric case}},
  author = {Krishna-murthy, H. R. and Wilkins, J. W. and Wilson, K. G.},
  journal = {Phys. Rev. B},
  volume = {21},
  issue = {3},
  pages = {1003--1043},
  numpages = {0},
  year = {1980},
  month = {Feb},
  publisher = {American Physical Society},
  doi = {10.1103/PhysRevB.21.1003},
  url = {https://link.aps.org/doi/10.1103/PhysRevB.21.1003}
}

@Article{Schaefer2016c,
  Title                    = {{Non-perturbative landscape of the Mott-Hubbard transition: Multiple divergence lines around the critical endpoint}},
  Author                   = {Sch\"afer, T. and Ciuchi, S. and Wallerberger, M. and Thunstr{\"o}m, P. and Gunnarsson, O. and Sangiovanni, G. and Rohringer, G. and Toschi, A.},
  Journal                  = {Phys. Rev. B},
  Year                     = {2016},

  Month                    = {Dec},
  Pages                    = {235108},
  Volume                   = {94},

  Doi                      = {10.1103/PhysRevB.94.235108},
  Issue                    = {23},
  Numpages                 = {25},
  Publisher                = {American Physical Society},
  Url                      = {http://link.aps.org/doi/10.1103/PhysRevB.94.235108}
}

@Article{Sordi2011,
  Title                    = {Mott physics and first-order transition between two metals in the normal-state phase diagram of the two-dimensional Hubbard model},
  Author                   = {Sordi, G. and Haule, K. and Tremblay, A.-M. S.},
  Journal                  = {Phys. Rev. B},
  Year                     = {2011},

  Month                    = {Aug},
  Pages                    = {075161},
  Volume                   = {84},

  Doi                      = {10.1103/PhysRevB.84.075161},
  Issue                    = {7},
  Numpages                 = {25},
  Publisher                = {American Physical Society},
  Url                      = {http://link.aps.org/doi/10.1103/PhysRevB.84.075161}
}

@Article{Stan2015,
  Title                    = {Unphysical and physical solutions in many-body theories: from weak to strong correlation},
  Author                   = {Stan, A. and Romaniello, P. and Rigamonti, S. and Reining, L. and Berger, J. A.},
  Journal                  = {New J. Phys.},
  Year                     = {2015},
  Number                   = {9},
  Pages                    = {093045},
  Volume                   = {17},

  Url                      = {http://stacks.iop.org/1367-2630/17/i=9/a=093045}
}

@Article{Tagliavini2018,
  Title                    = {Efficient Bethe-Salpeter equations' treatment in dynamical mean-field theory},
  Author                   = {Tagliavini, A. and Hummel, S. and Wentzell, N. and Andergassen, S. and Toschi, A. and Rohringer, G.},
  Journal                  = {arXiv:1803.03036},
  Year                     = {2018},

  Adsnourl                 = {https://arxiv.org/abs/1803.03036}
}

@PhdThesis{Krien2018,
  Title                    = {{Conserving dynamical mean-field approaches to strongly correlated systems}},
  Author                   = {Krien, Friedrich},
  School                   = {{Universit\"at Hamburg}},
  Year                     = {2018}
}

@article{Ward1950,
  title = {{An Identity in Quantum Electrodynamics}},
  author = {Ward, J. C.},
  journal = {Phys. Rev.},
  volume = {78},
  issue = {2},
  pages = {182--182},
  numpages = {0},
  year = {1950},
  month = {Apr},
  publisher = {American Physical Society},
  doi = {10.1103/PhysRev.78.182},
  url = {https://link.aps.org/doi/10.1103/PhysRev.78.182}
}

@article{Behn1978,
  title = {{On the Ward Identities for the Hubbard Model. Particle-Number Conservation and the Analytical Structure of the Self-Energy}},
  author = {Behn, U.},
  journal = {Phys. Stat. Sol. B},
  volume = {88},
  issue = {2},
  pages = {699-704},
  year = {1978},
  month = {Apr},
  publisher = {Wiley-VCH},
  doi = {10.1002/pssb.2220880237},
  url = {https://doi.org/10.1002/pssb.2220880237}
}

@article{Chalupa2021,
  title = {{Fingerprints of the Local Moment Formation and its Kondo Screening in the Generalized Susceptibilities of Many-Electron Problems}},
  author = {Chalupa, P. and Sch\"afer, T. and Reitner, M. and Springer, D. and Andergassen, S. and Toschi, A.},
  journal = {Phys. Rev. Lett.},
  volume = {126},
  issue = {5},
  pages = {056403},
  numpages = {9},
  year = {2021},
  month = {Feb},
  publisher = {American Physical Society},
  doi = {10.1103/PhysRevLett.126.056403},
  url = {https://link.aps.org/doi/10.1103/PhysRevLett.126.056403}
}

@Article{Vucicevic2018,
  Title                    = {{Practical consequences of the Luttinger-Ward functional multivaluedness for cluster DMFT methods}},
  Author                   = {Vu\v{c}i\v{c}evi\'c, J. and Wentzell, N. and Ferrero, M. and Parcollet, O.},
  Journal                  = {Phys. Rev. B},
  Year                     = {2018},

  Month                    = {Mar},
  Pages                    = {125141},
  Volume                   = {97},

  Doi                      = {10.1103/PhysRevB.97.125141},
  Issue                    = {12},
  Numpages                 = {24},
  Publisher                = {American Physical Society},
  Url                      = {https://link.aps.org/doi/10.1103/PhysRevB.97.125141}
}

@Book{Kopietz2010,
  Title                    = {{Introduction to the Functional Renormalization Group, Lecture Notes in Physics 798}},
  Author                   = {Kopietz, Peter and 
Bartosch, Lorenz and Sch{\"u}tz, Florian},
  Publisher                = {Springer-Verlag Berlin Heidelberg},
  Year                     = {2010},
  ISBN = {978-3-642-05093-0},
  DOI = {10.1007/978-3-642-05094-7}
}

@mastersthesis{Reitner2020DA,
  title  = {Dychotomy between local and uniform compressibility in correlated systems: the role of the irreducible vertex},
  author = {Reitner, Matthias},
  school = {Technischen Universit{\"a}t Wien},
  year   = {2020},
  type   = {Master thesis},
  doi = {10.34726/hss.2020.67262}
}

@article{Tarantino2018,
  doi       = {10.1088/1361-648x/aaaeab},
  url       = {https://doi.org/10.1088/1361-648x/aaaeab},
  year      = 2018,
  month     = {mar},
  publisher = {{IOP} Publishing},
  volume    = {30},
  number    = {13},
  pages     = {135602},
  author    = {Walter Tarantino and Bernardo S Mendoza and Pina Romaniello and J A Berger and Lucia Reining},
  title     = {Many-body perturbation theory and non-perturbative approaches: screened interaction as the key ingredient},
  journal   = {Journal of Physics: Condensed Matter}
}

@article{Reitner2025,
  title = {{Nonperturbative feats in the physics of correlated antiferromagnets}},
  author = {Reitner, M. and Del Re, L. and Capone, M. and Toschi, A.},
  journal = {Phys. Rev. Res.},
  volume = {7},
  issue = {3},
  pages = {033264},
  numpages = {20},
  year = {2025},
  month = {Sep},
  publisher = {American Physical Society},
  doi = {10.1103/ympr-9m73},
  url = {https://link.aps.org/doi/10.1103/ympr-9m73}
}

@article{Springer2020,
  title     = {Interplay between local response and vertex divergences in many-fermion systems with on-site attraction},
  author    = {Springer, D. and Chalupa, P. and Ciuchi, S. and Sangiovanni, G. and Toschi, A.},
  journal   = {Phys. Rev. B},
  volume    = {101},
  issue     = {15},
  pages     = {155148},
  numpages  = {14},
  year      = {2020},
  month     = {Apr},
  publisher = {American Physical Society},
  doi       = {10.1103/PhysRevB.101.155148},
  url       = {https://link.aps.org/doi/10.1103/PhysRevB.101.155148}
}

@article{Essl2024,
  title     = {{General Shiba mapping for on-site four-point correlation functions}},
  author    = {E\ss{}l, Herbert and Reitner, Matthias and Sangiovanni, Giorgio and Toschi, Alessandro},
  journal   = {Phys. Rev. Res.},
  volume    = {6},
  issue     = {3},
  pages     = {033061},
  numpages  = {23},
  year      = {2024},
  month     = {Jul},
  publisher = {American Physical Society},
  doi       = {10.1103/PhysRevResearch.6.033061},
  url       = {https://link.aps.org/doi/10.1103/PhysRevResearch.6.033061}
}

@article{Strand2011,
  title     = {{Dynamical mean field theory phase-space extension and critical properties of the finite temperature Mott transition}},
  author    = {Strand, Hugo U. R. and Sabashvili, Andro and Granath, Mats and Hellsing, Bo and \"Ostlund, Stellan},
  journal   = {Phys. Rev. B},
  volume    = {83},
  issue     = {20},
  pages     = {205136},
  numpages  = {10},
  year      = {2011},
  month     = {May},
  publisher = {American Physical Society},
  doi       = {10.1103/PhysRevB.83.205136},
  url       = {https://link.aps.org/doi/10.1103/PhysRevB.83.205136}
}

@article{Walsh2019,
  title     = {{Thermodynamic and information-theoretic description of the Mott transition in the two-dimensional Hubbard model}},
  author    = {Walsh, C. and S\'emon, P. and Poulin, D. and Sordi, G. and Tremblay, A.-M. S.},
  journal   = {Phys. Rev. B},
  volume    = {99},
  issue     = {7},
  pages     = {075122},
  numpages  = {20},
  year      = {2019},
  month     = {Feb},
  publisher = {American Physical Society},
  doi       = {10.1103/PhysRevB.99.075122},
  url       = {https://link.aps.org/doi/10.1103/PhysRevB.99.075122}
}
\end{document}